\documentclass[useAMS,usenatbib]{mn2e}
\usepackage{graphicx}    
\usepackage{subfig}
\usepackage{amssymb}

\DeclareGraphicsExtensions{.ps,.pdf,.png}
\makeatletter
\def\fps@figure{htbp}
\makeatother
\def \aap  {A\&A}

\def \aj  {AJ}
\def \apj  {ApJ}
\def \apjs  {ApJS}
\def \araa  {ARA\&A}

\def \mnras {MNRAS}
\def \nat {Nature}
\def\apjl {ApJL}
\begin{document}
\title[Evolution of Massive Galaxy Structural Properties and Sizes via Star Formation]{Evolution of Massive Galaxy Structural Properties and Sizes via Star Formation In the GOODS NICMOS Survey}
\author[Ownsworth et al.]{Jamie~R.~Ownsworth$^{1}$\thanks{E-mail: ppxjo1@nottingham.ac.uk}, Christopher~J.~Conselice$^1$, Alice~Mortlock$^{1}$, \newauthor
William~G.~Hartley$^{1}$, Fernando~Buitrago$^{1,2}$ \\
$^{1}$University of Nottingham, School of Physics and Astronomy, Nottingham, NG7 2RD, U.K. \\
$^{2}$SUPA\thanks{Scottish Universities Physics Alliance}, Institute for Astronomy, University of Edinburgh, Royal Observatory, Edinburgh, EH9 3HJ, U.K.}
\date{Accepted ??. Received ??; in original form ??}
\pagerange{\pageref{firstpage}--\pageref{lastpage}} \pubyear{2012}
\maketitle

\label{firstpage}

\begin{abstract}
We present a study of the resolved star-forming properties of a sample of distant massive ($M_{\*} > 10^{11}M_{\odot}$) galaxies in the GOODS NICMOS Survey (GNS), based on deep Hubble Space Telescope imaging from the GOODS North and South fields. We derive dust corrected UV star formation rates (SFRs) as a function of radius for 45 massive galaxies within the redshift range $1.5 < z < 3$ in order to measure the spatial location of ongoing star formation in massive galaxies. We find that the star formation rates present in different regions of a galaxy reflect the already existent stellar mass density, i.e. high density regions have higher star formation rates than lower density regions, on average. This observed star formation is extrapolated in several ways to the present day, and we measure the amount of new stellar mass that is created in individual portions of each galaxy to determine how the stellar mass added via star formation changes the observed stellar mass profile, the S\'{e}rsic index and effective radius over time. We find that these massive galaxies fall into three broad classifications of star formation distribution: 1) Total stellar mass added via star formation is insignificant compared to the stellar mass that is already in place at high redshift. 2) Stellar mass added via star formation is only significant in the outer regions ($R > 1\rm{kpc})$ of the galaxy. 3) Stellar mass added via star formation is significant in both the inner $(R < 1\rm{kpc})$ and outer regions of the galaxy.  These different star formation distributions increase the effective radii over time, which are on average a factor of $\sim16\pm5\%$ larger, with little change in the S\'{e}rsic index (average $\Delta n = -0.9\pm0.9$) after evolution. We also implement a range of simple stellar migration models into the simulated evolutionary path of these galaxies in order to gauge its effect on the properties of our sample. This yields a larger increase in the evolved effective radii than the pure static star formation model, with a maximum average increase of $\Delta R_{e}\sim 54\pm 19\%$, but with little change in the S\'{e}rsic index, $\Delta n \sim-1.1\pm1.3$. These results are not in agreement with the observed change in the effective radius and S\'{e}rsic index between $z\sim2.5$ and $z\sim0$  obtained via various observational studies. We conclude that star formation and stellar migration alone cannot account for the observed change in structural parameters for this galaxy population, implying that other mechanisms must additionally be at work to produce the evolution, such as merging.
\end{abstract}


\section{Introduction}
\label{sec:Intro}
One of the least understood aspects of galaxy evolution is the star formation rates in galaxies, how these vary across individual galaxies, and influence galaxy properties. A key way to address galaxy evolution directly is to understand how the nearby galaxy population was put into place and evolved from higher redshift galaxies, which we can now observe in near complete mass-selected samples up to $z=3$ (e.g. Daddi et al. 2007; Conselice et al. 2011). One major finding of high redshift studies is that massive galaxies $(M_{*} > 10^{11}M_{\odot})$ have significantly smaller effective radii than low redshift galaxies of similar mass (e.g. Daddi et al 2005; Trujillo et al. 2006a, 2006b, 2007, 2011; Buitrago et al. 2008, 2011; Cimatti et al. 2008; van Dokkum et al. 2008, 2010; Franx et al. 2008 ; van der Wel et al. 2008; Damjanov et al. 2009; Carrasco et al. 2010; Newman et al. 2010; Szomoru et al. 2011; Weinzirl et al. 2011). 

Several physical processes have been proposed to explain this strong size evolution within the massive galaxy population at $z < 2$. These can be divided into two distinct categories, external processes such as gas poor (“dry”) mergers (e.g. Khochfar \& Silk 2006; Naab et al. 2009) and cold gas flows along cosmic web filaments (e.g. Dekel et al. 2009; Conselice et al. 2012 submitted) as a means for puffing up the stellar components of these massive galaxies, or internal processes such as adiabatic expansion resulting from stellar mass loss and strong AGN-fuelled feedback (e.g. Fan et al. 2008, 2010; Hopkins et al. 2010A; Bluck et al. 2011). One process that has not been looked at in detail is the internal star formation distribution present within massive galaxies at high redshift, and whether this can account for the observed structural evolution. This can now be examined due to high resolution data from the GOODS NICMOS Survey taken with the ACS and NICMOS-3 instruments on the Hubble Space Telescope (Conselice at al. 2011). 

We know that galaxies evolve significantly in stellar mass from observational studies showing that half of the stellar mass of present day galaxies is already in place by $z \sim1$ (e.g. Brinchmann \& Ellis 2000; Drory et al. 2004; Bundy et al. 2006; P\'{e}rez-Gonz\'{a}lez et al. 2008; Mortlock et al. 2011). The most massive galaxies $(M_{*} > 10^{11}M_{\odot})$ appear on average to have red rest-frame colours which we expect to see for galaxies dominated by old stellar populations (Saracco et al. 2005; Labb\'{e} et al. 2006; Conselice et al. 2007; Gr\"{u}tzbauch et al. 2011). However, Bauer et al. (2011b) show that $\sim80\%$ of these massive red galaxies likely harbour dusty star formation. This star formation over cosmic time could contribute large amounts of stellar mass to massive galaxies, and depending on where this mass is created could affect their observable structural properties as they evolve.

In the merger scenario, estimates for the total number of major mergers experienced by massive galaxies since $z=3$ is $N_{m} = 1.7 \pm 0.5$ (Bluck et al. 2009). This would imply an average stellar mass increase of, at best, a factor of two due to major mergers. However over the same period of time the effective radius of massive galaxies has increased on average by a factor of three for disk\--like galaxies, and a factor of five for spheroid\--like galaxies, effectively building up stellar mass in the outer regions of galaxies (see e.g. Buitrago et al. 2008, 2011; Trujillo et al. 2007; Carrasco et al. 2010; van Dokkum et al. 2010). This additional stellar mass could arise from star formation already present at high redshift within these outer regions.

To date studies have only looked at the total star formation rates of these galaxies as a whole (e.g. P\'{e}rez-Gonz\'{a}lez et al. 2008; Cava et al. 2010; van Dokkum et al. 2010; Bauer et al. 2011; Gr\"{u}zbauch et al. 2011; Viero et al. 2011; Hilton et al. 2012), but have not examined the locations of the star formation within these galaxies. Thus, we combine the observed stellar mass profiles with the observed star formation profiles of high redshift massive galaxies in order to measure the effect stellar mass added via star formation over $\sim 10$ Gyr has on different spatial regions, and to the total stellar mass profile. We also ascertain whether this star formation can account for the observed size evolution.

Along with size evolution within the massive galaxy population there is also a change in overall morphology. The present day universe is populated by massive galaxies with early\--type morphologies (e.g. Baldry et al 2004, Conselice et al. 2006). At earlier epochs, $z>1.5$, observational studies have found that the massive galaxy population is dominated by galaxies with late\--type morphologies (e.g. Buitrago et al. 2011; Cameron et al. 2011; van der Wel et al. 2011; Weinzirl et al. 2011). This morphological shift can be seen via a change in S\'{e}rsic index from low values, $n < \sim2.5$ denoting a possible late\--type morphology, to high values, $n > \sim2.5$ denoting a possible early\--type morphology. In the hierarchical model of galaxy evolution there are many methods that can drive morphological evolution. These methods include in situ star formation producing disk\--like systems (e.g. Dekel et al. 2009; Oser et al. 2010; Ricciardelli et al. 2010; Wuyts et al. 2010; Bournaud et al. 2011), and/or mergers with satellite galaxies producing a more spheroid\--like system (e.g. Khochfar \& Silk 2006; Hopkins et al. 2009; Feldmann et al. 2010; Oser et al. 2010). We therefore also investigate how in situ star formation over cosmic time changes the S\'{e}rsic index of the massive galaxies, and ascertain whether this process can account for the observed morphological changes.

This paper is set out as follows: Section 2 discusses the GOODS NICMOS Survey, the galaxy sample, and how the data used in this paper was obtained. Section 3.1 examines the stellar mass radial density distributions of the massive galaxies. Section 3.2 describes how the stellar mass density added via star formation is calculated. In Section 3.3 we examine the evolved galaxy profiles. Section 4.1 presents the findings of how the structure and size of the massive galaxies is altered by star formation. In Section 4.2 we introduce a simple stellar migration model to the stellar mass added by star formation in order to gauge the effect this has on structures and sizes. Section 5 and 6 contain the discussion and summary of our findings, respectively. Throughout this paper we assume $\Omega_{M}=0.3$, $\Omega_{\Lambda}=0.7$ and $H_{0}=70$ km s$^{-1}$ Mpc$^{-1}$. AB magnitudes and a Salpeter IMF are
used throughout.

\section{Data and Analysis}
\label{sec:Data}

In this section we describe the survey we  use in this study, the GOODS NICMOS Survey (GNS), as well as the measurements of photometric redshifts, stellar masses, rest-frame colours and star formation rates for our galaxies.

\subsection{The GOODS NICMOS Survey}
\label{sec:GNS}

The data used in this paper is obtained through the GOODS NICMOS Survey (GNS). The GNS is a 180 orbit Hubble Space Telescope survey consisting of 60 single pointings with the NICMOS-3 near-infrared camera, with an imaging depth of three orbits per pointing (Conselice et al. 2011).

These pointings were optimised to contain the maximum number of massive galaxies $(M>10^{11}M_{\odot})$ in the redshift range $1.7<z<3$, identified in the two GOODS fields by their optical-to-infrared colours (see Conselice et al 2011). The survey covers a total area of about 45 arcmin$^{2}$ with a pixel scale of $\sim0.1$ arcsec/pixel, corresponding to $\sim$0.9 kpc at the redshift range of interest $(1.5<z<3)$. The target selection, survey characteristics and data reduction are fully described in Conselice et al. (2011). Other analyses of the GNS data set can be found in Bauer et al. (2011), Bluck et al. (2010), Buitrago et al. (2008), Gr\"{u}tzbauch et al. (2011) and Mortlock et al. (2011).

The GNS has a $5\sigma$ limiting magnitude of $H_{AB}= 26.8$, which is significantly deeper than ground based near-infrared imaging of the GOODS fields carried out with e.g. ISAAC on the VLT, which reaches a $5\sigma$ depth of $H_{AB}=24.5$ (Retzlaff et al 2010). Sources were extracted from the NICMOS $H_{160}$-band image and matched to the optical HST-ACS bands B,V,$i$ and $z$, which are available down to a AB limiting magnitude of $B = 28.2$. The matching is done within a radius of $2$ arcsec, however the average separation between optical and $H_{160}$-band coordinates is much better with $\sim0.28 \pm 0.4$ arcsec, roughly corresponding to the NICMOS resolution (see also Bauer et al. 2011b).
 
The photometric catalogue covering the Bv$iz$H bands comprises 8298 galaxies, and is used to compute photometric redshifts, rest-frame colours and stellar masses described in the following sections (see also Conselice et al. 2011 for more details). Along with this, each galaxy has imaging data in the Bv$iz$ ACS bands (Giavalisco et al. 2004).

\begin{figure*}
\includegraphics[scale=0.27]{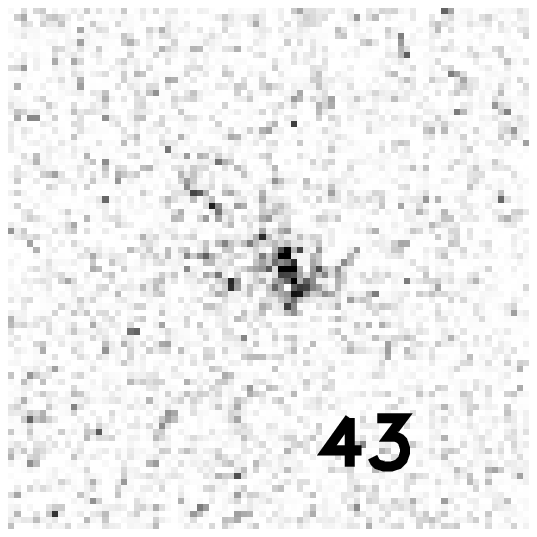}
\includegraphics[scale=0.27]{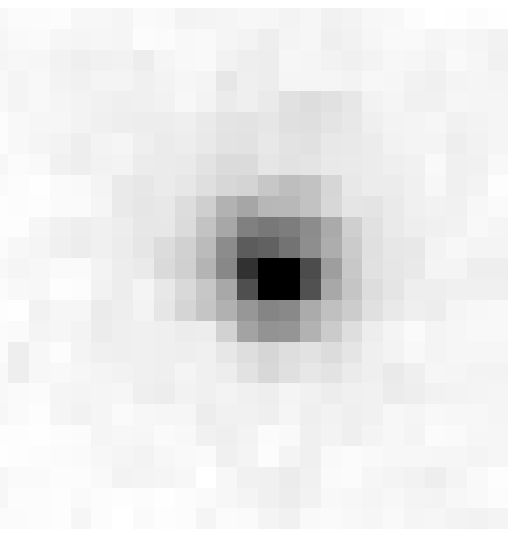}
\vline 
\includegraphics[scale=0.27]{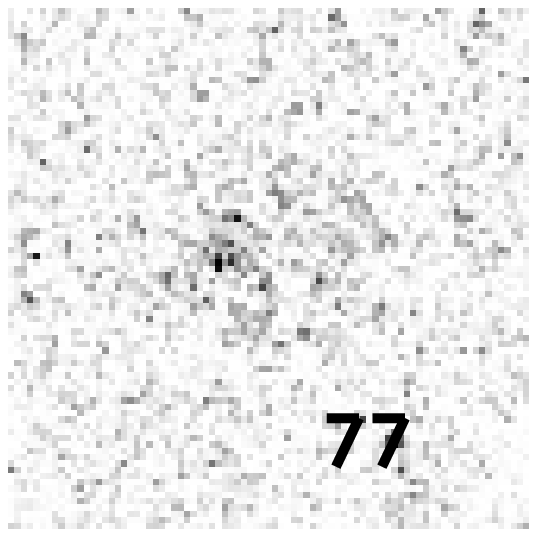}
\includegraphics[scale=0.27]{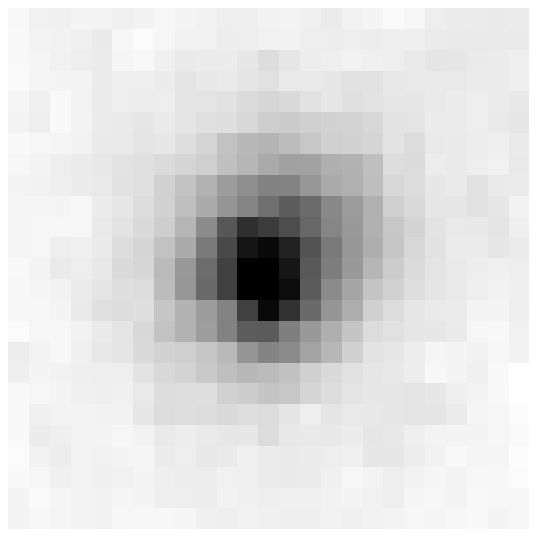}
\vline 
\includegraphics[scale=0.27]{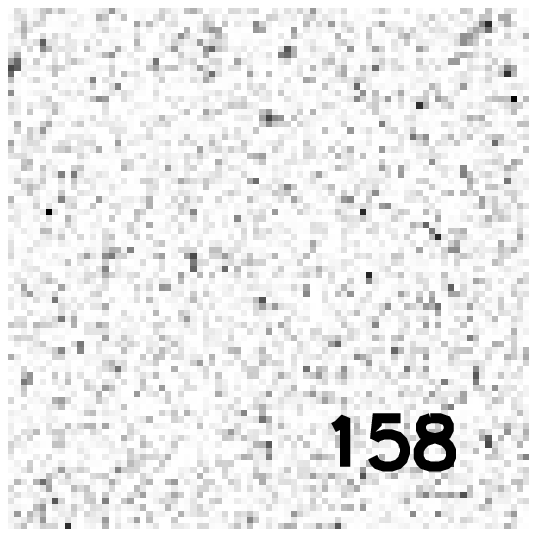}
\includegraphics[scale=0.27]{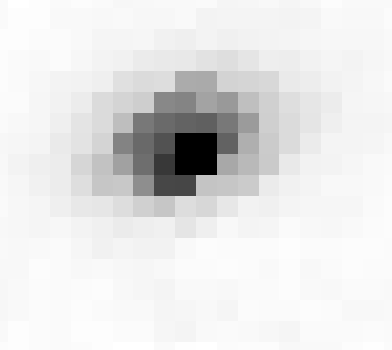}
\vline 
\includegraphics[scale=0.27]{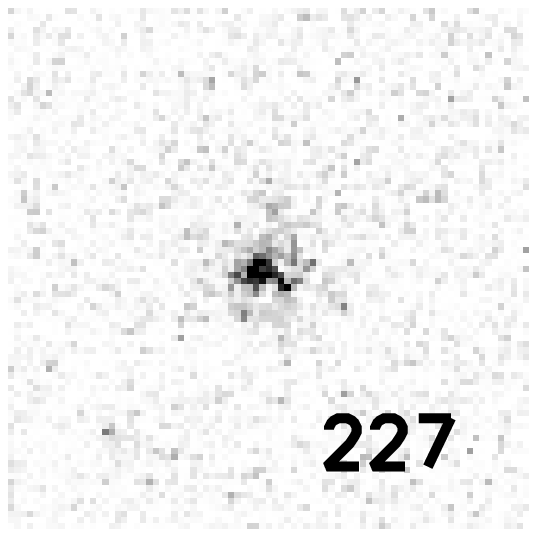}
\includegraphics[scale=0.27]{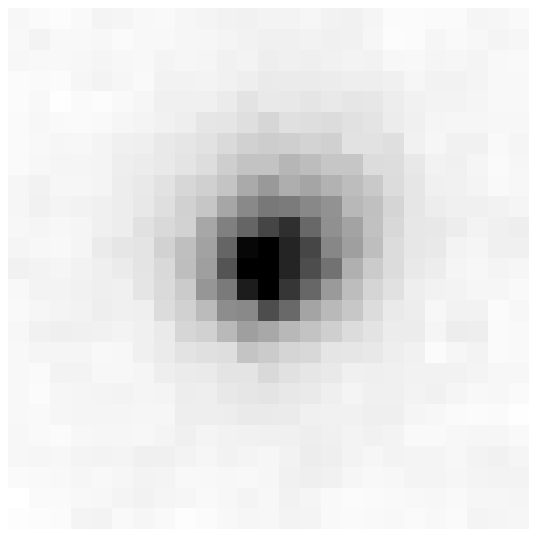}
\vline 
\includegraphics[scale=0.27]{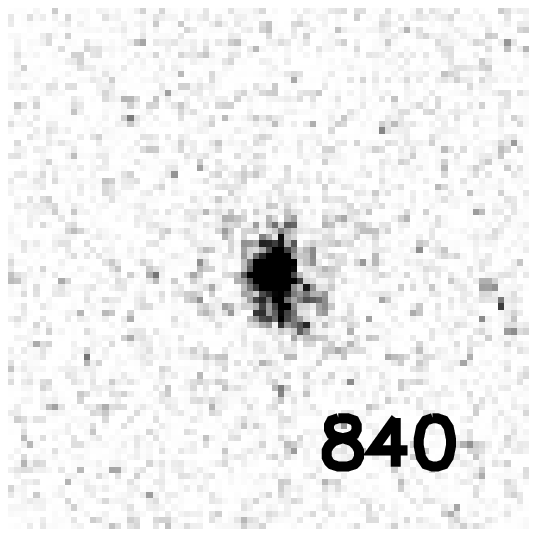}
\includegraphics[scale=0.27]{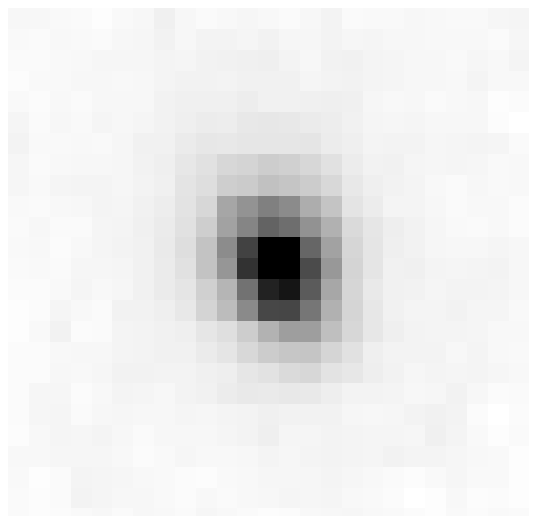}
\hrule
\includegraphics[scale=0.27]{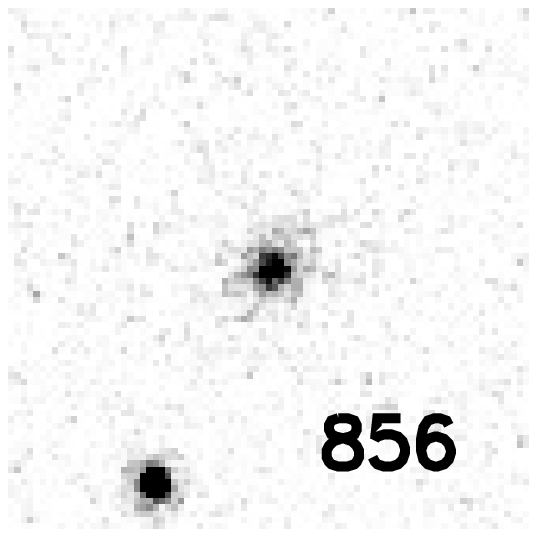}
\includegraphics[scale=0.27]{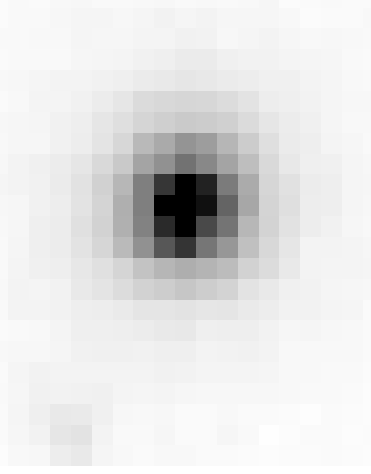}
\vline 
\includegraphics[scale=0.27]{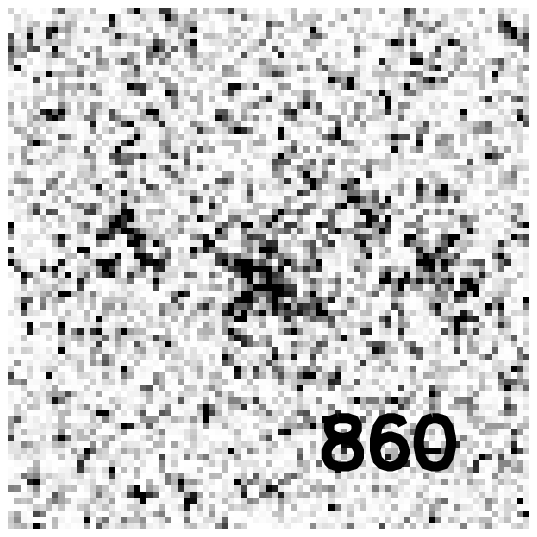}
\includegraphics[scale=0.27]{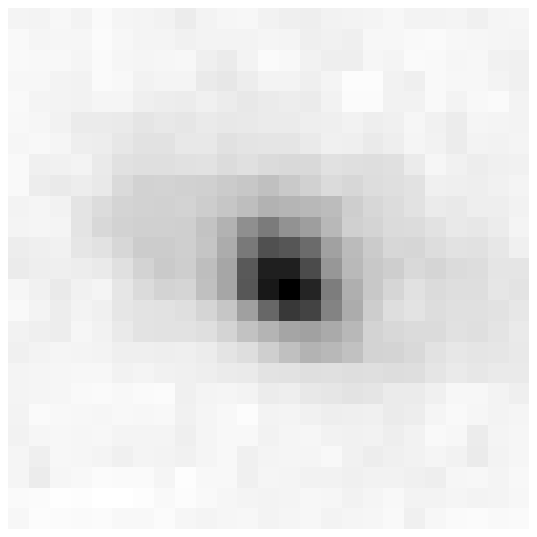}
\vline 
\includegraphics[scale=0.27]{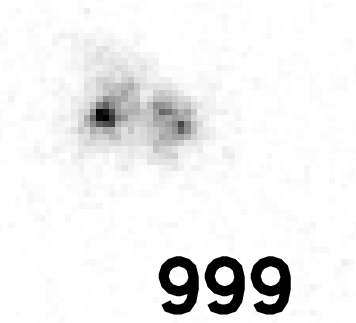}
\includegraphics[scale=0.27]{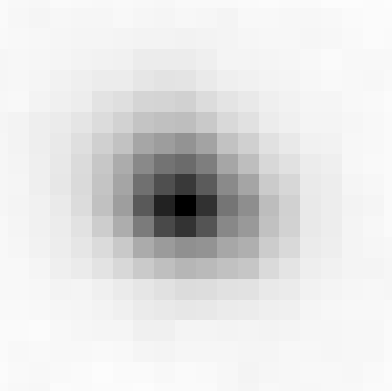}
\vline 
\includegraphics[scale=0.27]{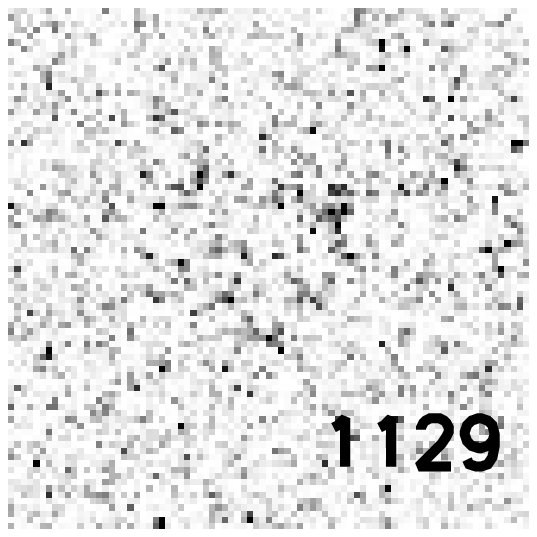}
\includegraphics[scale=0.27]{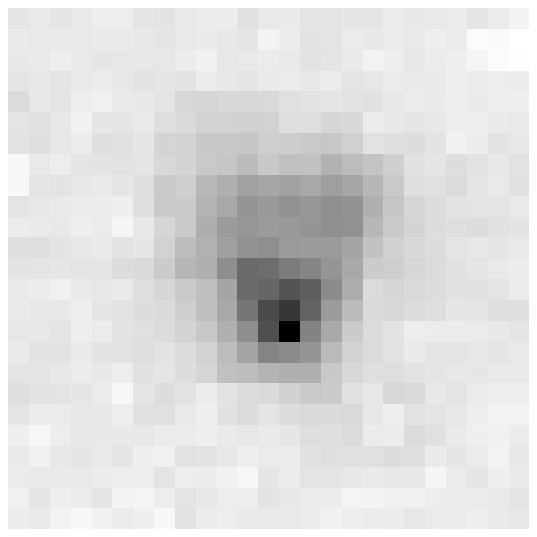}
\vline 
\includegraphics[scale=0.27]{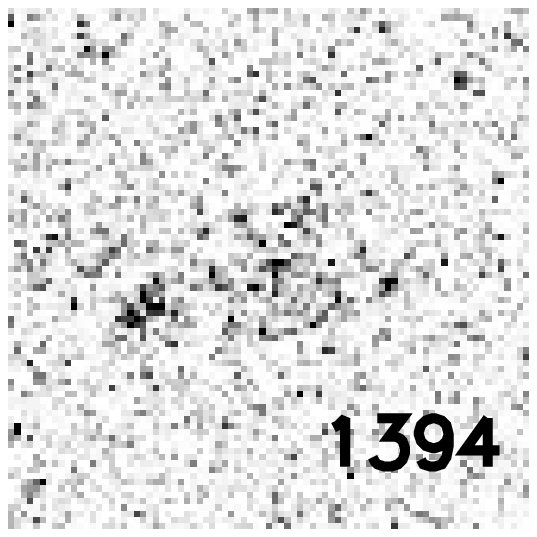}
\includegraphics[scale=0.27]{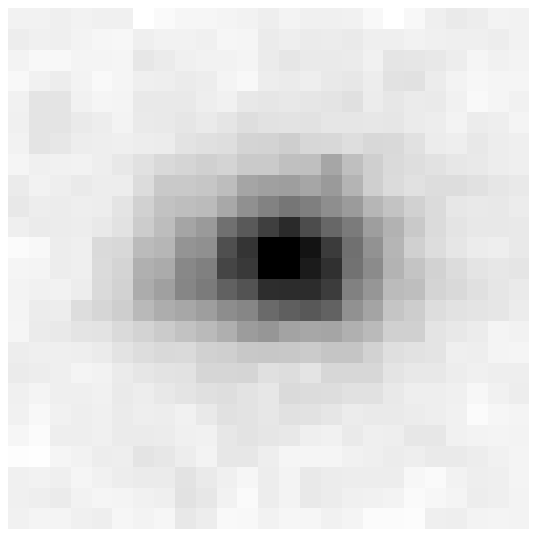}
\hrule
\includegraphics[scale=0.27]{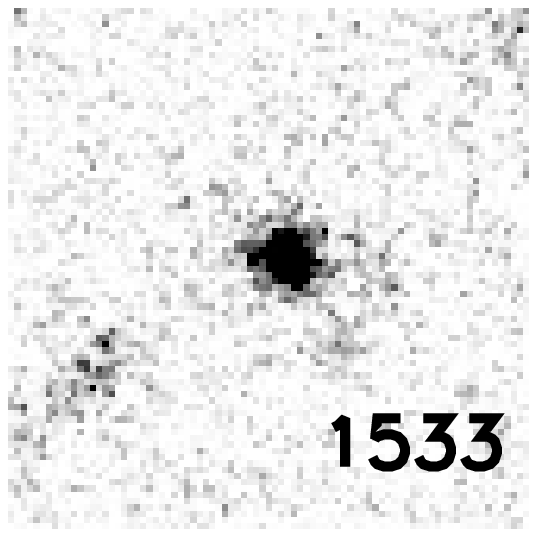}
\includegraphics[scale=0.27]{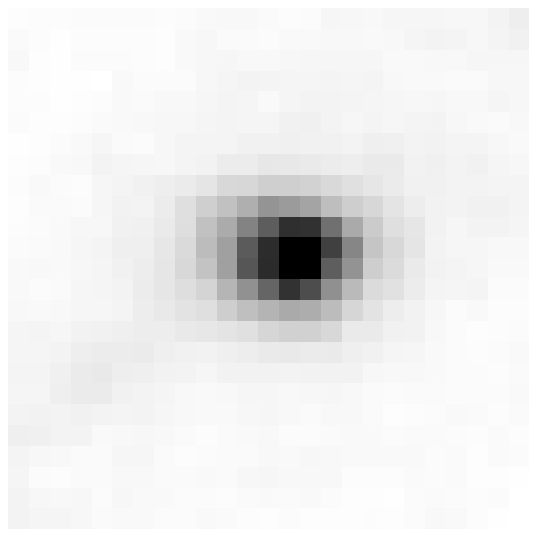}
\vline 
\includegraphics[scale=0.27]{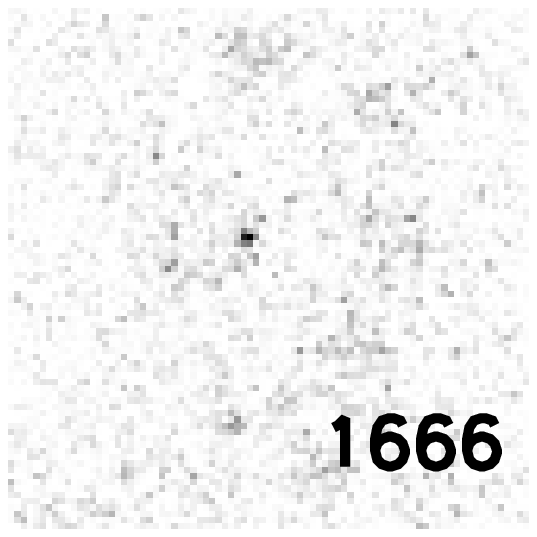}
\includegraphics[scale=0.27]{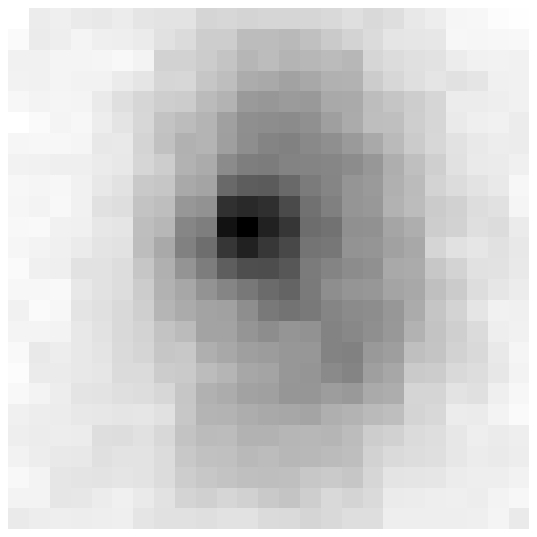}
\vline 
\includegraphics[scale=0.27]{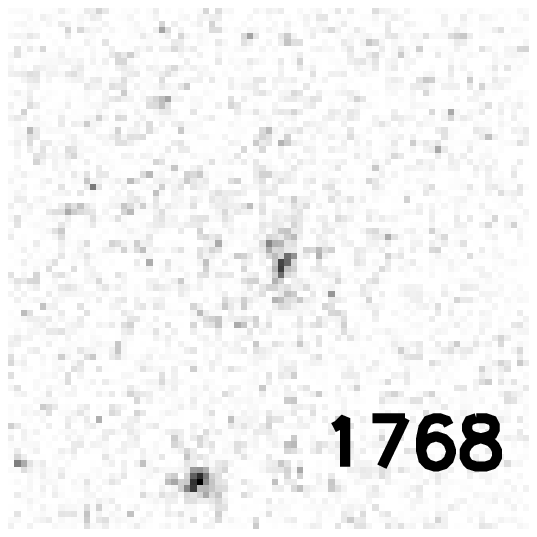}
\includegraphics[scale=0.27]{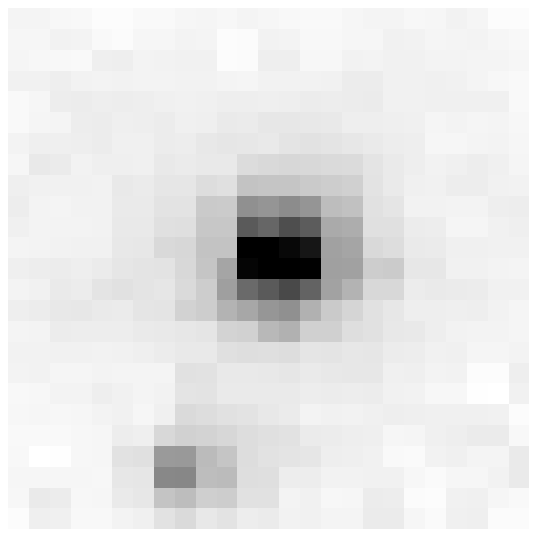}
\vline 
\includegraphics[scale=0.27]{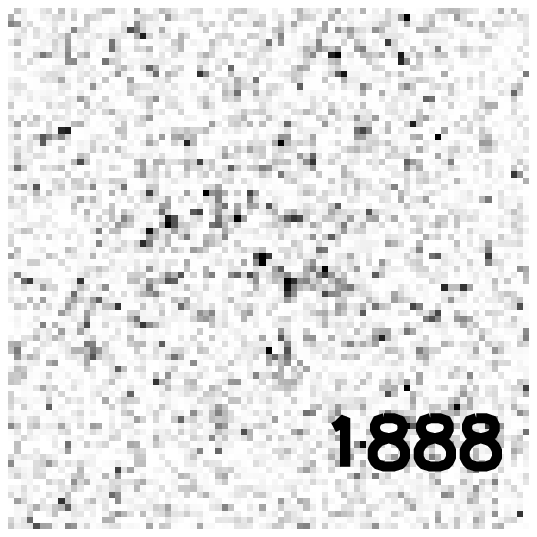}
\includegraphics[scale=0.27]{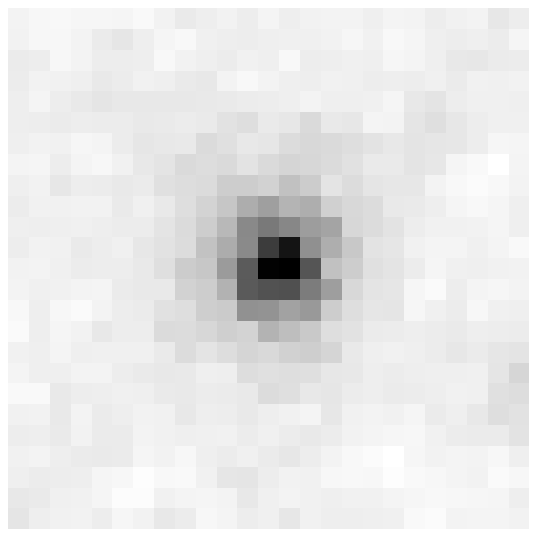}
\vline 
\includegraphics[scale=0.27]{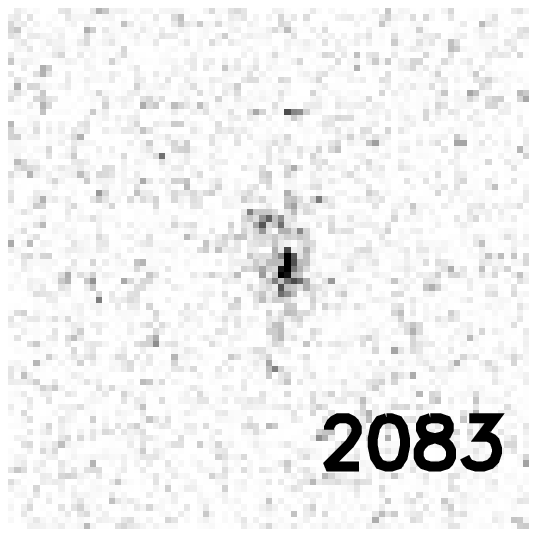}
\includegraphics[scale=0.27]{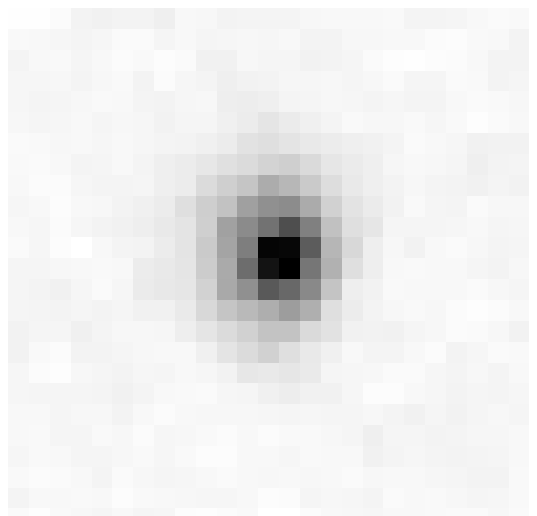}
\hrule
\includegraphics[scale=0.27]{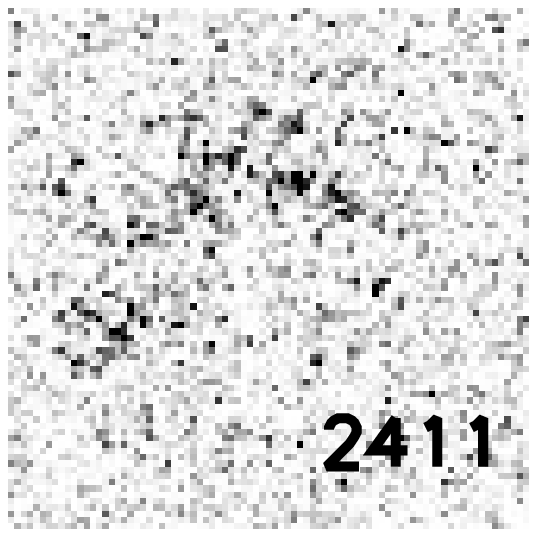}
\includegraphics[scale=0.27]{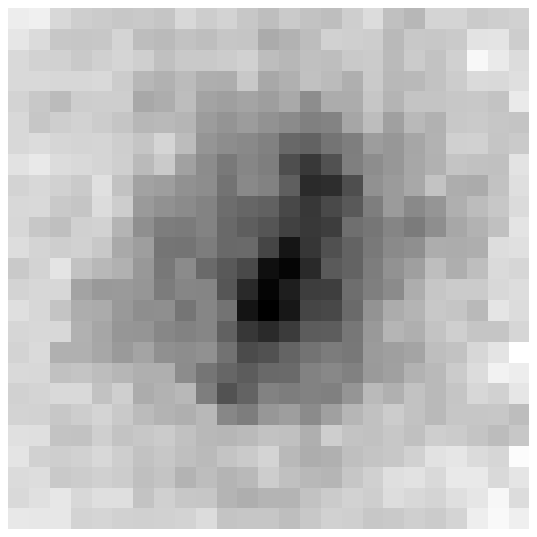}
\vline 
\includegraphics[scale=0.27]{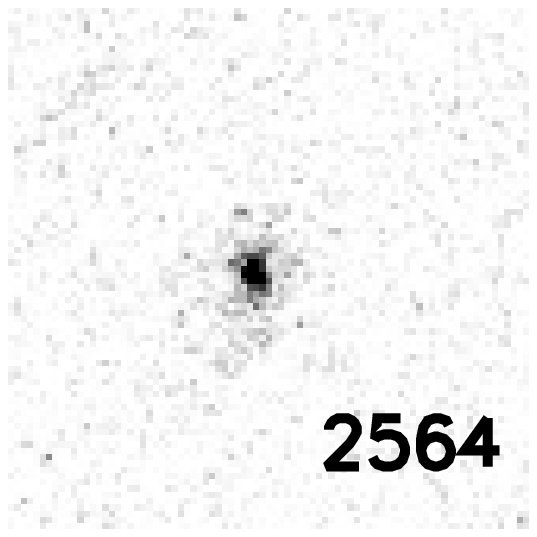}
\includegraphics[scale=0.27]{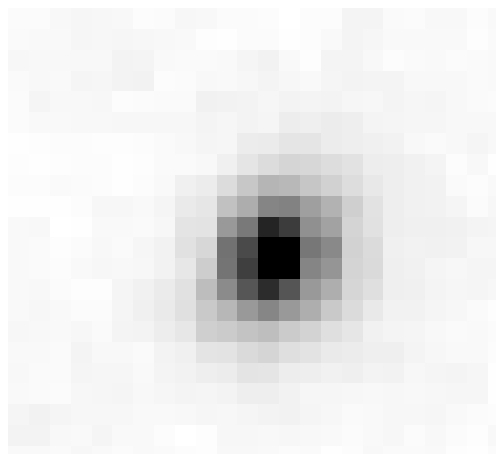}
\vline 
\includegraphics[scale=0.27]{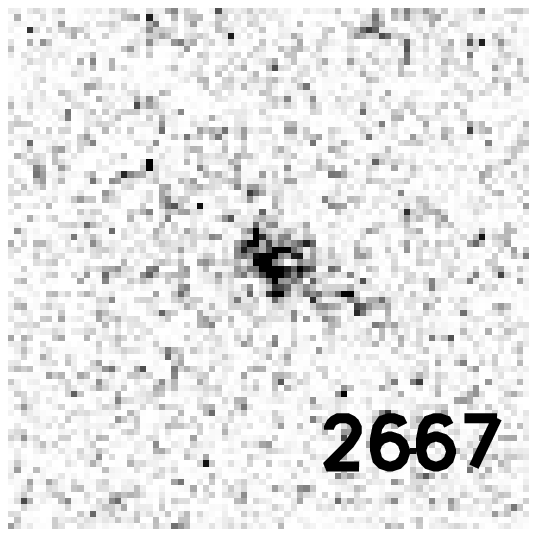}
\includegraphics[scale=0.27]{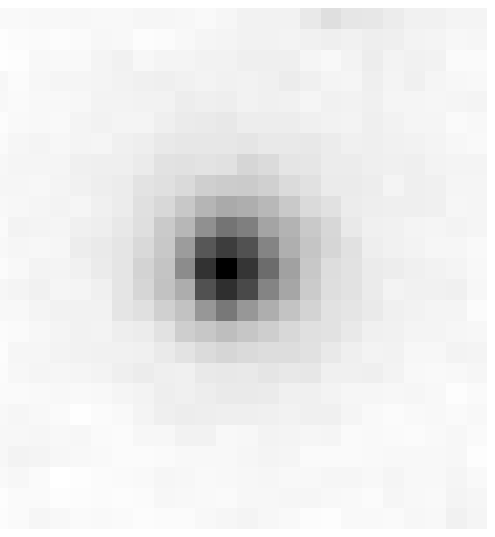}
\vline 
\includegraphics[scale=0.27]{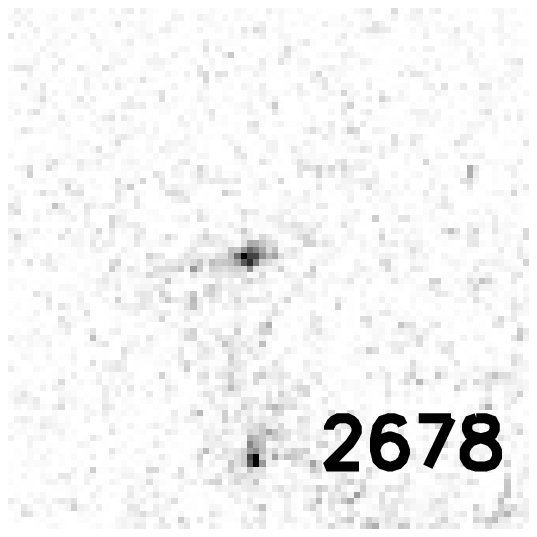}
\includegraphics[scale=0.27]{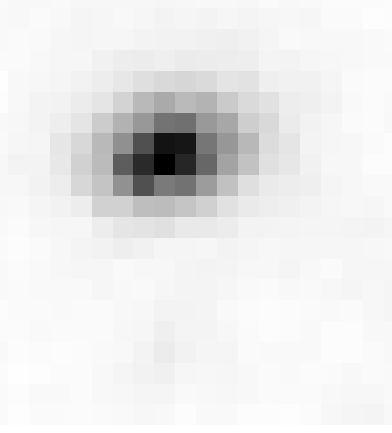}
\vline 
\includegraphics[scale=0.27]{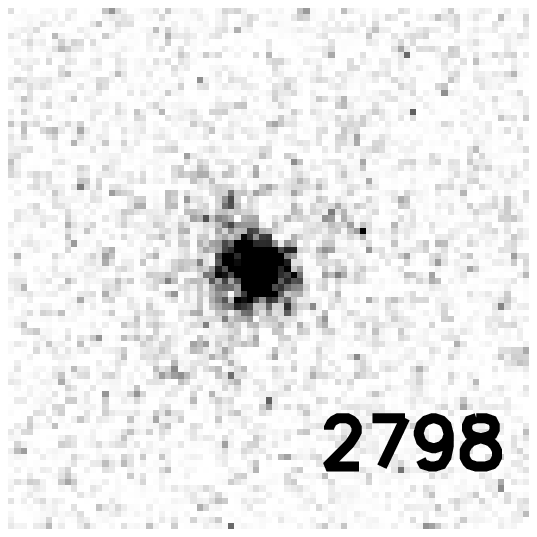}
\includegraphics[scale=0.27]{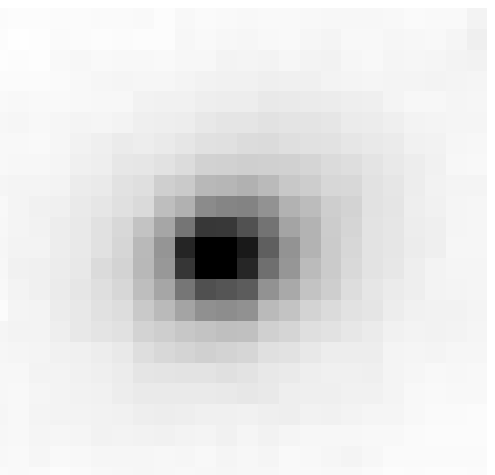}
\hrule
\includegraphics[scale=0.27]{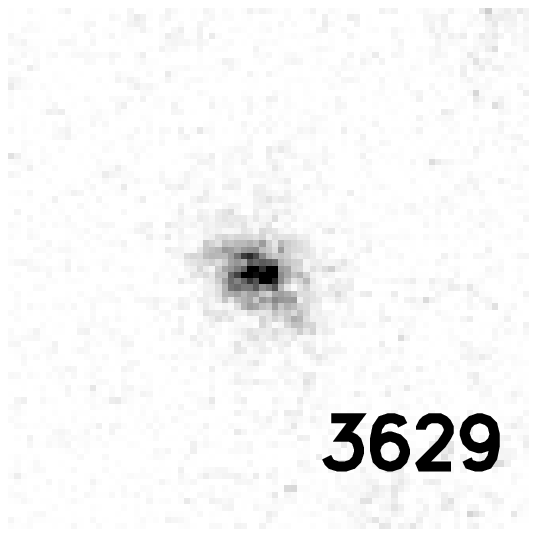}
\includegraphics[scale=0.27]{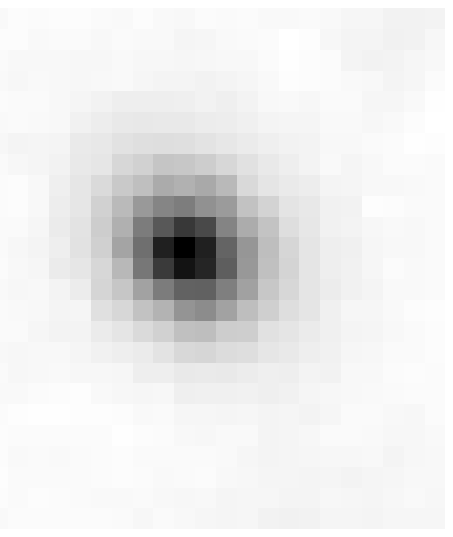}
\vline 
\includegraphics[scale=0.27]{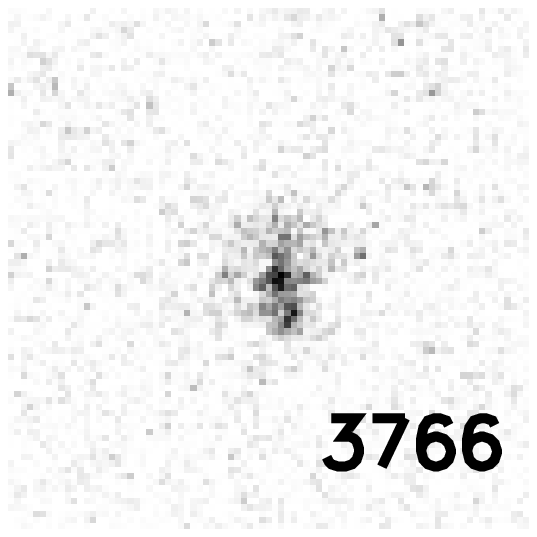}
\includegraphics[scale=0.27]{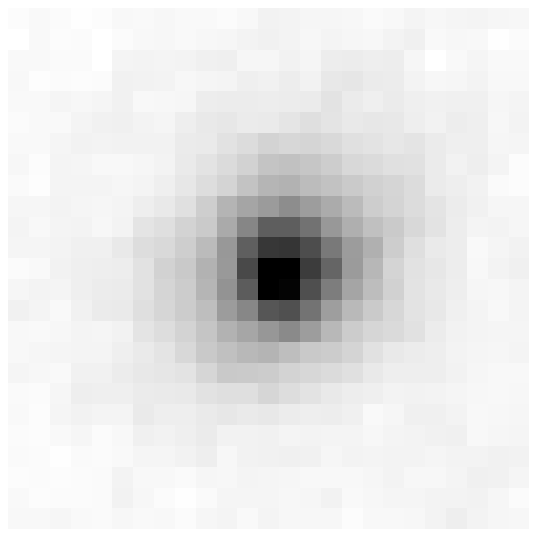}
\vline 
\includegraphics[scale=0.27]{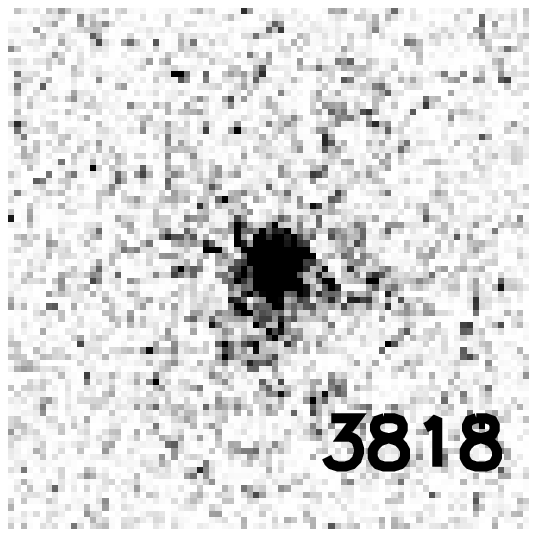}
\includegraphics[scale=0.27]{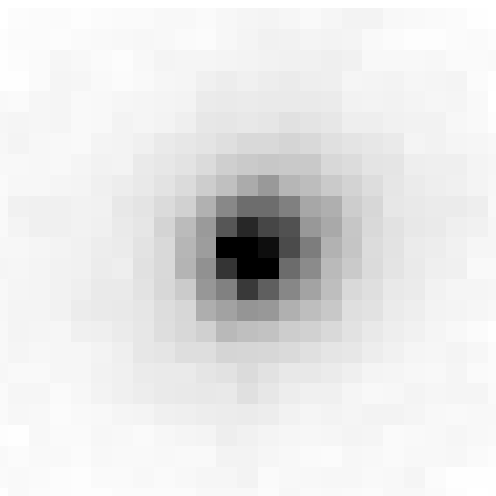}
\vline 
\includegraphics[scale=0.27]{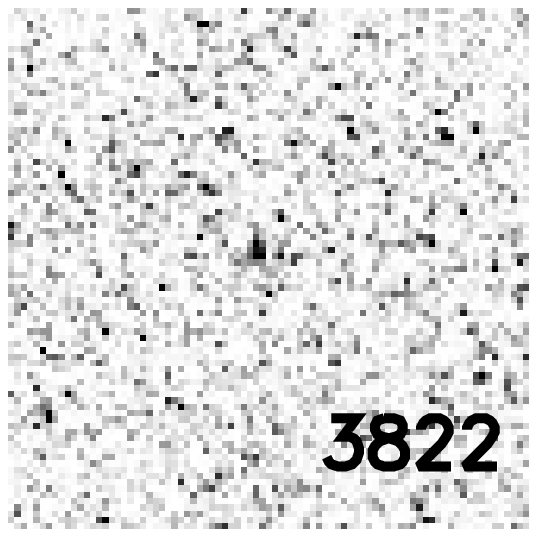}
\includegraphics[scale=0.27]{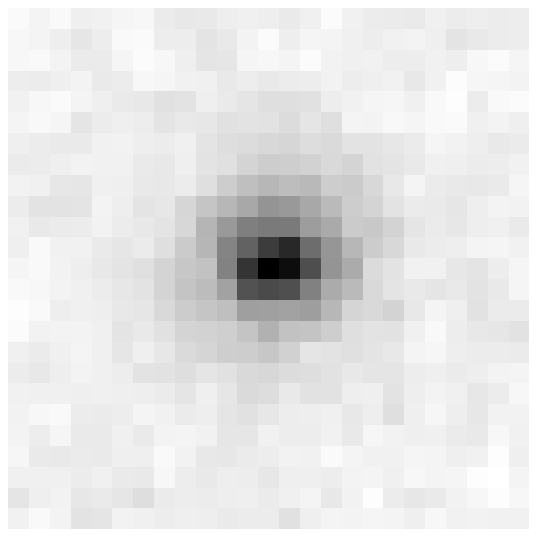}
\vline 
\includegraphics[scale=0.27]{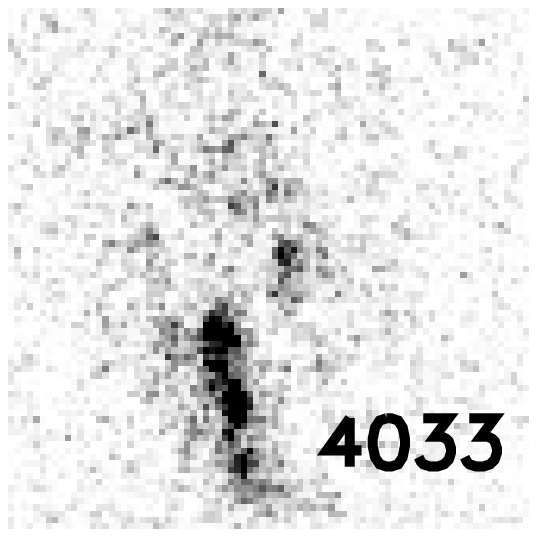}
\includegraphics[scale=0.27]{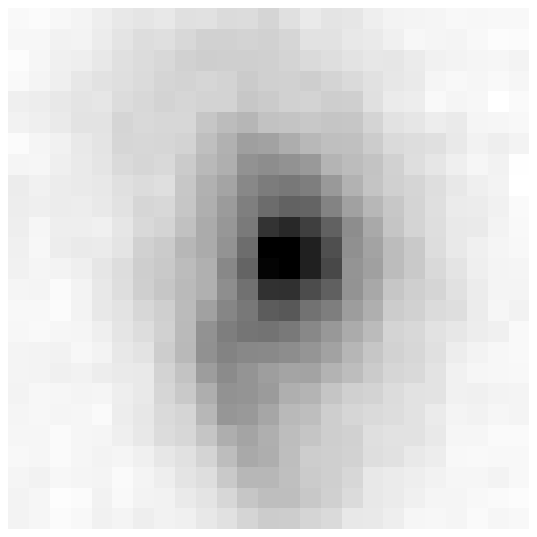}
\hrule
\includegraphics[scale=0.27]{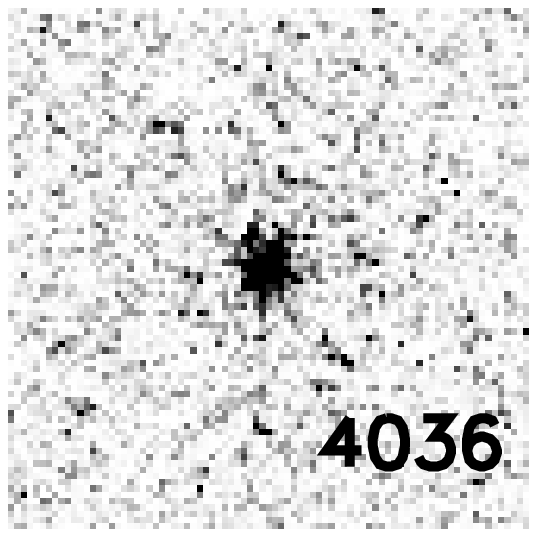}
\includegraphics[scale=0.27]{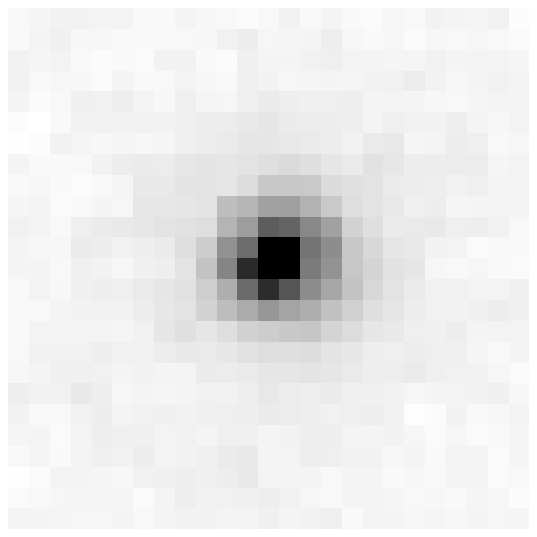}
\vline 
\includegraphics[scale=0.27]{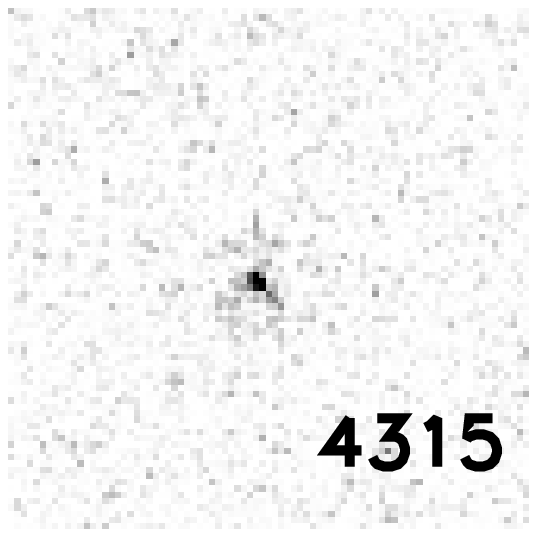}
\includegraphics[scale=0.27]{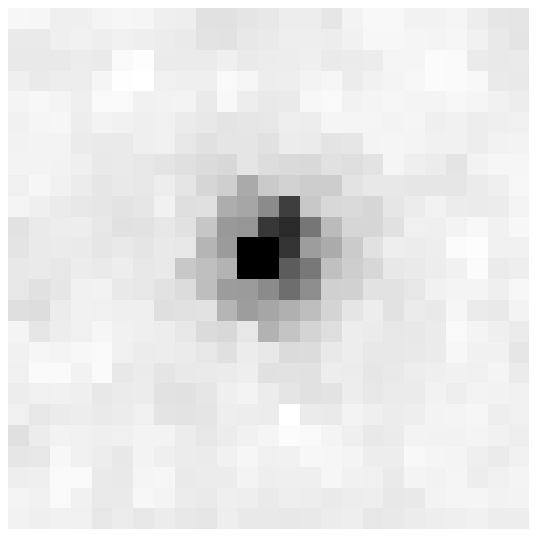}
\vline 
\includegraphics[scale=0.27]{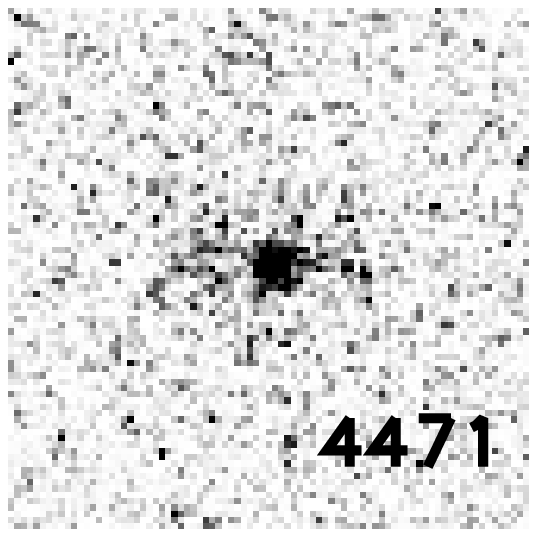}
\includegraphics[scale=0.27]{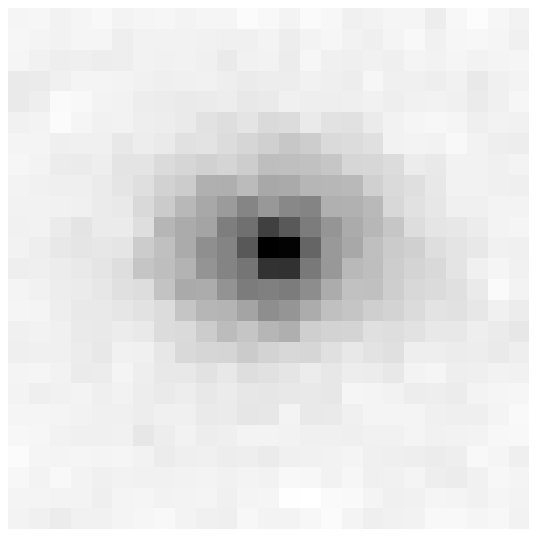}
\vline 
\includegraphics[scale=0.27]{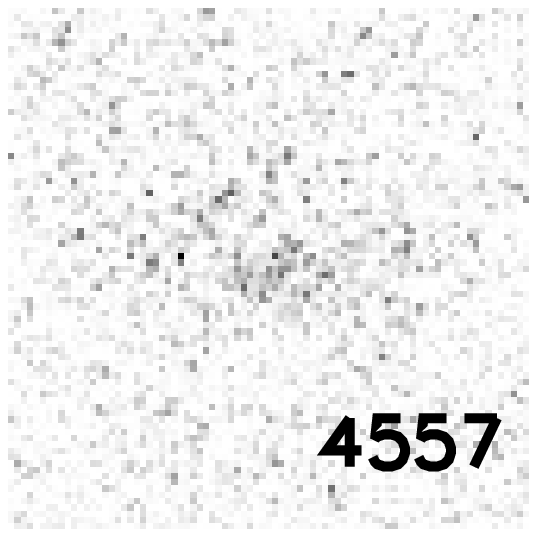}
\includegraphics[scale=0.27]{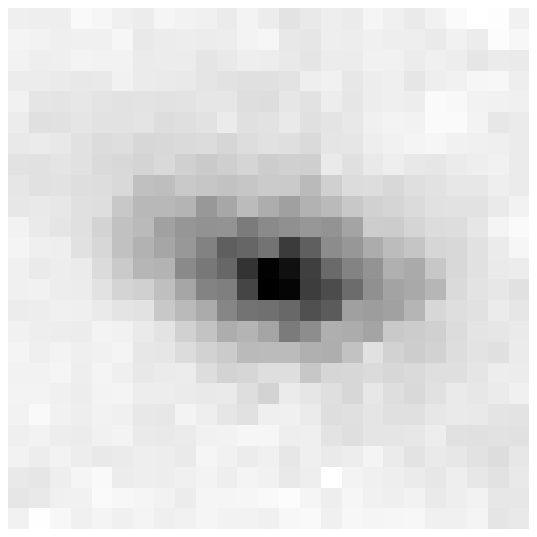}
\vline 
\includegraphics[scale=0.27]{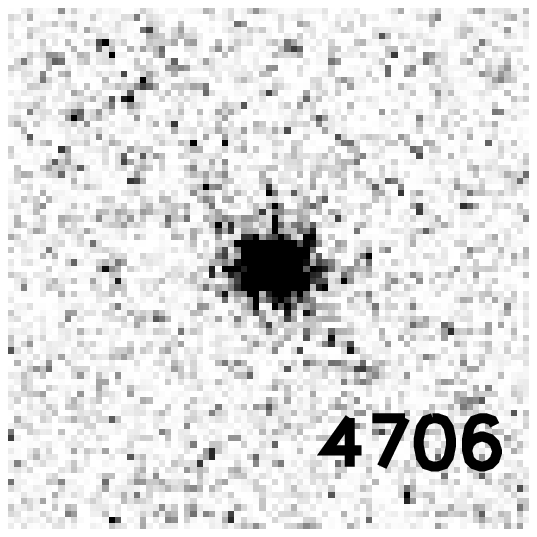}
\includegraphics[scale=0.27]{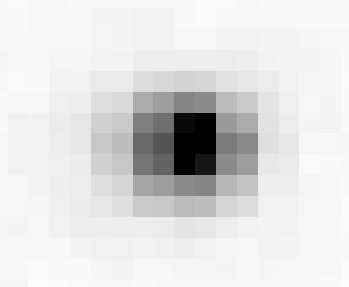}
\hrule
\includegraphics[scale=0.27]{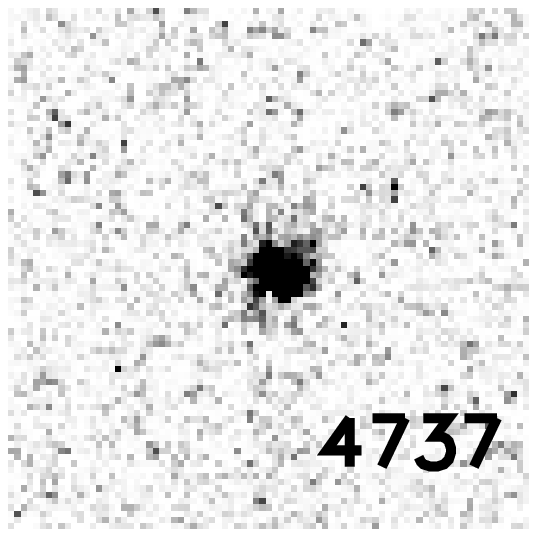}
\includegraphics[scale=0.27]{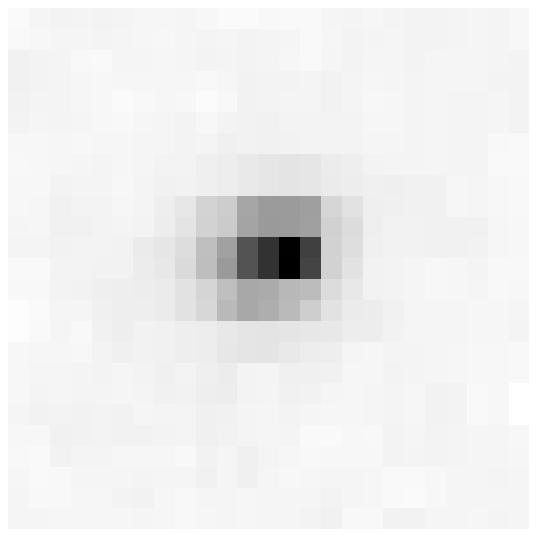}
\vline 
\includegraphics[scale=0.27]{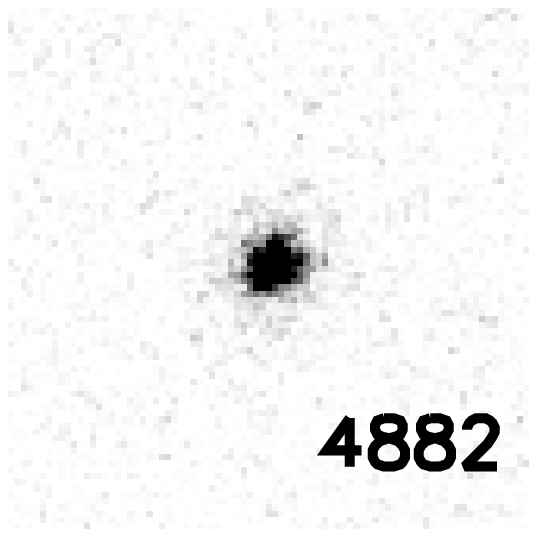}
\includegraphics[scale=0.27]{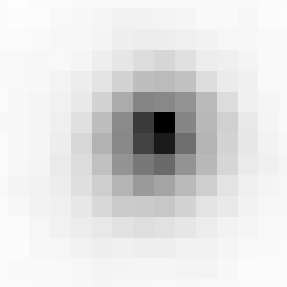}
\vline 
\includegraphics[scale=0.27]{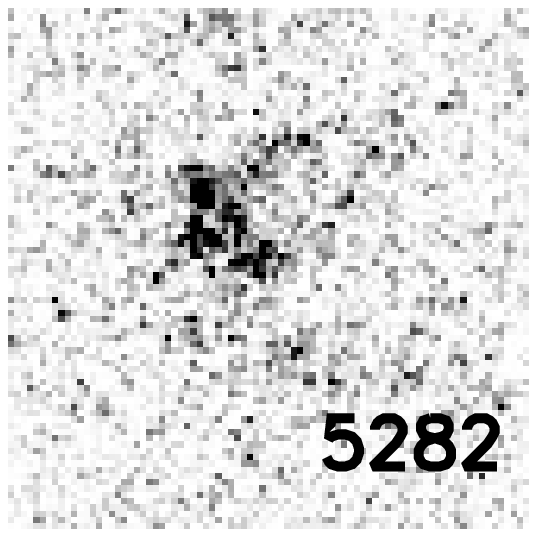}
\includegraphics[scale=0.27]{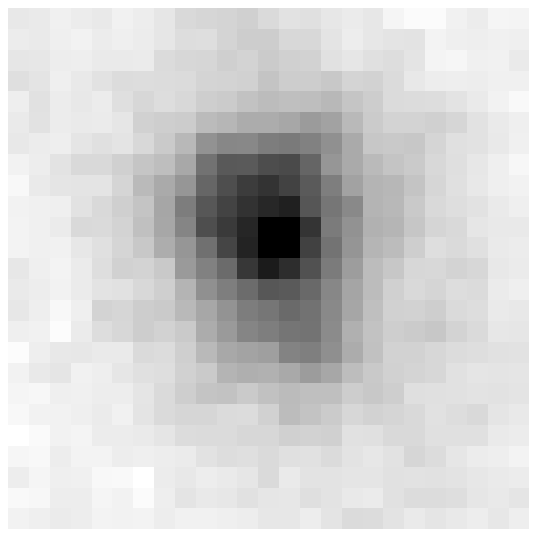}
\vline 
\includegraphics[scale=0.27]{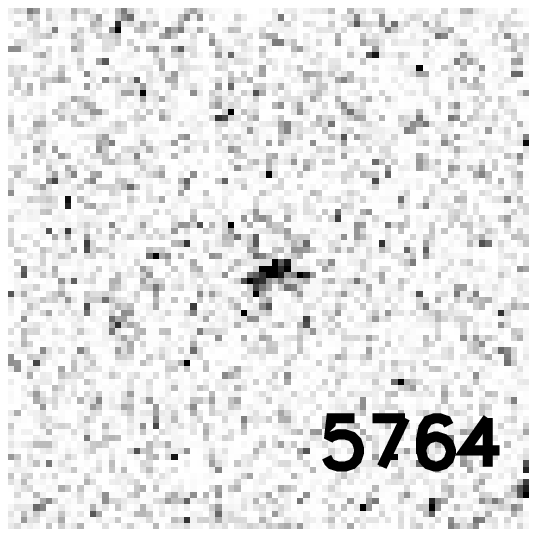}
\includegraphics[scale=0.27]{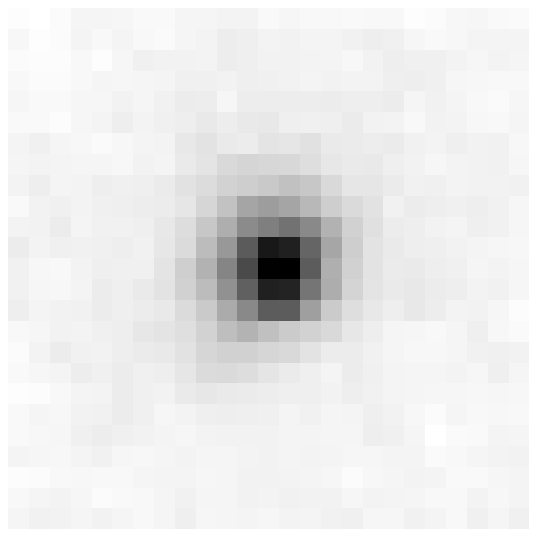}
\vline 
\includegraphics[scale=0.27]{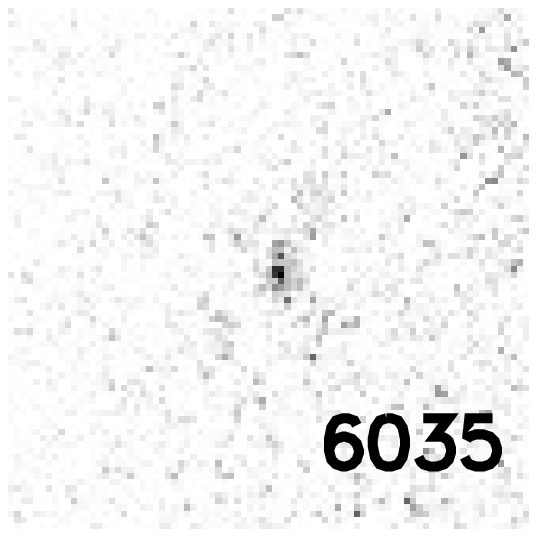}
\includegraphics[scale=0.27]{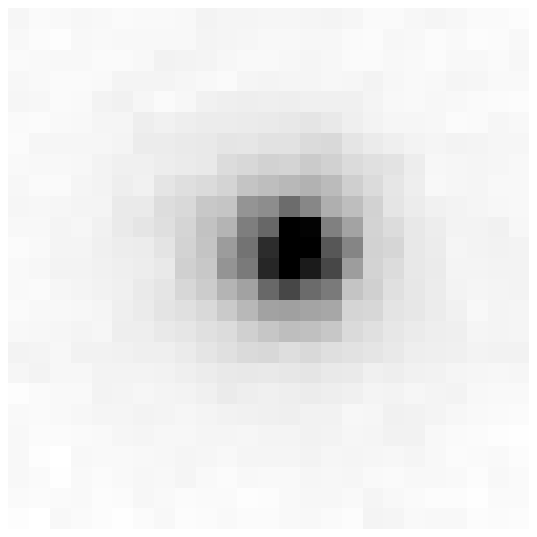}
\hrule
\includegraphics[scale=0.27]{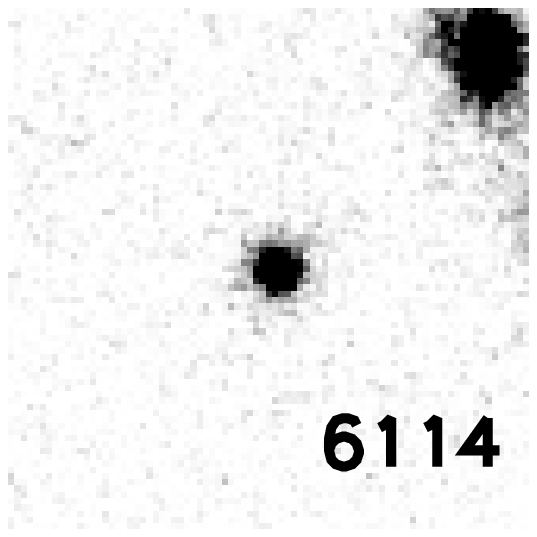}
\includegraphics[scale=0.27]{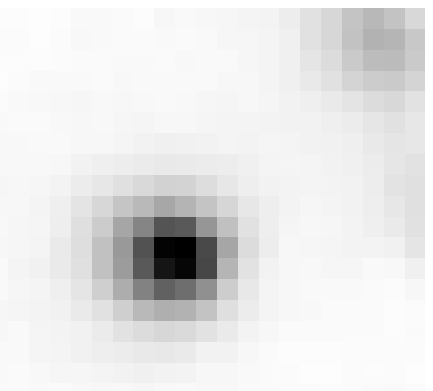}
\vline 
\includegraphics[scale=0.27]{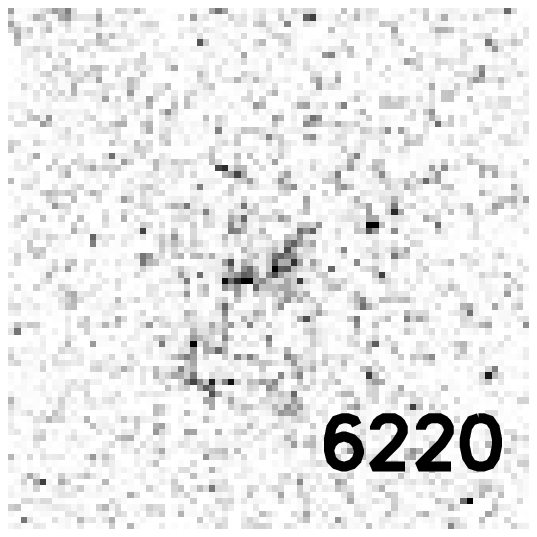}
\includegraphics[scale=0.27]{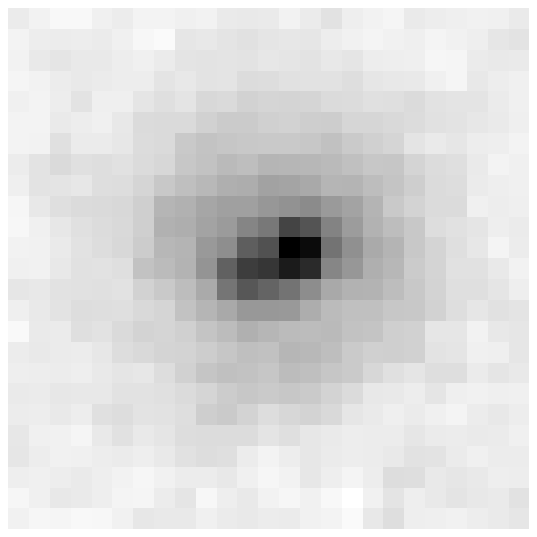}
\vline 
\includegraphics[scale=0.27]{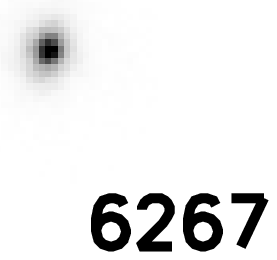}
\includegraphics[scale=0.27]{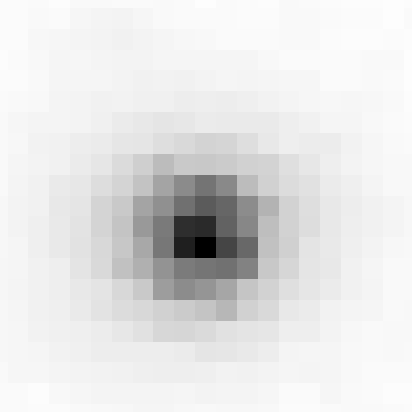}
\vline 
\includegraphics[scale=0.27]{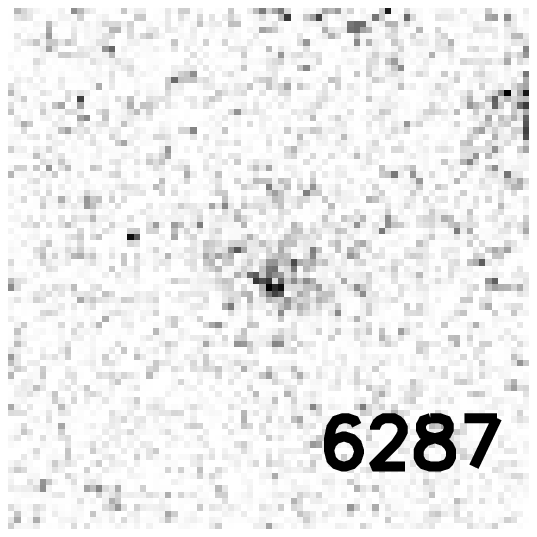}
\includegraphics[scale=0.27]{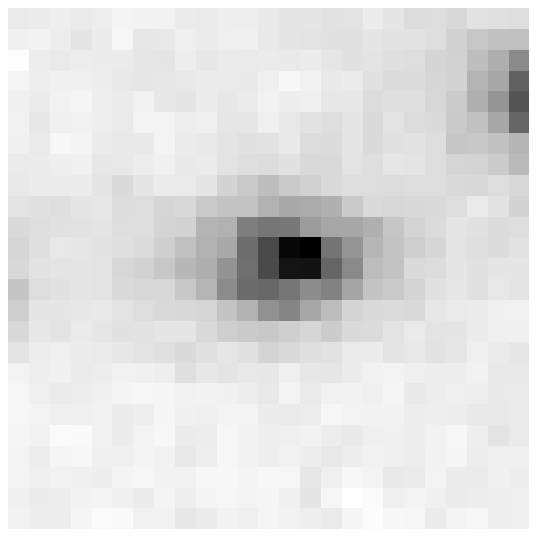}
\vline 
\includegraphics[scale=0.27]{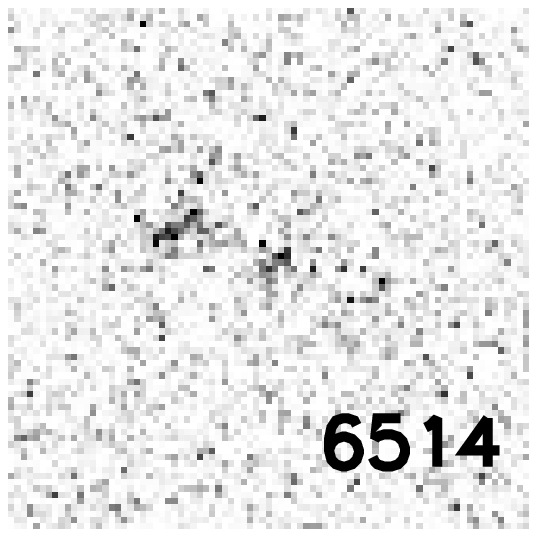}
\includegraphics[scale=0.27]{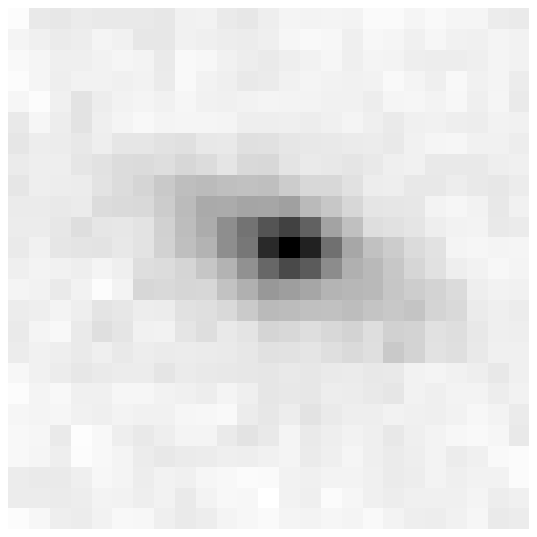}
\hrule
\includegraphics[scale=0.27]{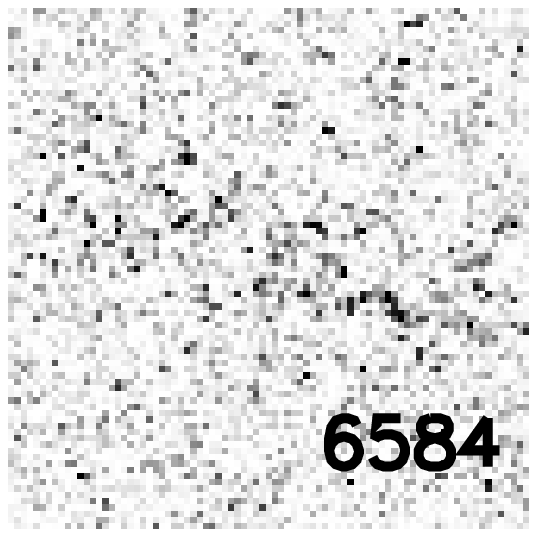}
\includegraphics[scale=0.27]{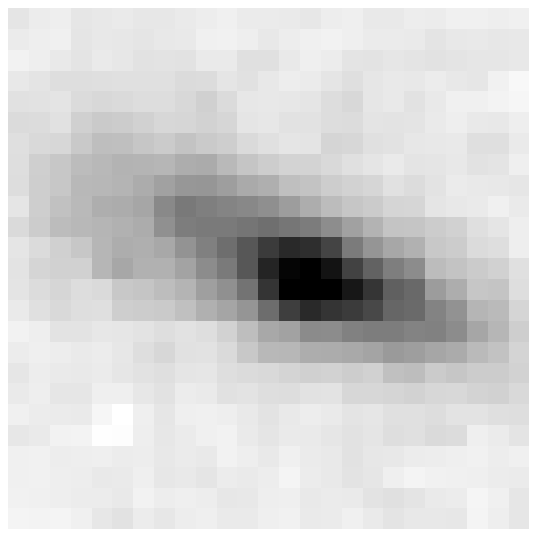}
\vline 
\includegraphics[scale=0.27]{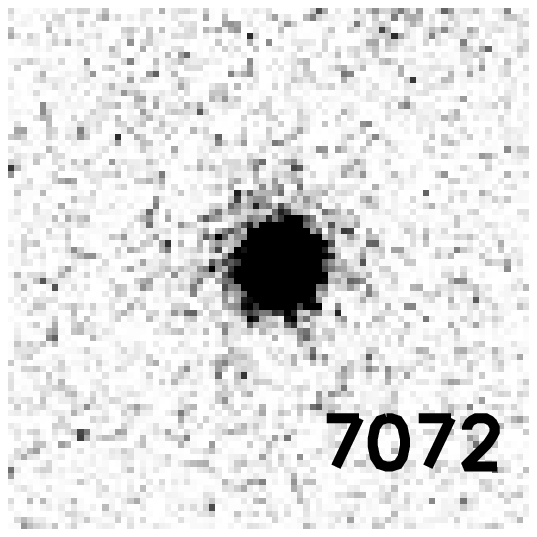}
\includegraphics[scale=0.27]{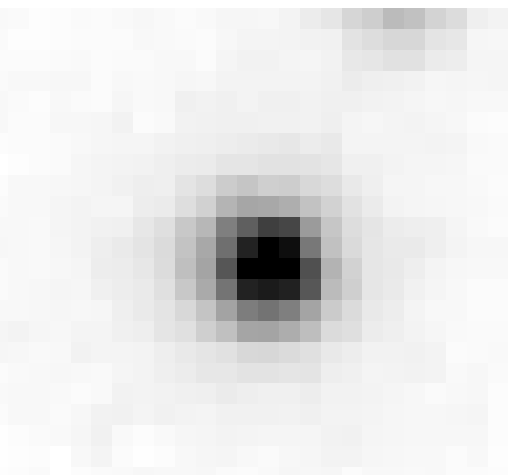}
\vline 
\includegraphics[scale=0.27]{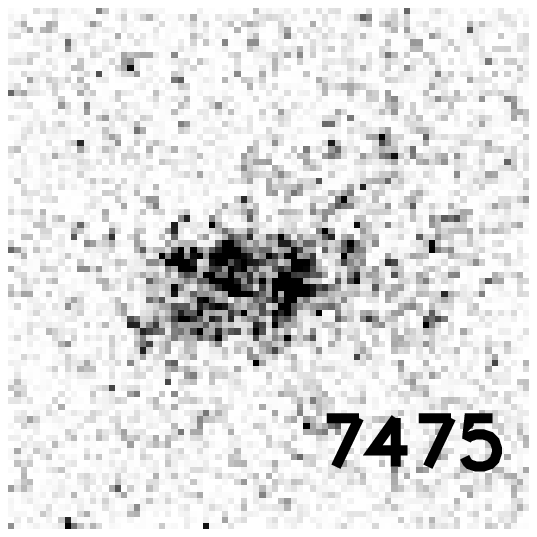}
\includegraphics[scale=0.27]{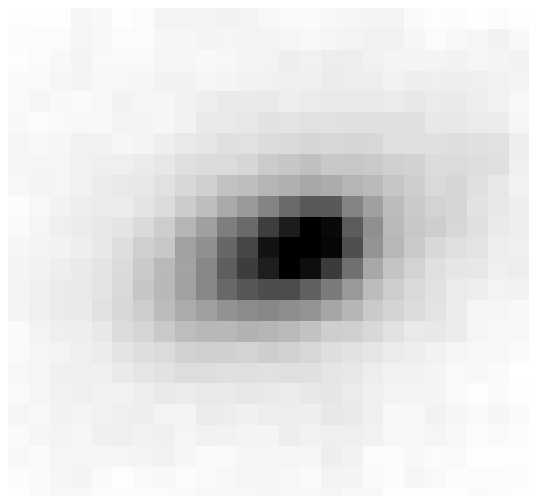}
\vline 
\includegraphics[scale=0.27]{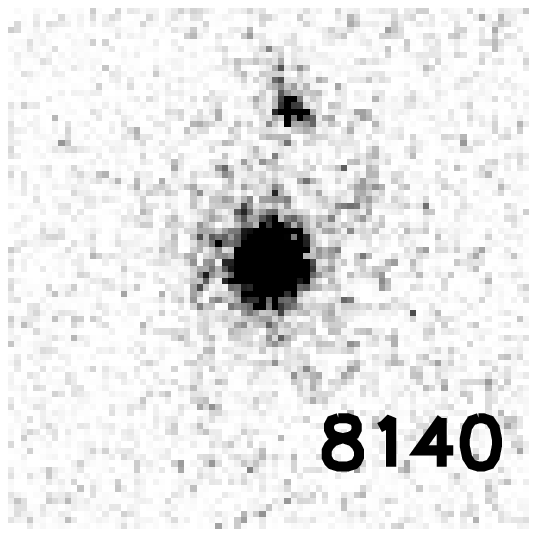}
\includegraphics[scale=0.27]{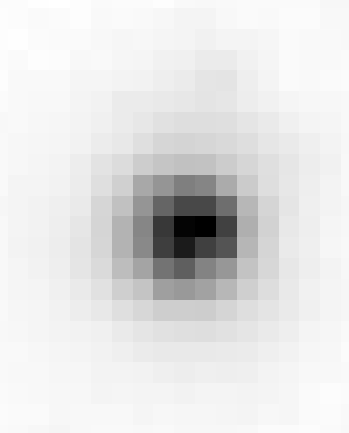}
\vline 
\includegraphics[scale=0.27]{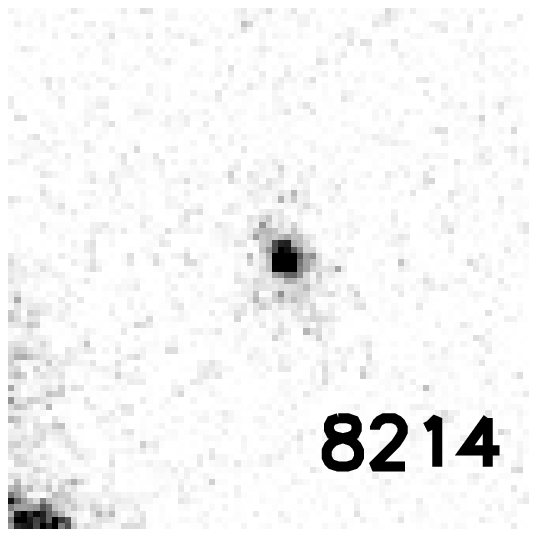}
\includegraphics[scale=0.27]{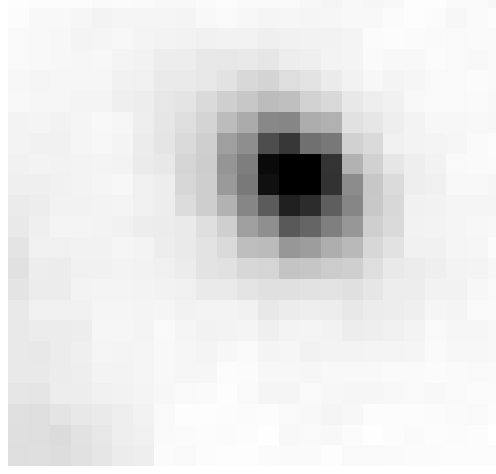}

\caption{Images in the $z_{850}$ band (left) and $H_{160}$ band (right) of the 45 massive galaxies in this study. The ID number of each galaxy is shown in the lower right hand corner of the $z_{850}$ band image with the corresponding image of the same region in the $H_{160}$ band to the right. All images are 2.5 by 2.5 arcsecond cutouts for galaxies between $1.5 < z < 3$ centred on the $H_{160}$ band detection. The properties of each galaxy are listed in Table 1.}
\label{images}
\end{figure*}

Within our NICMOS fields we find a total of 81 galaxies with stellar masses larger than $10^{11}M_{\odot}$ with photometric and spectroscopic redshifts in the range  $1.5<z<3$. This $H_{160}$-band sample of massive galaxies is reduced to 52 due to optical band non\--detections where we are unable to calculate accurate ultraviolet (UV) dust extinction corrections (Bauer et al. 2011). The sample is further reduced to 45 galaxies due to  removing those galaxies with S\'{e}rsic fits to the $H_{160}$ light profiles with high uncertainties (Buitrago et al. 2008). We examined the UV surface brightness profiles of the excluded galaxies in the bands they were detected in, and found them to be consistent with the profiles of the remaining galaxies. Figure \ref{images} shows the $z_{850}$ and $H_{160}$ band images of the 45 galaxies used in this study.

\subsection{Redshifts}
\label{sec:redshifts}

Where possible, we use spectroscopic redshifts published in the literature for our GNS galaxies, otherwise we use our own measured photometric redshifts. Spectroscopic redshifts of sources in the GOODS-N field were compiled by Barger et al. (2008), whereas the GOOD-S field spectroscopic redshifts are taken from the FIREWORKS compilation (Wuyts et al. 2008). In the full GNS sample, there are 537 spectroscopic redshifts for sources in GOODS-N and 369 in GOODS-S. In the massive galaxy sample used in this paper there are however only six galaxies with spectroscopic redshifts.

Photometric redshifts  are therefore crucial for this study.   These photo-zs were obtained by fitting template spectra to the Bv$iz$H photometric data points using the HYPERZ code (Bolzonella et al. 2000). The method is described in more detail in Gr\"{u}tzbauch et al. (2011b). The synthetic spectra used by HYPERZ are constructed with the Bruzual $\&$ Charlot evolutionary code (Bruzual $\&$ Charlot 1993) representing roughly the different morphological types of galaxies found in the local universe. We use five template spectra corresponding to the spectral types of E, Sa, Sc and Im, as well as a single starburst scenario. The reddening law is taken from Calzetti et al. (2000). HYPERZ computes the most likely redshift solution in the parameter space of age, metallicity and reddening. The best fit redshift and corresponding probability are then output together with the best fit parameters of spectral type, age, metallicity, $A_{V}$and secondary solutions.

To assess the reliability of our photometric redshifts we compare them to available spectroscopic redshifts in the GOODS fields. We matched the two catalogues to our photometric catalogue with a matching radius of $2$ arcsec, obtaining 906 secure spectroscopic redshifts. The reliability of photometric redshifts measures we use is defined by $\Delta z/(1+z) \equiv (z_{spec}-z_{photo})/(1+z_{spec})$. In the following we compare the median offset from the one-to-one relationship between photometric and spectroscopic redshifts, $\langle \Delta z /(1+z)\rangle$, and the RMS scatter around this relation, $\sigma_{\Delta z / (1+z)}=0.061$. We then investigate the performance of HYPERZ at different redshifts, at low redshift $(z<1.5)$ and at $1.5 \le z \le 3$, which is the redshift range of the galaxy sample we use. For the high redshift complete sample we obtain an average offset $\langle \Delta z /(1+z)\rangle = 0.06$ and a RMS of  $\sigma_{\Delta z / (1+z)}=0.1$, with a fraction of catastrophic outliers of $20\%$, where catastrophic outliers are defined as galaxies with  $ | \Delta z /(1+z) | > 0.3$, which corresponds to $\sim 3$ times the RMS scatter. This error has been folded into the results.

\subsection{Stellar Masses and e-folding Star formation Times}
\label{sec:mass}

Stellar masses and rest-frame colours of our sample are determined from multicolour stellar population fitting techniques using the same catalogue of five broad band data points used to determine photometric redshifts for all GNS galaxies. A detailed description of how stellar masses and rest-frame $(U-B)$ colours are derived can be found in Conselice et al. (2011) and Gr\"{u}tzbauch et al.(2011), and is summarised in the following.

To calculate these masses and colours we construct a grid of model spectral energy distributions (SEDs) from Bruzual $\&$ Charlot (2003) stellar population synthesis models, assuming a Salpeter initial mass function and a varying star formation history, age, metallicity and dust extinction. The star formation history is characterised by an exponentially declining model of the form 

\begin{equation}
 SFR (t) = SFR_{0} \times e^{-t/ \tau}.
\end{equation}
\noindent

The parameters in Equation 1 are varied over a wide range of values within the ranges; $\tau = 0.01$ to $10$ Gyr - with values (all in Gyr): 0.01 0.025 0.05 0.08 0.1 0.12 0.65 1.21 1.37 1.73 1.99 2.52 2.71 2.84 2.99 3.09 3.34 3.7 4.23 4.33 4.45 4.69 5.16 5.23 5.47 5.68 5.83 6.21 6.61 6.95 7.03 7.27 7.37 7.95 8.38 8.76 8.80 8.94 9.57 9.80 9.88, and the time since the onset of star formation ranging from $t = 0$ to $ 10$ Gyr,
with a condition that ages are not older than the universe itself at the redshift of observation. 
The dust content is parametrised by the V-band optical depth with values $\tau_{V} = 0.0, 0.5, 1, 2 $ and the metallicity ranges from 0.0001 to 0.05 (Bruzual $\&$ Charlot 2003).

The magnitudes obtained from the model SEDs are fit to the observed photometric data of each galaxy using a Bayesian approach. A likelihood distribution stellar mass, age and absolute magnitude at each possible star formation history is computed for each galaxy. 
The stellar mass is determined based on this distribution, where the most likely stellar mass produces a peak in the distribution, and the uncertainty is the width. The final error, as a result of the models used, lie within the range of 0.2 to 0.3 dex.

We use negative $\tau$ models in this paper to fit our stellar masses, although other forms of the 
star formation are included later in the paper when considering the addition of stellar mass, 
including constant star formation rate measures, and an exploration of possible forms of the star formation, 
including those which are maximal and would exceed the observed stellar mass density evolution (\S 5.4.1).
While there is some evidence that the star formation history actually increases from $z = 8$ to $z = 3$ 
(e.g., Papovich et al. 2011), there is also
evidence that at redshifts lower than this, and particularly for high mass galaxies, that the star formation
rate is starting to decline (e.g., Conselice et al. 2007; 2011).  We investigate this in detail for our
sample by holding at a constant co-moving volume, as Papovich et al. (2011) does, and seeing how our
star formation rate changes for the same co-moving density.  Doing this, we find that the star formation 
declines over our epoch using this method, although the star formation history at redshifts $z > 3$ is more
complicated than this.  However, this does show that our values of $\tau$ that we use here are mimicking
the form of the empirical star formation history.  

It is possible that the stellar masses are an over estimate due to the poor treatment of the TP-AGB phase in a star's life. The effects of this phase are less important at the rest frame wavelengths used in this study, especially in the infra-red $H_{160}$ band. Using newer models by Bruzual \& Charlot (2010) which have an improved treatment of the TP-AGB phase we find that this lowers the stellar masses of the massive galaxy sample by $<0.07$ dex. This effect from the new models is smaller than the stellar mass error, and the effects of cosmic variance, and is therefore negligible. Table \ref{tab:values} contains the full list of values for all variables used in this study.

\subsection{Star Formation Rates}
\label{sec:SFRexplan}

The star formation rates (SFRs) used in this paper are measured from rest-frame UV luminosities, using the methods described in Bauer et al. (2011). The rest-frame UV provides a direct measurement of ongoing SFR, since the UV luminosity is directly related to the presence of young and short-lived stellar populations produced by recent star formation. However, UV light is very susceptible to dust extinction and a careful dust-correction has to be applied. The correction we use here is based on the rest-frame UV slope (Bauer at al. 2011). We briefly describe the method in the following.

We determine the $SFR_{UV,uncorrected}$ from the observed optical ACS $z_{850}$-band flux density (with a 5 $\sigma$ limit of 27.5 in the AB system) spanning wavelengths of 2125 \-- 3400\AA\, for $z = 1.5$ \-- 3 galaxies. After applying an SED based k\--correction using the IDL \textsc{kcorrect} package (Blanton \& Roweis 2007, v4.2).   This is done by using the full SEDs of these galaxies.   This result of this is a k-correction at rest-frame UV wavelengths $\sim2800$\AA\, which we use throughout this paper. 

To measure the SFR  we first derive the UV luminosity of the massive galaxies, then use the Kennicutt (1998) conversion from 2800\AA\, luminosity to SFR assuming a Salpeter IMF:

\begin{equation}
SFR_{UV} (M_{\odot}\mathrm{yr^{-1}}) = 1.4 \times 10^{-28} L_{2800} (\mathrm{ergs\, s^{-1}\, Hz^{-1})} 
\end{equation}
 \noindent
Before dust extinction is taken into account we find a limiting $SFR_{UV,obs} = 0.3 \pm 0.1 M_{\odot}\mathrm{yr^{-1}}$ at $z=1.5$, and a limiting $SFR_{UV,obs} = 1.0 \pm 0.3 M_{\odot}\mathrm{yr^{-1}}$ at $z=3$. The errors quoted here take into account photometric errors and the conversion from a luminosity. The error for individual SFRs are around 30\%. This error is dominated by the dust correction applied.

We compare our total integrated SFR for each galaxy in this sample to the same sample used in Bauer et al. (2011) which included both SED determined and UV SFR. We find on average that our total SFRs are slightly higher due to the use of a larger aperture but within the quoted error.

Several studies (e.g. Bauer et al. 2011; Reddy et al. 2012) have found that in when comparing the IR derived SFRs plus UV derived SFRs ($SFR_{IR+UV}$) against dust corrected UV SFRs ($SFR_{UV,corr}$) that the $SFR_{IR+UV}$ is on average a factor of 3 larger than $SFR_{UV,corr}$. This overestimation has been seen in other studies looking at luminous galaxies (e.g. Papovich et al. 2007). Results from the Herschel Space Telescope (e.g. Elbaz et al 2010; Nordon et al 2010; Hilton et al. 2012) suggest that at $z>1.5$, the 24 $\mu m$ flux may overestimate the true SFR due to a rise in the strength of polycyclic aromatic hydrocarbon (PAH) features, changes in the SEDs, or AGN contamination. 

Recent work on the same sample of massive galaxies used in this paper by Hilton et al. (2012) using Herschel Space Telescope data and new fitting methods of the same sample of massive galaxies found that the scatter of these results can be reduced to $<1\sigma$ between the IR+UV and UV corrected SFRs by taking into account new templates in the FIR that account for these issues.   We however are unfortunately forced to only use the $SFR_{UV,corr}$ in this paper since the {\em Spitzer} IR 24$\mu$m and {\em Herschel} images are not resolved.

\subsection{Dust Corrections}
\label{sec:dust}
To obtain reliable star formation rates in the rest-frame ultraviolet, we need to account for the obscuration due to dust along the line of sight. Meurer et al. (1999) found a correlation between attenuation due to dust and the rest-frame UV slope, $\beta$, for a sample of local starburst galaxies (where $F_{\lambda} \sim \lambda^{\beta}$). More recent studies of local galaxies using the \emph{Galaxy Evolution Explorer} (GALEX) near-ultraviolet band show that the UV slope from the local starburst relation can be used to recover the dust attenuation of moderately luminous galaxies at $z \sim 2 $ (Buat et al. 2005; Seibert et al. 2005; Reddy et al. 2010).

A method for determining dust extinction uses the reddening parameter extracted from the best-fitting SED template as described in \S 2.3. We fit sets of template stellar population synthesis models to derive the stellar masses (Gr\"{u}tzbauch et al. 2011; Conselice et al. 2011). This method has some limitations when using it to correct for dust as this approach assumes that the UV slope is due to dust reddening instead of other sources, such as evolved stellar populations (Gr\"{u}tzbauch et al. 2011).

We apply a method for determining a UV dust attenuation, $A_{2800}$, in terms of the UV slope. The UV slope is determined using an SED-fitting procedure described in Bauer et al. (2011). To summarise, we fit an SED to the multi-wavelength observations from optical-to-infrared. The SEDs obtained for all sources in the GOODS fields were fit with stellar population synthesis models. The best fitting templates were then used to obtain a synthetic estimate of the UV emission at 1600\AA\, and 2800\AA. From the model-derived UV luminosities at 1600\AA\, and 2800\AA\, we calculate the spectral slope, $\beta$. The Calzetti et al. (2000) law is then used to derive $A_{2800}$ from the UV spectra slope, which we apply to the UV-derived star formation rates. Using this method we find an average extinction value of $A_{2800} = 3.2 \pm 1.0$ magnitudes for our sample. 

Bauer et al. (2011) find in a comparison between an SED determined, and an observationally derived $A_{2800}$, that the values obtained from these two methods are in relatively good agreement for $M_{*} > 10^{11} M_{\odot}$ across the whole redshift range with an average offset of $\delta A_{2800} = 0.86$. This corresponds to a $\sim 27\%$ error in the average dust attenuation and this is folded into the following results.

\section{Stellar Mass Density Profiles}
\label{sec:dens}
\subsection{Stellar Mass Radial Density Distributions}
\noindent

\begin{figure*}
\subfloat[]{\includegraphics[scale=0.5]{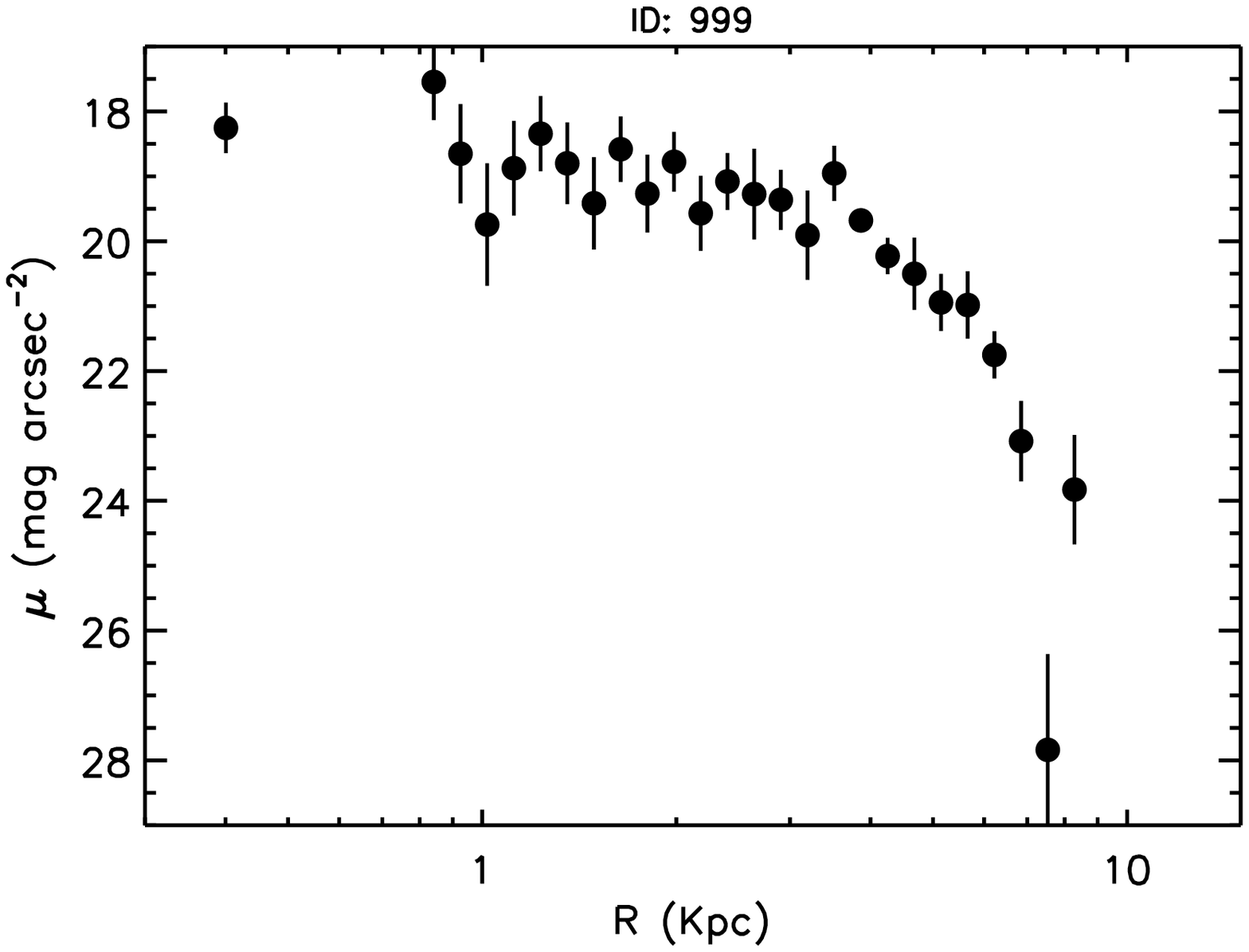}}                
  \subfloat[]{\includegraphics[scale=0.5]{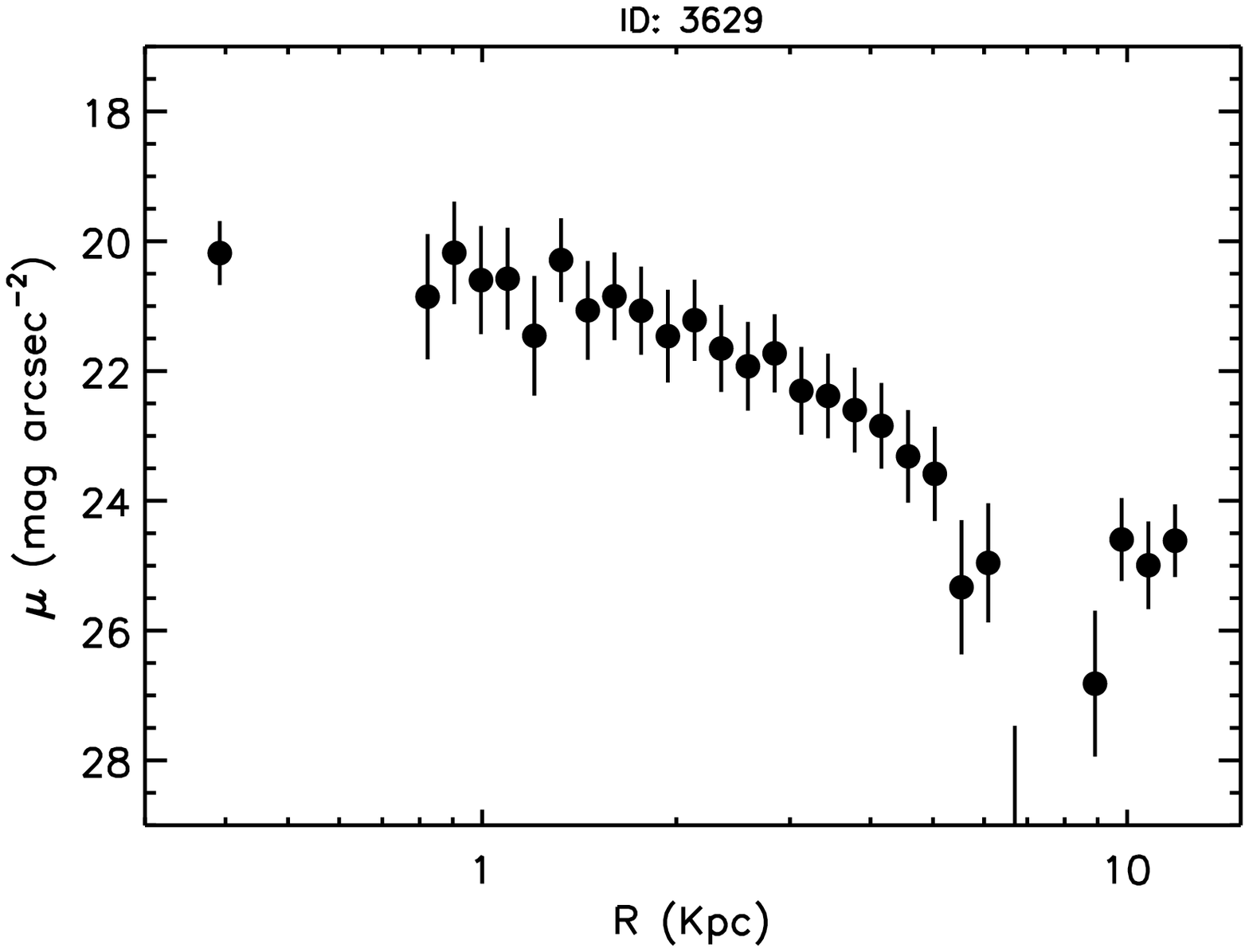}} 
\caption{Example rest frame UV surface brightness profiles from ACS $z_{850}$\--band imaging. (a) Galaxy ID: 999, Initial stellar mass: $1.5\times10^{11}M_{\odot}$, Rest frame optical effective radius: 2.0kpc, Rest frame optical S\'{e}rsic index:$n=1.42$, SF growth classification: Inner SF growth.  (b) Galaxy ID: 3629, Stellar mass: $2.6\times10^{11}M_{\odot}$, Rest frame optical effective radius: 1.8kpc, Rest frame optical S\'{e}rsic index:$n=1.26$, SF growth classification: Non\--significant SF growth.}
\label{sb}
\end{figure*}

We construct our sample galaxy's stellar mass density distributions by examining the distribution of the $H_{160}$, rest-frame optical light profiles for our sample (for $1.5 < z < 3$ this corresponds to rest $\lambda  = ~640 \-- ~400$nm). We base this determination on the S\'{e}rsic fits to the light distribution. A detailed description of how the S\'{e}rsic indices were measured can be found in Buitrago et al. (2008, 2011) and is summarised in the following.

The S\'{e}rsic profiles were measured using GALFIT (Peng et al. 2002, 2010). GALFIT uses  $r^{1/n}$ 2D models of the form  (S\'{e}rsic 1968):
\begin{equation}
 \Sigma(r) = \Sigma_{e}\times\rm{exp}(-b_{n}[(R/R_{e})^{1/n}-1])
\end{equation}

\noindent  Where $R_{e}$ is the effective radius of the galaxy, $\Sigma_{e}$ is the surface brightness at $R_{e}$, $n$ is the S\'{e}rsic index and $b_{n} = 1.9992n - 0.3271$. This model is convolved with the Point Spread Function (PSF) of the images, and GALFIT determines the best fit by comparing the  convolved model with the observed galaxy surface brightness distribution using a Levenberg-Marquardt algorithm to minimise the $\chi^{2}$ of the fit. We used single S\'{e}rsic models to compare our size estimations with previous work. Neighbouring galaxies are masked out before the fitting, and in the case of overlapping isophotes the objects are fit simultaneously. Due to variations of the shape of the NICMOS-3 PSF in our images, we select five non- saturated bright stars to sample the PSF within our imaging and with which to gauge the accuracy of the parameter measurements. The structural parameters of each galaxy are measured five times for each unique star. The uncertainty ($1\sigma$) on the structural parameters due to changes in the PSF is $\sim 15\%$ for the effective radius $r_{e}$, and $\sim 20\%$ for the S\'{e}rsic index $n$.

A concern when measuring sizes and S\'{e}rsic indices at high redshift is surface brightness dimming which in principal could bias our measured sizes. Previous studies have examined this issue, and have conducted many simulations in order to check the importance of surface brightness dimming in HST observations (e.g. Trujillo et al. 2006a, 2007; Buitrago et al. 2011). In Appendix A of Buitrago et al. (2011) one can find the descriptions of the extensive simulations we conducted in order to asses the reliability of the galaxy structural parameters within GNS. The median observable characteristics of our massive galaxies ($H_{AB} = 22.5$, n$\sim$2, r$_{e}\sim$2 kpc) allow us to retrieve their structural properties without any significant bias. However it is worth noting that, for individual galaxies, the parameters are not as well constrained for galaxies which display higher S\'ersic indices.

Using the total stellar mass for the galaxies and the H-band light (see \S 2.3) we equate the total stellar mass to the total rest-frame optical light received from the individual galaxies. We then convert the $H_{160}$-band S\'{e}rsic profile to a stellar mass profile. With the total initial stellar mass defined by:

\begin{equation}
 M_{*} = \rho_{e}\int_0^{R_{\mathrm{max}}}\rm{exp}{(-b_{n}[(R/R_{e})^{1/n}-1])}2\pi R \mathrm{d}R
\end{equation}

\noindent  The radius within which the total stellar mass is contained, $R_{\mathrm{max}}$, is taken to be 20kpc in all cases. The effective radius, $R_{e}$ and S\'{e}rsic index, $n$, come from the $H_{160}$\--band S\'{e}rsic profile as described in this section. From this, the stellar mass density at the effective radius $\rho_{e}$ is calculated, and the full stellar mass density profile is constructed via:

\begin{equation}
 \rho_{observed}(R) = \rho_{e}\times\rm{exp}(-b_{n}[(R/R_{e})^{1/n}-1])
\end{equation}

\noindent with the implicit assumption that the mass to light ratio is constant over the galaxy, as used in other works studying surface brightness profiles (e.g. Szomoru et al. 2011).

\subsection{Stellar Mass Density Added Via Star Formation}

\begin{center}
\begin{table*}
\begin{center}
\begin{tabular}{ c c c c c c c c c c c }
  \hline
  ID & SF Class & $\rm{SFR_0}$ & $z$  & Mass & $A_{2800}$ & $R_{e}$ & $n$ & $\tau$ $(yr)$  \\
  \hline             
43 	& OG & 126.7$\pm$34.2 & 1.79	 & 11.0$\pm$0.4 	 & 3.8$\pm$1.0 & 1.8$\pm$0.3  & 2.5$\pm$0.8  & $6.5\times10^{8}$ \\  
77 	& NG & 173.8$\pm$46.9 & 2.33	 & 11.1$\pm$0.2	 & 3.5$\pm$0.9 & 3.2$\pm$0.1 & 1.4$\pm$0.1  & $1.2\times10^{8}$ \\
158 	& NG & 28.8$\pm$7.8 & 1.84	 & 11.2$\pm$0.4  	 & 2.0$\pm$0.5 & 1.4$\pm$0.3  & 2.0$\pm$2.0  & $6.5\times10^{8}$ \\
227 	& NG & 23.1$\pm$6.2 & 2.48	 & 11.2$\pm$0.3	 & 0.8$\pm$0.2 & 2.1$\pm$0.1 & 2.1$\pm$0.1  & $1.2\times10^{8}$  \\
840 	& NG & 87.8$\pm$23.7 & 2.31   & 11.1$\pm$0.3	 & 2.4$\pm$0.6 & 1.6$\pm$0.2 & 2.3$\pm$0.5  & $1.2\times10^{8}$ \\
856 	& NG & 257.8$\pm$69.6&2.32   & 11.2$\pm$0.2 	 & 2.6$\pm$0.7 & 1.7$\pm$0.1 & 3.7$\pm$0.3  & $1.2\times10^{8}$ \\
860 	& OG & 367.0$\pm$81.0 &1.79   & 11.2$\pm$0.3 	 & 3.9$\pm$1.1 & 3.8$\pm$0.3  & 3.5$\pm$0.6  & $6.5\times10^{8}$ \\
999 	& IG & 672.6$\pm$181.6 &1.58   & 11.2$\pm$0.4	 & 4.0$\pm$1.1 & 2.0$\pm$0.1  & 1.4$\pm$0.3  & $1.4\times10^{9}$\\   
1129	& NG & 464.7$\pm$125.5 &2.61    & 11.3$\pm$0.4   & 5.0$\pm$1.3 & 3.1$\pm$0.1 & 0.2$\pm$0.1 & $1.2\times10^{8}$  \\
1394	& NG & 208.4$\pm$56.3 &2.29    & 11.5$\pm$0.3 	 & 3.5$\pm$0.9 & 3.3$\pm$0.1 & 1.1$\pm$0.1 & $1.2\times10^{8}$ \\
1533 	& NG & 97.4$\pm$26.3 &2.45   & 11.6$\pm$0.3    & 1.8$\pm$0.5 & 1.3$\pm$0.1  & 0.9$\pm$0.5  & $6.5\times10^{8}$\\      
1666 	& NG & 244.5$\pm$66.0 &*1.76    & 11.9$\pm$0.4 	 & 4.7$\pm$1.3 & 6.2$\pm$0.1 & 0.6$\pm$0.1 & $1.2\times10^{8}$   \\
1768 	& NG & 314.7$\pm$85.0 &2.22   & 11.4$\pm$0.3 	 & 3.6$\pm$1.0 & 1.4$\pm$0.1  & 2.1$\pm$0.6  & $1.2\times10^{8}$ \\
1888	& NG & 133.4$\pm$36.0 &2.75   & 11.2$\pm$0.4	 & 3.5$\pm$0.9 & 2.0$\pm$0.5  & 5.0$\pm$1.2  & $1.2\times10^{8}$  \\
2083 	& NG & 160.6$\pm$43.4 &2.31    & 11.1$\pm$0.4	 & 3.5$\pm$0.9 & 1.2$\pm$0.2  & 1.3$\pm$0.6  & $1.2\times10^{8}$  \\
2411	& OG & 241.7$\pm$65.3 &2.09    & 11.0$\pm$0.3    & 3.8$\pm$1.0 & 5.2$\pm$0.1 & 0.6$\pm$0.1 & $6.5\times10^{8}$   \\
2564 	& NG & 127.8$\pm$34.5 &2.10 	 & 11.1$\pm$0.3     & 3.5$\pm$0.9 & 1.4$\pm$0.1  & 1.8$\pm$0.3  & $1.2\times10^{8}$\\
2667    & OG & 284.8$\pm$76.9 &1.79    & 11.2$\pm$0.4     & 3.5$\pm$0.9 & 2.0$\pm$0.1  & 3.5$\pm$0.9  & $6.5\times10^{8}$  \\   
2678	& NG & 51.8$\pm$14.0 &2.30    & 11.1$\pm$0.3     & 1.2$\pm$0.3 & 1.4$\pm$0.1  & 0.8$\pm$0.1 & $1.2\times10^{8}$  \\
2798	& NG & 196.2$\pm$53.0 &1.72    & 11.6$\pm$0.3    & 3.5$\pm$0.9 & 4.5$\pm$0.6  & 4.6$\pm$0.2  & $6.5\times10^{8}$ \\
3629	& NG & 140.6$\pm$37.0 & 2.17    & 11.4$\pm$0.2     & 2.5$\pm$0.7 & 1.8$\pm$0.2  & 1.3$\pm$0.2  & $6.5\times10^{8}$  \\
3766	& NG & 76.0$\pm$20.5 & 1.87    & 11.2$\pm$0.2    & 2.2$\pm$0.6 & 2.8$\pm$0.1  & 1.9$\pm$0.3  & $1.4\times10^{9}$  \\ 
3818	& NG & 255.9$\pm$69.1 &1.82   & 11.5$\pm$0.4     & 3.5$\pm$0.9 & 3.7$\pm$0.4  & 5.1$\pm$1.3  & $6.5\times10^{8}$    \\
3822    & NG & 115.3$\pm$31.1 &2.41   & 11.1$\pm$0.5     & 3.5$\pm$0.9 & 1.9$\pm$0.1  & 2.1$\pm$0.5  & $1.2\times10^{8}$  \\
4033	& OG & 81.5$\pm$22.0 &1.72    & 11.2$\pm$0.3    & 1.0$\pm$0.3 & 6.1$\pm$0.8  & 1.2$\pm$0.1 & $1.7\times10^{9}$  \\
4036    & OG & 134.3$\pm$36.3 &1.72   & 11.0$\pm$0.3     & 3.5$\pm$0.9 & 1.7$\pm$0.4  & 7.5$\pm$2.0  & $6.5\times10^{8}$ \\
4315    & NG& 231.1$\pm$62.4 &2.85   & 11.0$\pm$0.4     & 4.0$\pm$1.1 & 0.9$\pm$0.5  & 1.8$\pm$1.3  & $1.2\times10^{8}$ \\
4471	& OG& 401.0$\pm$108.3 &*2.29    & 11.2$\pm$0.4     & 3.5$\pm$0.9 & 2.8$\pm$0.1  & 1.4$\pm$0.1 & $6.5\times10^{8}$ \\
4557	& NG& 232.6$\pm$62.8 &2.09    & 11.3$\pm$0.4     & 3.1$\pm$0.8 & 4.6$\pm$0.1  & 1.4$\pm$0.1  & $1.2\times10^{8}$ \\
4706    & NG & 236.2$\pm$63.8 &*2.35    & 11.1$\pm$0.2     & 3.4$\pm$0.9 & 0.9$\pm$0.1 & 1.7$\pm$1.2  & $1.2\times10^{8}$   \\
4737    & NG& 34.4$\pm$9.2 &2.52   & 11.0$\pm$0.3    & 1.3$\pm$0.4 & 1.1$\pm$0.3 & 3.3$\pm$3.7  & $1.2\times10^{8}$ \\
4882    & IG& 448.4$\pm$121.1 &1.67   & 11.1$\pm$0.3    & 4.2$\pm$1.1 & 1.2$\pm$0.1 & 2.2$\pm$0.2  & $6.5\times10^{8}$ \\
5282    & IG& 276.8$\pm$74.7 &1.64   & 11.0$\pm$0.3    & 3.5$\pm$0.9 & 4.9$\pm$0.1 & 1.1$\pm$0.1 & $2.7\times10^{9}$\\
5764 	& NG & 170.2$\pm$41.6 &2.54   & 11.5$\pm$0.3     & 3.5$\pm$0.9 & 1.7$\pm$0.1  & 2.9$\pm$0.4  & $1.2\times10^{8}$ \\
6035    & NG & 208.6$\pm$56.3 &1.60   & 11.3$\pm$0.4     & 3.5$\pm$0.9 & 2.0$\pm$0.1  & 4.2$\pm$0.6  & $6.5\times10^{8}$\\
6114    & NG & 108.0$\pm$29.2 &1.96   & 11.3$\pm$0.3    & 2.5$\pm$0.7 & 1.5$\pm$0.6  & 0.3$\pm$0.3 & $6.5\times10^{8}$\\
6220    & IG& 558.3$\pm$150.7 &1.71  & 11.0$\pm$0.4     & 5.1$\pm$1.4 & 3.8$\pm$0.1  & 1.5$\pm$0.1 & $6.5\times10^{8}$ \\
6267    & OG& 136.2$\pm$36.8 &*1.54   & 11.4$\pm$0.2    & 1.6$\pm$0.4 & 2.3$\pm$0.1  & 2.3$\pm$0.5  & $1.7\times10^{9}$ \\
6287    & NG& 102.9$\pm$27.8 & 1.84  & 11.0$\pm$0.4     & 3.5$\pm$0.9 & 2.5$\pm$0.1 & 2.4$\pm$0.4  & $6.5\times10^{8}$\\
6514    & NG& 199.6$\pm$53.9 &2.49    & 11.2$\pm$0.3    & 3.2$\pm$0.9 & 2.7$\pm$0.2  & 2.5$\pm$0.1 & $1.2\times10^{8}$  \\
6584    & NG& 180.4$\pm$48.7 &1.62   & 11.2$\pm$0.3    & 3.5$\pm$0.9 & 6.2$\pm$0.1 & 1.1$\pm$0.1 & $1.2\times10^{8}$ \\
7072    & OG& 101.0$\pm$27.3 &1.74   & 11.1$\pm$0.2    & 1.9$\pm$0.5 & 1.4$\pm$0.1  & 2.3$\pm$0.4  & $1.4\times10^{9}$ \\
7475    & OG& 377.9$\pm$102.0 &*1.61   & 11.2$\pm$0.3    & 3.5$\pm$0.9 & 4.9$\pm$0.2  & 2.1$\pm$0.2  & $6.5\times10^{8}$ \\
8140    & OG& 319.6$\pm$86.3 &*1.90    & 11.4$\pm$0.4     & 3.0$\pm$0.8 & 1.8$\pm$0.1  & 2.7$\pm$0.5  & $6.5\times10^{8}$   \\
8214    & OG& 817.0$\pm$220.6 &2.05   & 11.3$\pm$0.2    & 3.8$\pm$1.0 & 1.6$\pm$0.1 & 1.7$\pm$0.2  & $6.5\times10^{8}$\\

  \hline  
\end{tabular}
\caption{(col. 1) ID number of the galaxy; (col.2) The classification of the galaxy based on the location of the star formation (see \S3.3); Non\--significant star formation growth (NG), outer star formation growth (OG) and inner star formation growth (IG) ; (col.3) Total observed UV star formation rate in solar masses per year ; (col. 4) Best redshift of the object, spectroscopic redshifts are denoted by * ; (col. 5) Stellar Mass with error in units of log$_{10}M_{\odot}$ calculated from multi colour stellar population fitting techniques  ; (col. 6) $A_{2800}$ Dust correction and error in magnitudes, determined from UV slope fitting ; (col. 7) Effective radius and error in units of kpc from S\'{e}rsic $r^{1/n}$ 2D models fits of the $H_{160}$ band data using GALFIT. ; (col. 8) S\'{e}rsic index and error from S\'{e}rsic $r^{1/n}$ 2D models fits of the $H_{160}$ band data using GALFIT. ; (col. 9) e-folding star formation time in years calculated from multi\--colour stellar population fitting techniques (see \S2.3).}
\label{tab:values}
\end{center}
\end{table*}
\end{center}

We measure star formation profiles for our sample using the IRAF program \emph{ellipse} by fitting a series of isophotal ellipses to the $z_{850}$\--band data with the $\rm{H_{160}}$\--band determined centre of the massive galaxies. This isophotal fitting returns the $z_{850}$\--band  flux binned in a series of increasing radii. This is then converted to a dust corrected star-formation rate in each radius bin via the procedure described in \S 2.5. Examples of such surface brightness profiles are shown in Figure \ref{sb}.

The total galaxy magnitudes obtained from this isophotal fitting are checked against a previous catalogue of the $z_{850}$\--band magnitudes for these galaxies in Bauer et al (2011) measured with SExtractor, and are found to be consistent.
 
In order to measure how this star formation affects the local mass density of the galaxy we simulate the amount of stellar mass added via star formation in each radius bin by assuming that the same global star formation history we use in \S 2.3 applies throughout. We also apply several other star formation histories that these galaxies could experience such as constant SFR to $z=0$, constant SFR to $z=1.5$ and variations on the derived tau model. These are disused in \S 5.4.2. We find very similar results as discussed below for the tau models.

The exponentially declining model of star formation uses the observed dust-corrected rest-frame UV star formation as the initial rate, $SFR_{0}$, and the values of the e-folding time, $\tau$, which we obtained from  the $M_{*}$ fitting (see \S 2.3). To obtain the total amount of stellar mass added via star formation, $M_{SF}$, we integrate Equation 1 over time from the total look back time derived from the redshift of the galaxy, ranging from $\sim9.7$\, Gyr at $z=1.7$ to $\sim11.5$\, Gyr at $z=3$ to the present day. We experimented with evolving the massive galaxies only until $z=1$ but found very similar results as the evolution to $z=0$ as the majority of the evolution in both size and structural properties of these massive galaxies seems to occur within the first $\sim2$ Gyr of our simulation.

\begin{figure}
\includegraphics[scale=0.45]{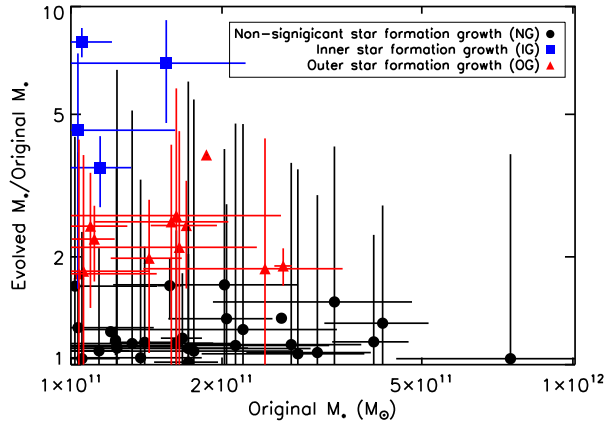}
\caption{Total stellar mass before ($\rm{Original\,M_{*}}$) and after evolution ($\rm{Evolved\,M_{*}}$) from the derived tau model of star formation evolution. The black circle represent the non\--significant star formation growth galaxies. The blue squares represent the inner star formation growth galaxies. The red triangles represent the outer star formation growth galaxies (see \S 3.3).}
\label{mass}
\end{figure}

In Figure \ref{mass} we show the change in the total stellar mass of each of the galaxies within our sample as measured through the SFR. We find that the total stellar mass on average increases by $90\%$ via this modelled star formation. The evolved total stellar masses of our galaxies do not exceed constraints placed upon the observed total stellar mass evolution from other studies (e.g. Cole et al. 2001; Bell et al. 2003; Conselice et al. 2007; Brammer et al. 2011; Mortlock et al, 2011). Brammer et al. (2011) show that the total stellar mass growth for massive galaxies from $z\sim2$ to $0$ is on the order of $100\%$. We see that there is an anticorrelation between the original stellar mass and the evolved stellar mass, with some of the lower mass galaxies increasing substantially in stellar mass, while the higher mass galaxies have a much smaller change in mass over cosmic time. This is a sign of galaxy downsizing (e.g. Cowie et al. 1996;  Bundy et al. 2006), such that the most massive galaxies are less affected by star formation at $z<3$ than the lower mass galaxies.

This total stellar mass added via star formation is converted to a stellar mass projected  density via, $\rho_{SR}=M_{SF}/A_{an}$, were $A_{an}$ is the area of the annulus the star formation is contained within. From this we construct a new stellar mass profile by including the stellar mass added via star formation to the original profile via:

\begin{equation}
 M_{*}(R,t)= M_{*}(R,t=0)+SFR_{0}(R)\int_0^{t}e^{-t/\tau} dt
\end{equation}

\noindent Where $M_{*}(R,t=0)$ is the initial stellar mass at radius $R$, $SFR_{0}(R)$ is the observed initial SFR at the same radius and $M_{*}(R,t)$ is the stellar mass at radius $R$ after time $t$. We consider other cases of SF histories in \S5.4.

\subsection{Profiles}

In this section we examine the profiles of the stellar mass density distributions at high redshift, the stellar mass density added via star formation, and for the combination of the two - an evolved stellar mass density profile for each of the 45 massive galaxies in our sample. The evolved profiles are then fit with a new S\'{e}rsic profile.
The S\'{e}rsic profiles for the evolved stellar mass profiles are obtained by the best fitting S\'{e}rsic (1968) function to the new profile,

\begin{equation}
\rho(R) = \rho_{e}\times\rm{exp}(-b_{n}[(R/R_{e})^{1/n}-1]) 
\end{equation}

\noindent
We find that the galaxies in our sample can be classified into three distinct groups based on the location of star formation regions and the effect they have on our sample galaxy's evolved stellar mass density profile. 

To examine the results we first divide the galaxies into two regions. An inner region, at $R \le 1 $kpc with an observed initial stellar mass density in the inner region, $\rho_{observed,inner}$ and a stellar mass density added via star formation in the inner region, $ \rho_{SF,inner}$. An outer region, $R > 1 $kpc with a observed stellar mass density, $\rho_{observed,outer}$ and a stellar mass density added via star formation in the outer region, $ \rho_{SF,outer}$. We chose 1kpc as the boundary for our inner region based on stellar surface brightness comparisons at high and low redshifts (e.g. Hopkins et al. 2009; Carrasco et al. 2010; Szomoru et al. 2011).  We discuss the three types below.

\underline{Non\--significant Star Formation Growth}  (NG) : This category is for galaxies in which the stellar mass density added via star formation is smaller then the galaxy's initial stellar mass density present over both the inner and outer regions. $\rho_{observed,inner} > \rho_{SF,inner}$ and $\rho_{observed,outer} > \rho_{SF,outer}$.

\underline{Outer Star Formation Growth} (OG) : In this category the stellar mass density added via star formation is greater than the initial stellar mass density present in the outer region, but the initial stellar mass density in the inner region is greater than the stellar mass density added via star formation; $\rho_{observed,inner} > \rho_{SF,inner}$ but $\rho_{observed,outer} < \rho_{SF,outer}$

\underline{Inner Star Formation Growth} (IG) : This category is for galaxies in which the initial stellar mass added via star formation is greater over both regions than the stellar mass density present, $\rho_{observed,inner} < \rho_{SF,inner}$ and $\rho_{observed,outer} < \rho_{SF,outer}$

\begin{figure*}
\subfloat[]{\includegraphics[scale=0.5]{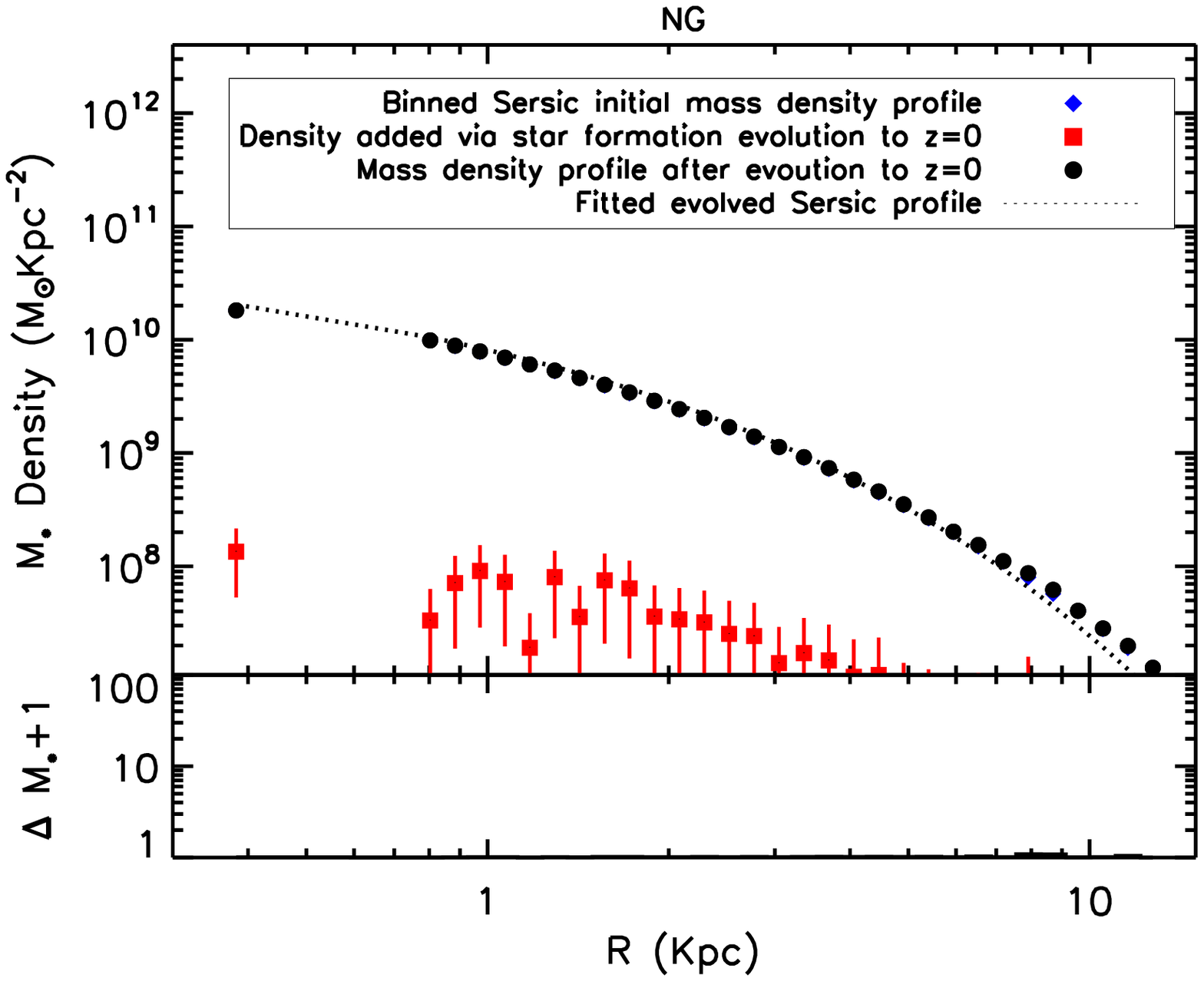}}                
  \subfloat[]{\includegraphics[scale=0.5]{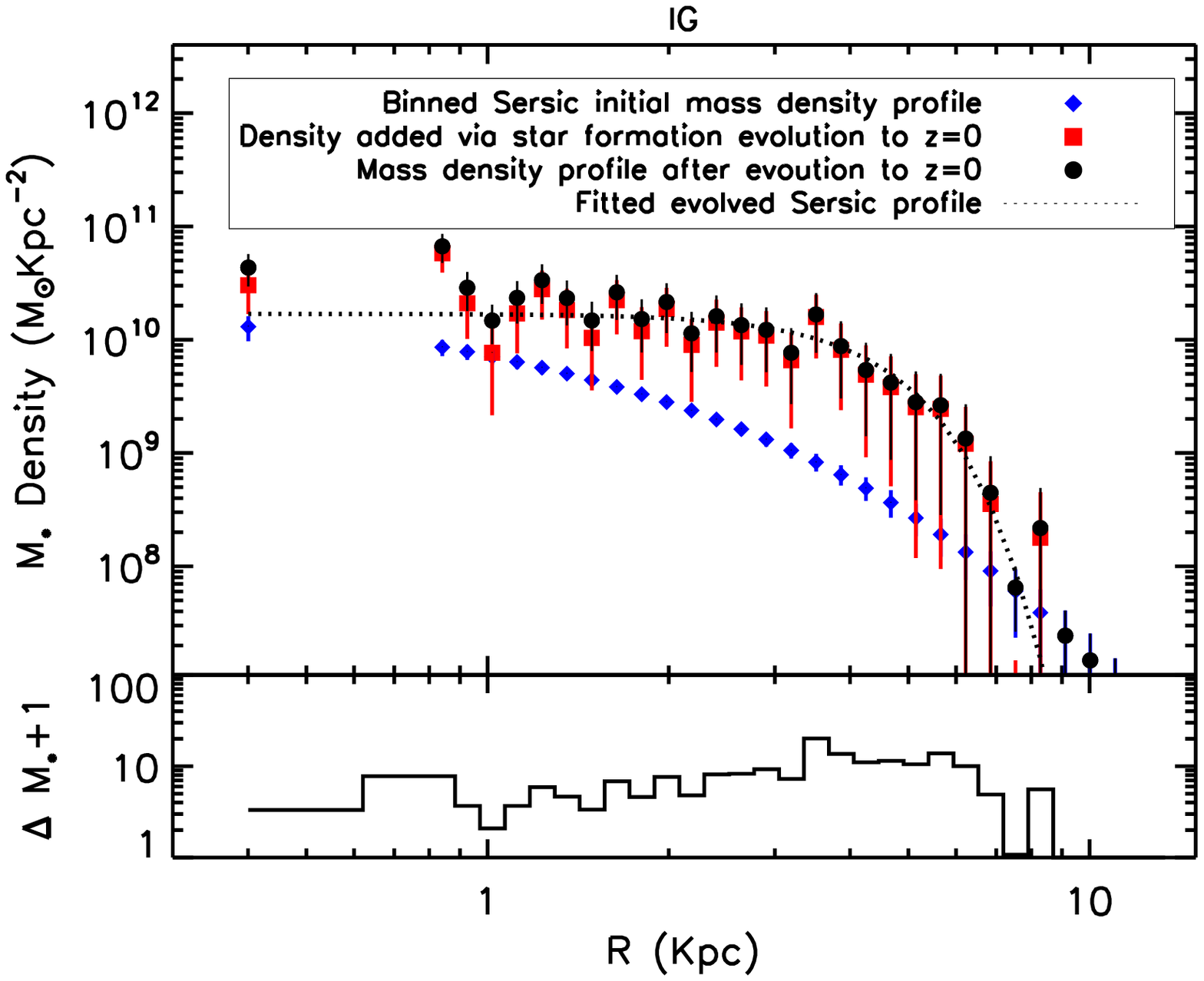}} 
\\       
  \subfloat[]{\includegraphics[scale=0.5]{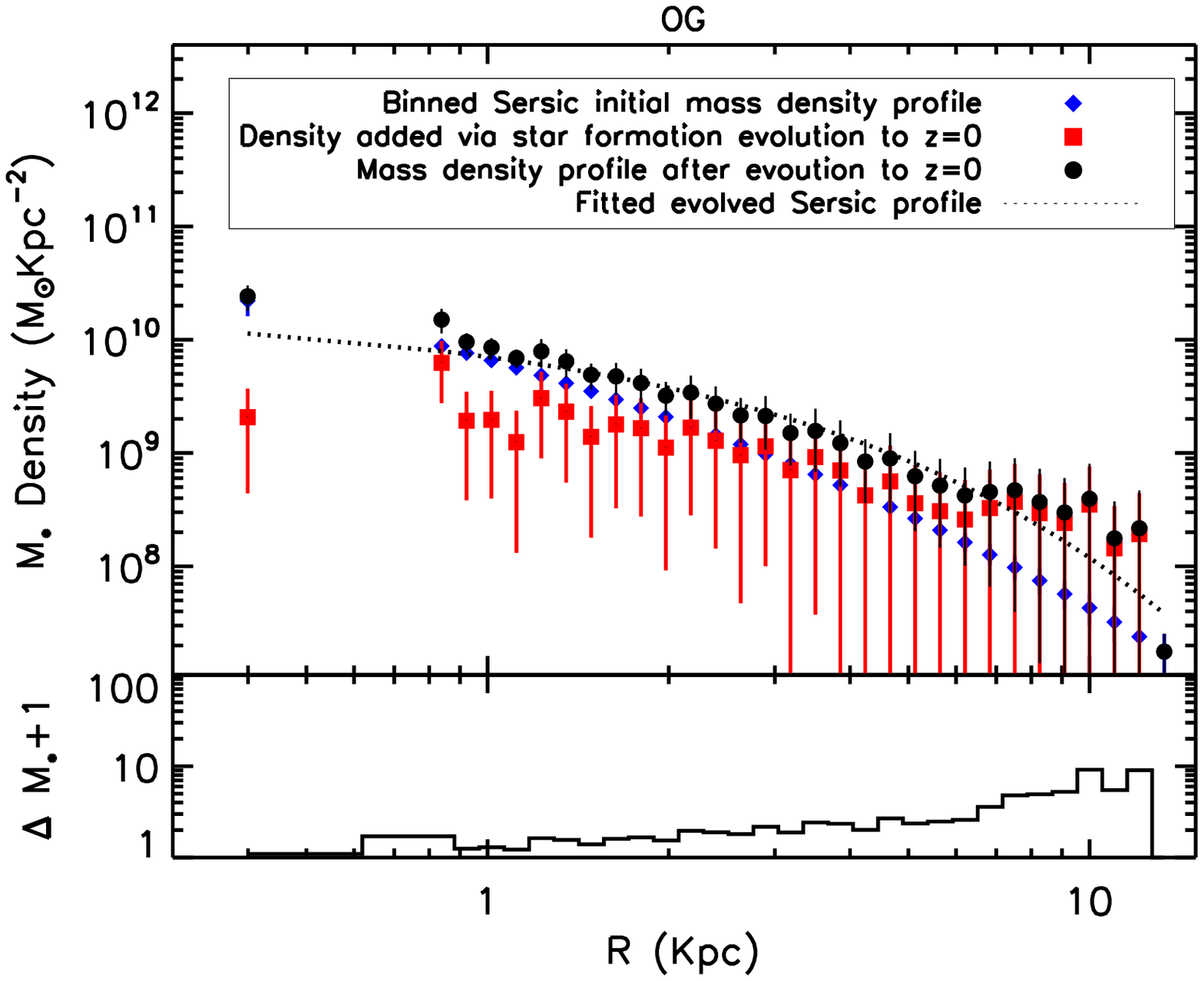}} 
\caption{Example of the three star formation growth classifications. a)Non\--significant star formation growth  (NG) b)Inner star formation growth (IG) c) Outer star formation growth (OG). The blue diamonds represent the observed stellar mass density present at high redshift based on the H$_{160}$-band S\'{e}rsic profile. The red squares represent the stellar mass density added via star formation to $z=0$. The black circles represent the combined (evolved) profiles of both the stellar mass density present at high redshift and the stellar mass density added via star formation to $z=0$. The black dotted line is the best fit S\'{e}rsic profile to the evolved stellar mass density profile. The sub\--plot shows the change in the stellar mass density profile from the stellar mass density present at high redshift compared to the evolved stellar mass density profile. }
\label{class}
\end{figure*}

In Figure \ref{class} we show examples of the three different galaxies classes and in Table \ref{tab:numbers} we list the numbers of each class we have in our sample.

\begin{center}
\begin{table}
\begin{center}
\begin{tabular}{| c | c | c | }
  \hline
  Type & No. of galaxies & $\%$ of sample \\
  \hline             
  Non\--significant SF Growth (NG) & $29$ & $64.4^{+4.5}_{-33.3}$\\
  Inner SF Growth (IG) & $4$ &  $8.9^{+2.2}_{-2.2}$\\
  Outer SF Growth (OG)& $12$ &  $26.7^{+31.1}_{-2.3}$\\
  \hline  
\end{tabular}
\caption{Evolved massive galaxies using the derived tau model of SF evolution separated into the three classifications. Insignificant Star Formation Growth (NG), Inner Star Formation Growth (IG), Outer Star Formation Growth (OG). We see that  nearly half of the sample resides in the NG class with a significant fraction in the OG class.  }
\label{tab:numbers}
\end{center}
\end{table}
\end{center}

\section{Results}

The profiles for the stellar mass already in place at high redshift (\S 3.1) and the stellar mass added via star formation (\S 3.2) are combined to give an evolved modelled stellar mass density profile of the galaxy after evolving for  $\sim 10$ Gyr from $z = \sim2.5$ until the present day. Using the new stellar mass density profiles we fit a new S\'{e}rsic profile of the same form as Equation 5 to examine how the stellar mass added via star formation would change the structure and sizes of our massive galaxies over time.

\subsection{Stellar Mass}

Figure \ref{mass} shows the growth in total stellar mass for all of the galaxies within our sample. As stated before the average growth for the sample is  $91 \pm 22 \%$. The evolved total stellar masses of our sample of galaxies does not exceed constraints placed upon the observed total stellar mass evolution from other studies (e.g. Conselice et al. 2007; Brammer et al. 2011; Mortlock et al. 2011). This represents the maximal stellar mass increase, negating the effects of supernova and other types of feedback that would impede star formation and reduce the total amount of stellar mass created. From this figure we can also see that there is a clear divide between the three classes of galaxies in our sample (see \S3.3). The NG class galaxies have the smallest change in total stellar mass of $22 \pm 34\%$, and do not come close to doubling in stellar mass. The IG class has the largest change in stellar mass of $474 \pm 89\%$, and lie exclusively in the top region of the figure. The OG class of galaxies in this sample have an intermediate mass change of $129 \pm 90\%$. This is a clear segregation in stellar mass build up via star formation between the three classes based upon the star formation locations, showing that the three different classes also have differing specific star formation rates, with IG galaxies having the highest and NG having the lowest. This divide is also present in all of the other models of SF we applied to this sample.

\subsection{Structure and Size Evolution}

\begin{figure*}
\subfloat[]{\includegraphics[scale=0.5]{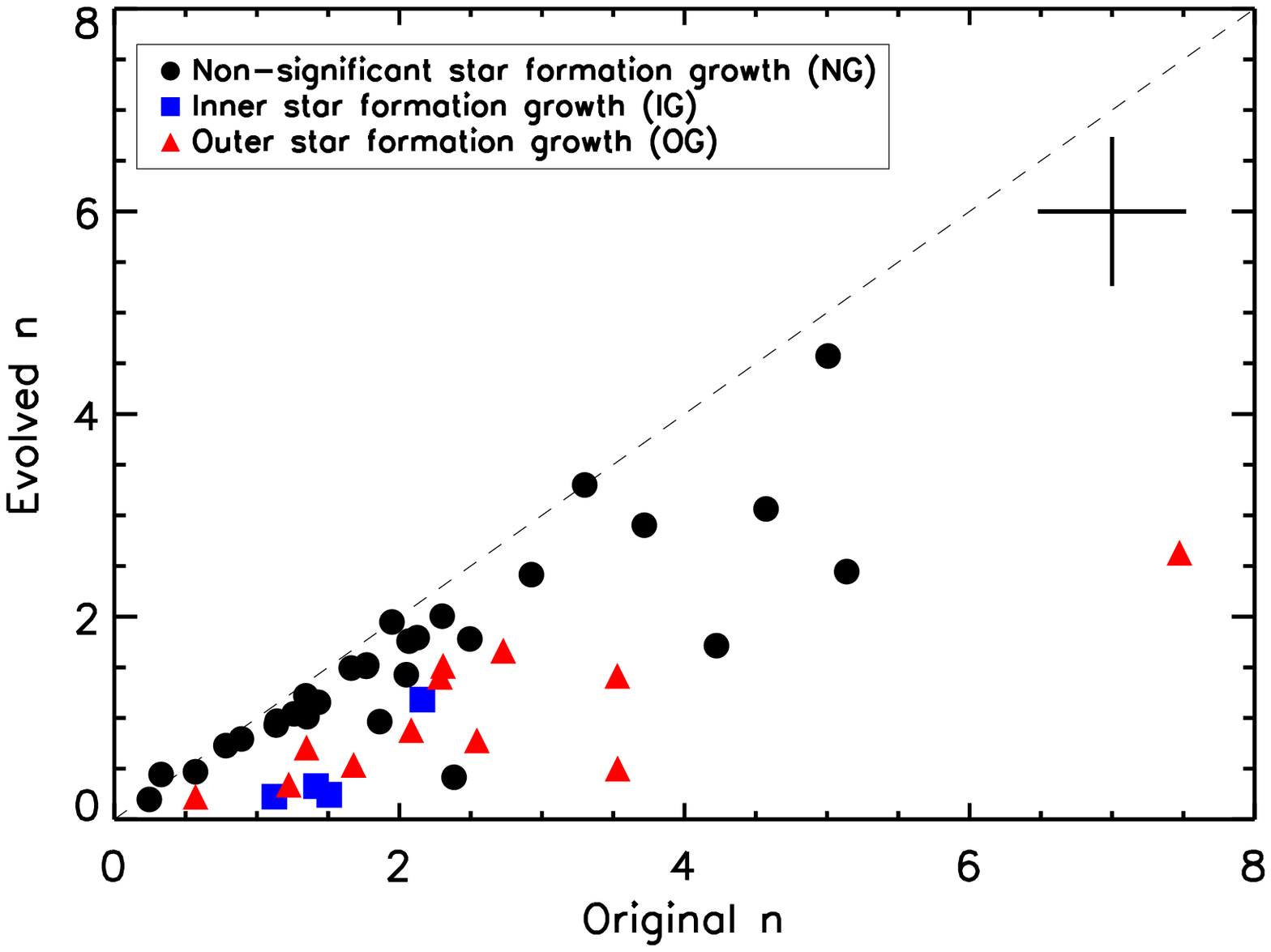}}                
  \subfloat[]{\includegraphics[scale=0.5]{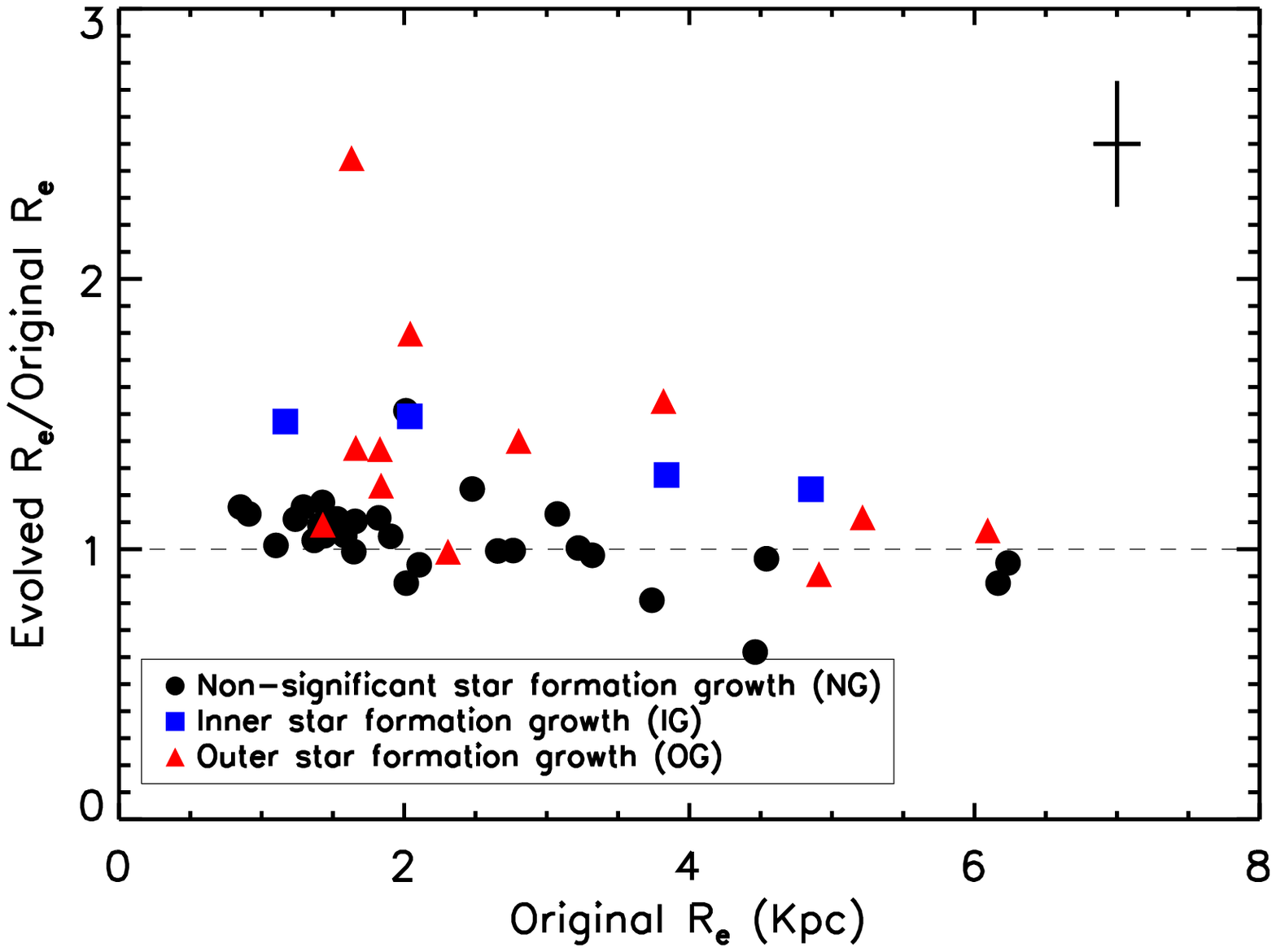}} 
 \caption{The evolution of the S\'{e}rsic index (a), and the effective radius (b) via star formation to $z=0$. The black circles denote galaxies classified as NG. The red triangles denote galaxies classified as OG and the blue squares denote galaxies classified as IG (see \S3.3). The dashed line in both cases shows a $1:1$ relation. The typical error bars are shown in the top right hand corner.}
\label{nomig}
\end{figure*}

We find from the S\'{e}rsic fits to the evolved profiles that the average change in $n$ over the whole massive galaxy sample is such that the S\'{e}rsic index goes slightly down, $n_{evolved}-n_{original} = \Delta n = -0.9 \pm 0.9$. This is consistent with a small change with the profile shape over time.

In Figure \ref{nomig}a we show that the change in  S\'{e}rsic index $n$ differs for the three profile classifications. The NG galaxies lie nearly completely along the 1:1 line, denoting a small change from the observed to the evolved $n$, $\Delta n = -0.6 \pm 0.1$. This is expected as these galaxies are classified as having a small amount of stellar mass density added via star formation compared to the observed stellar mass density. However, we note that the NG galaxies with a high original $n$ do not fall upon the 1:1, and have a lower $n$ after evolution due to small amounts of star formation in the outer regions.

Also, we find that the majority of the OG and IG galaxies lie below the 1:1 line. This reveals that these galaxies have a lowered $n$ value after star formation evolution. Over the whole OG class there is a change of $\Delta n = -1.6 \pm 0.4$, and $\Delta n = -1.1 \pm 0.3$ for the IG class.  The result is expected for the OG class as these are defined as galaxies where there is a disparity in the amounts of star formation between the two regions, inner and outer. This disparity results in the outer regions of the galaxy increasing in stellar mass density, while the inner region does not.

The small changes in S\'{e}rsic $n$ after star formation evolution shows that the star formation density within these galaxies largely follows the underlying stellar mass density profile. In the case of IG galaxies where the stellar mass added via star formation dominates over the entire galaxy, the initial stellar mass density profile is almost completely negligible after evolution but the new profile retains the same general shape.

The effective radius, $R_{e}$, for our entire sample increases by $16 \pm 5\%$ averaged over the entire sample after our simulation. Separating the galaxies into our different classifications we find that the NG class has a very minor increase in size of $4 \pm3\%$.  
We find the OG class has an increase in $R_{e}$ of $37 \pm 12\%$. This increase in the effective radius is due to the addition of stellar mass in the outer regions of these galaxies.
The IG class has an increase in $R_{e}$ of $36 \pm 16\%$. This small increase is most likely related to the non-changing $n$ we find for this class.

Figure \ref{nomig}b shows the evolution of the effective radius before and after star formation evolution. We find that galaxies in the NG class all lie close to their original effective radius with the other two classes having a larger change. We also see that the massive galaxies with smaller original effective radii have a larger growth in size after star formation evolution compared to systems with larger original effective radii.

\subsection{Stellar Migration}

\begin{figure*}
\subfloat[]{\includegraphics[scale=0.5]{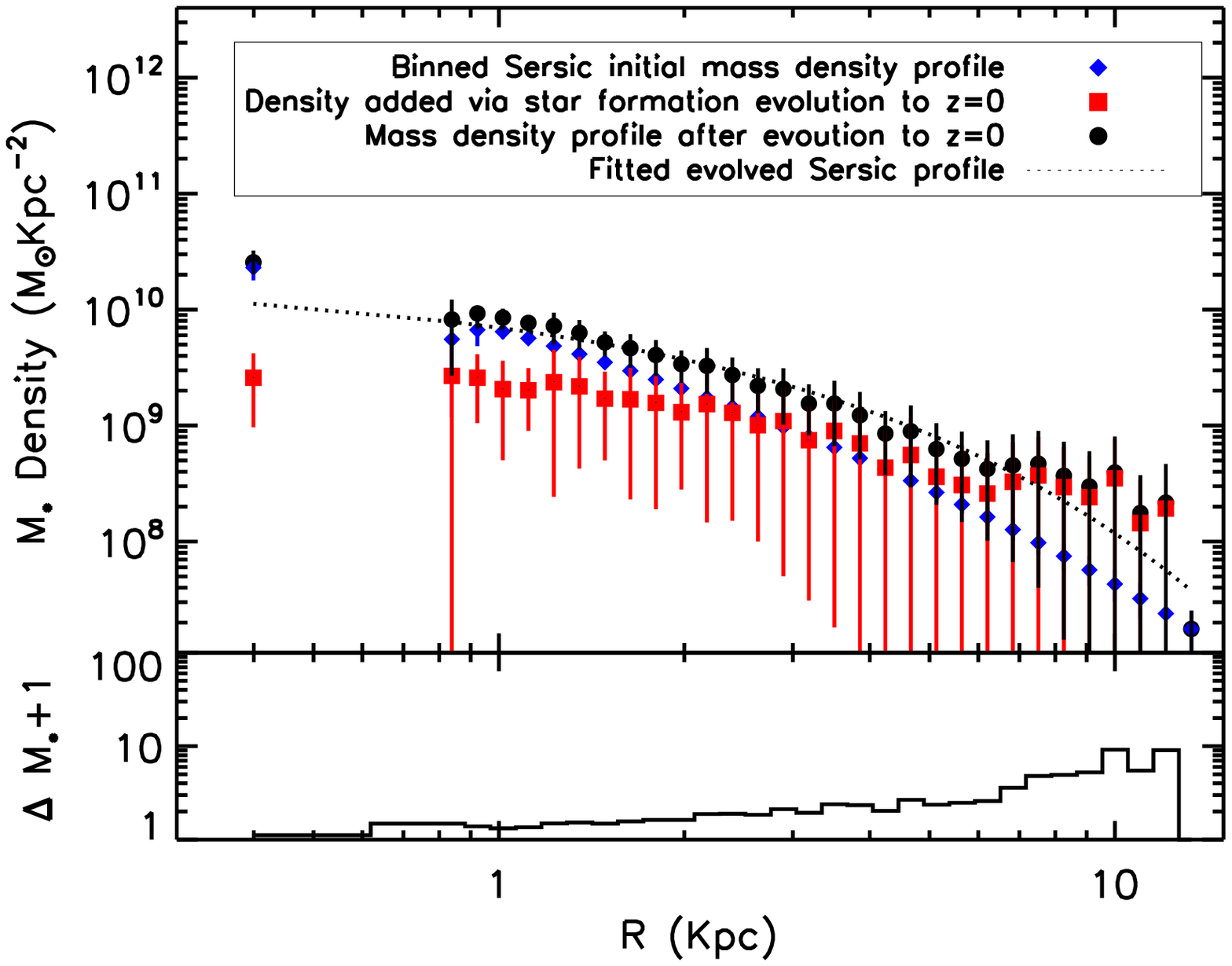}}                
  \subfloat[]{\includegraphics[scale=0.5]{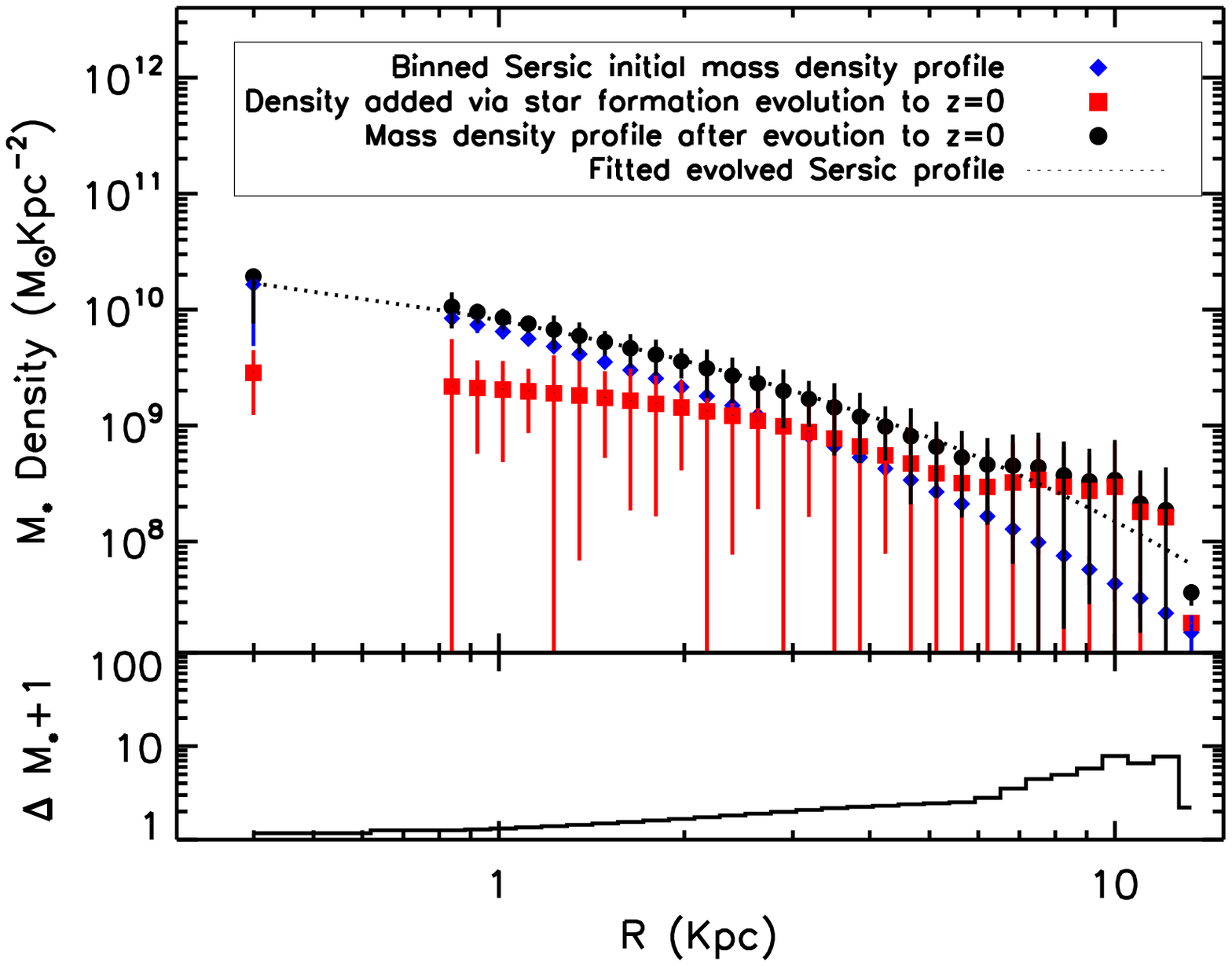}} 
\\       
  \subfloat[]{\includegraphics[scale=0.5]{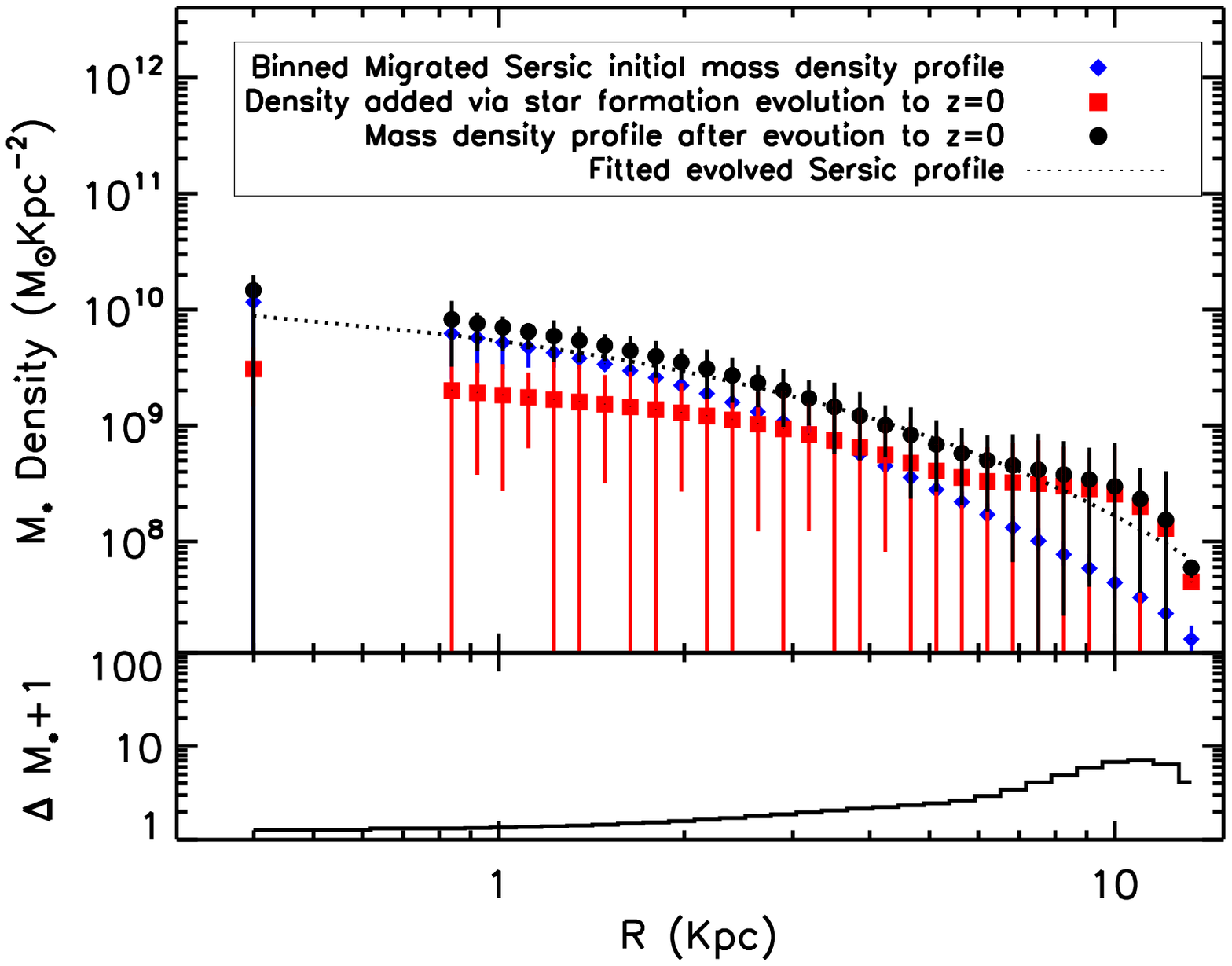}} 
\caption{Example of the effect of the stellar migration models on one example galaxy density profiles with increasing Gaussian widths of $\sigma=$ (a)0.1kpc, (b) 0.5kpc, (c) 1.0kpc. The non\--migration profile can be seen in Figure 4 (c) which is the galaxy we use in this example. The sub\--plot shows the change in the stellar mass density profile from the stellar mass density present at high redshift compared to the evolved stellar mass density profile. }
\label{mig}
\end{figure*}

\begin{figure*}
\subfloat[]{\includegraphics[scale=0.5]{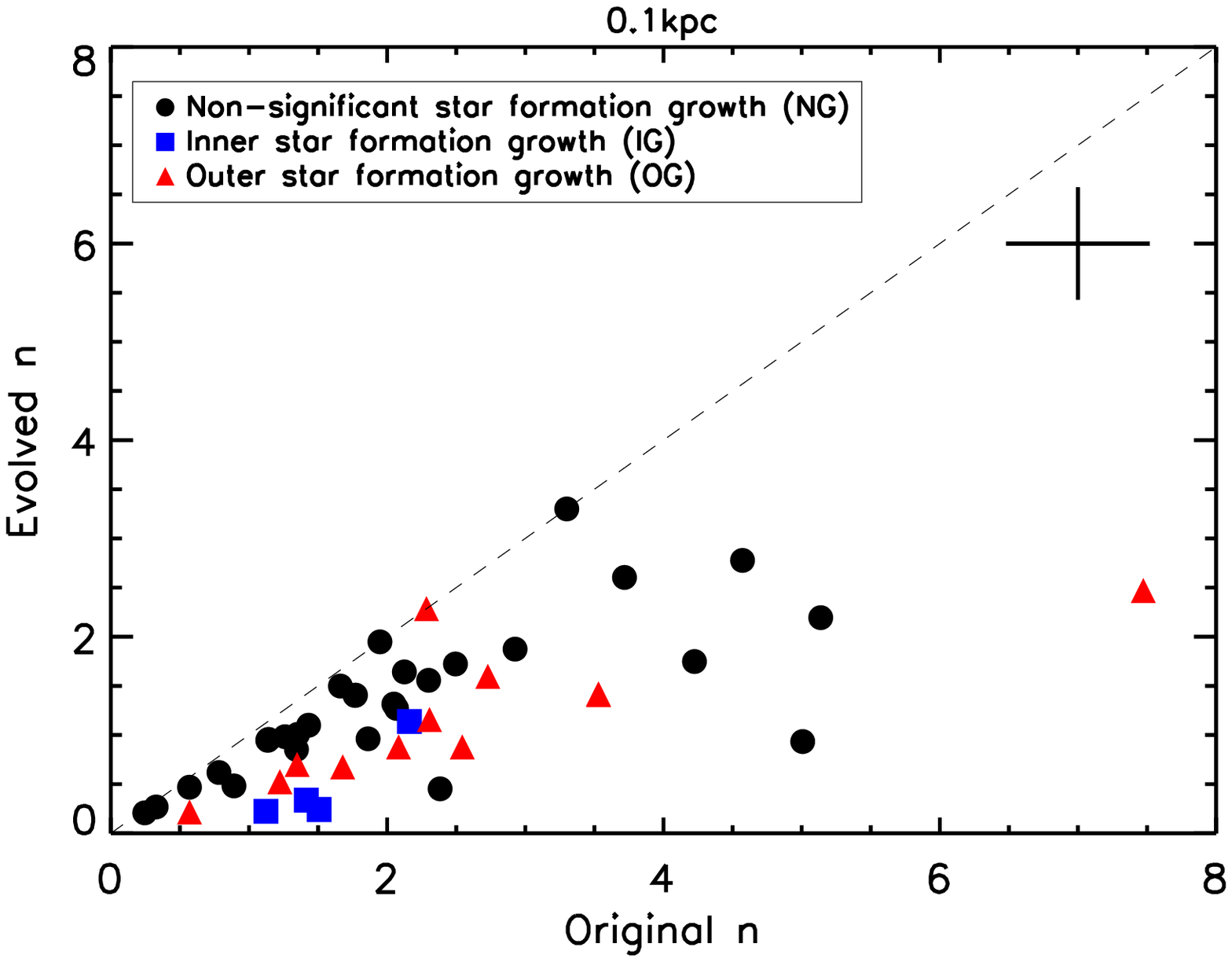}}                
  \subfloat[]{\includegraphics[scale=0.5]{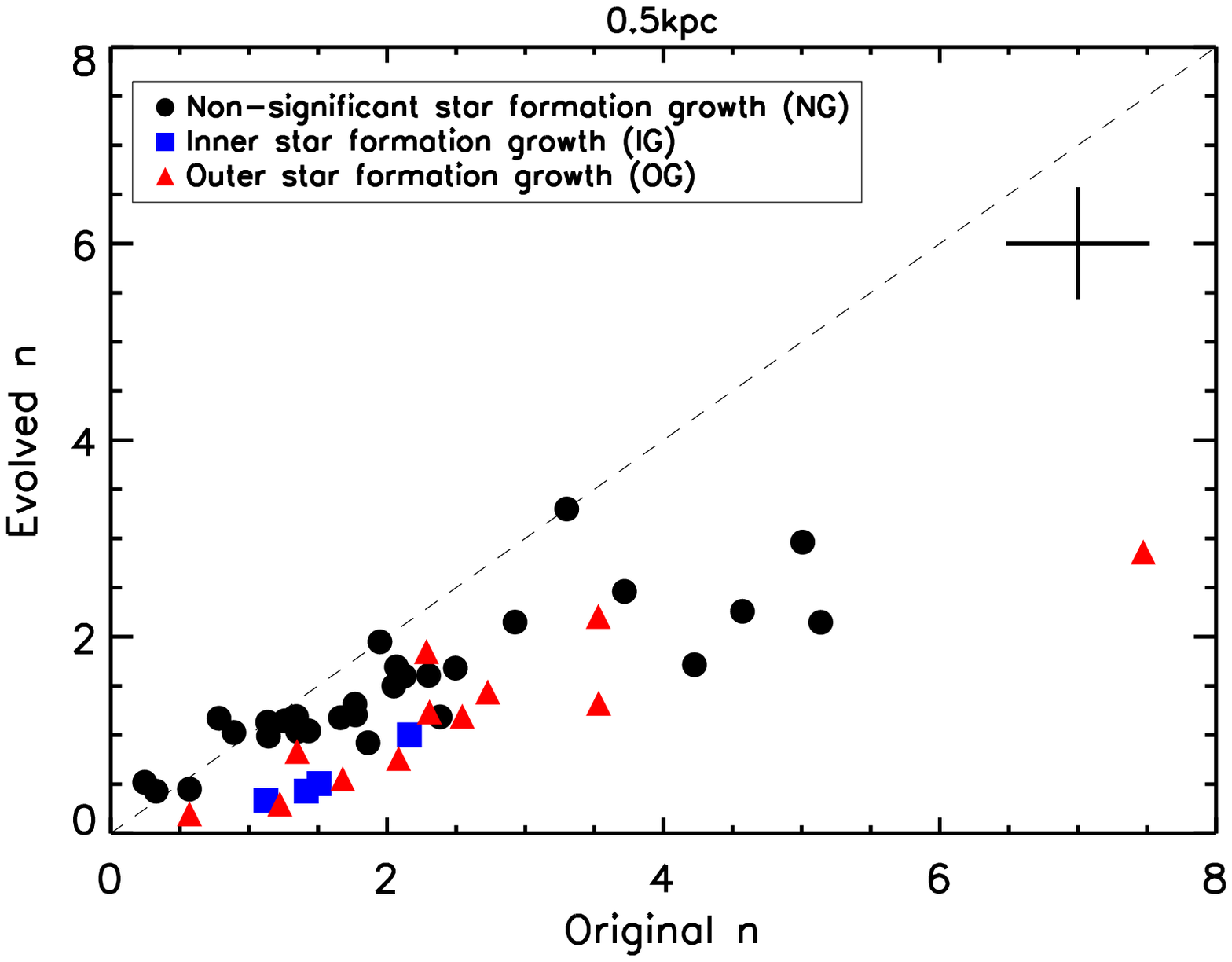}} 
\\
 \subfloat[]{\includegraphics[scale=0.5]{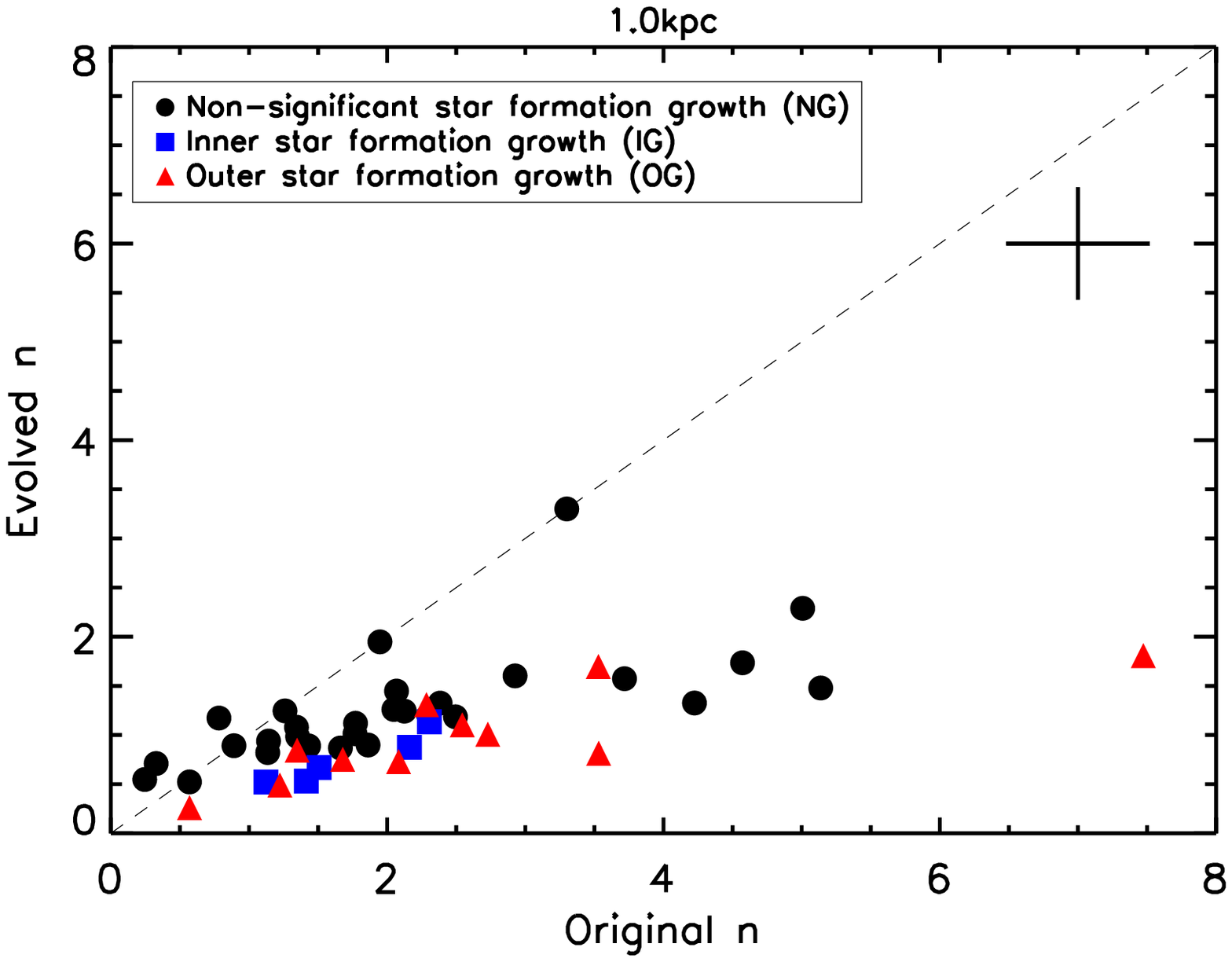}} 
\caption{The evolution of the S\'{e}rsic index with due to various stellar migration models, the Gaussian form for stellar migration is shown in Equation 8. The values of $\sigma$ used are (a) 0.1kpc, (b) 0.5kpc (c) 1.0kpc. The black circles denote galaxies classified as NG. The red triangles denote galaxies classified as OG and the blue squares denote galaxies classified as IG. The typical error bars are shown in the top right hand corner.}
\label{mign}
\end{figure*}

\begin{figure*}
\subfloat[]{\includegraphics[scale=0.5]{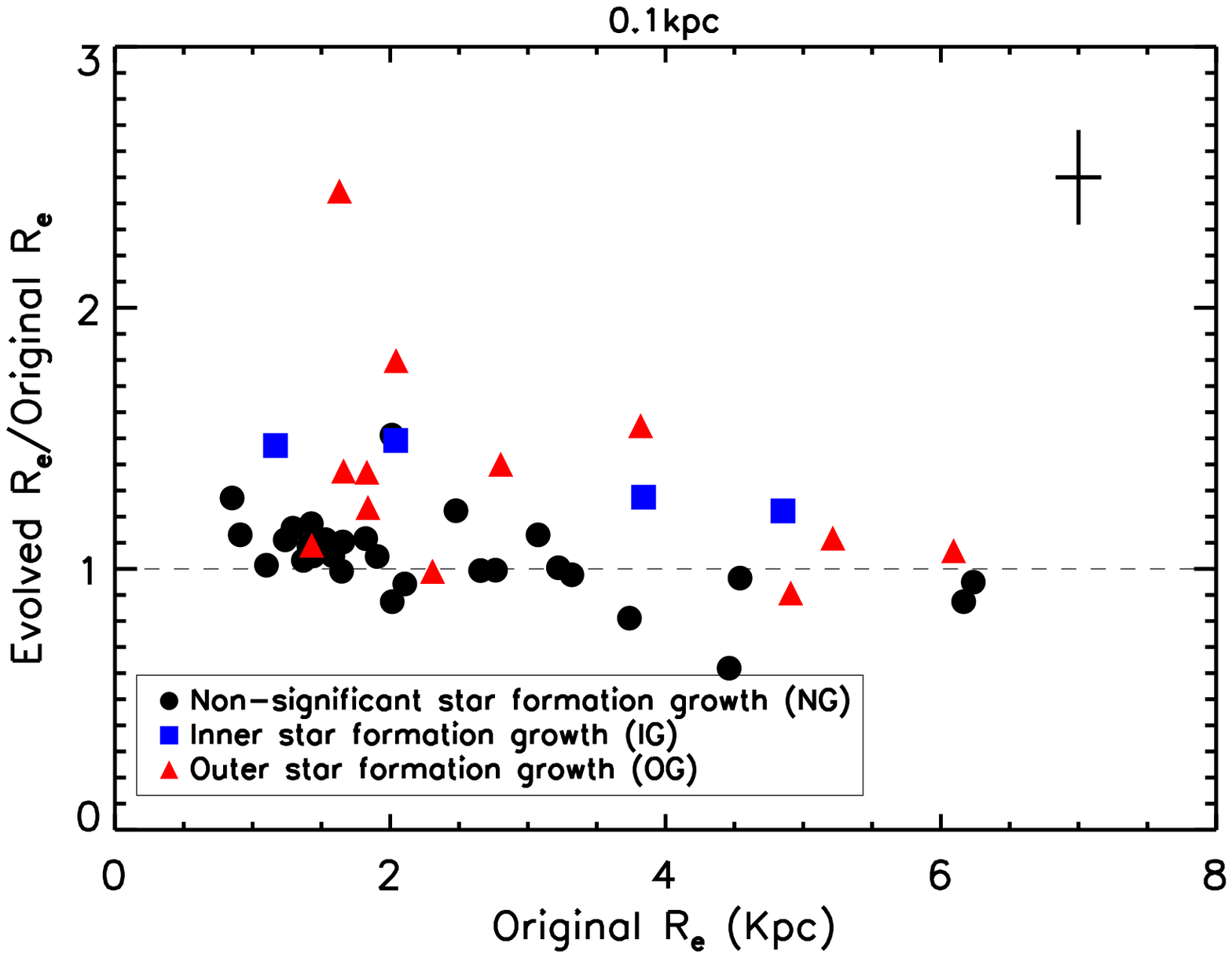}}                
  \subfloat[]{\includegraphics[scale=0.5]{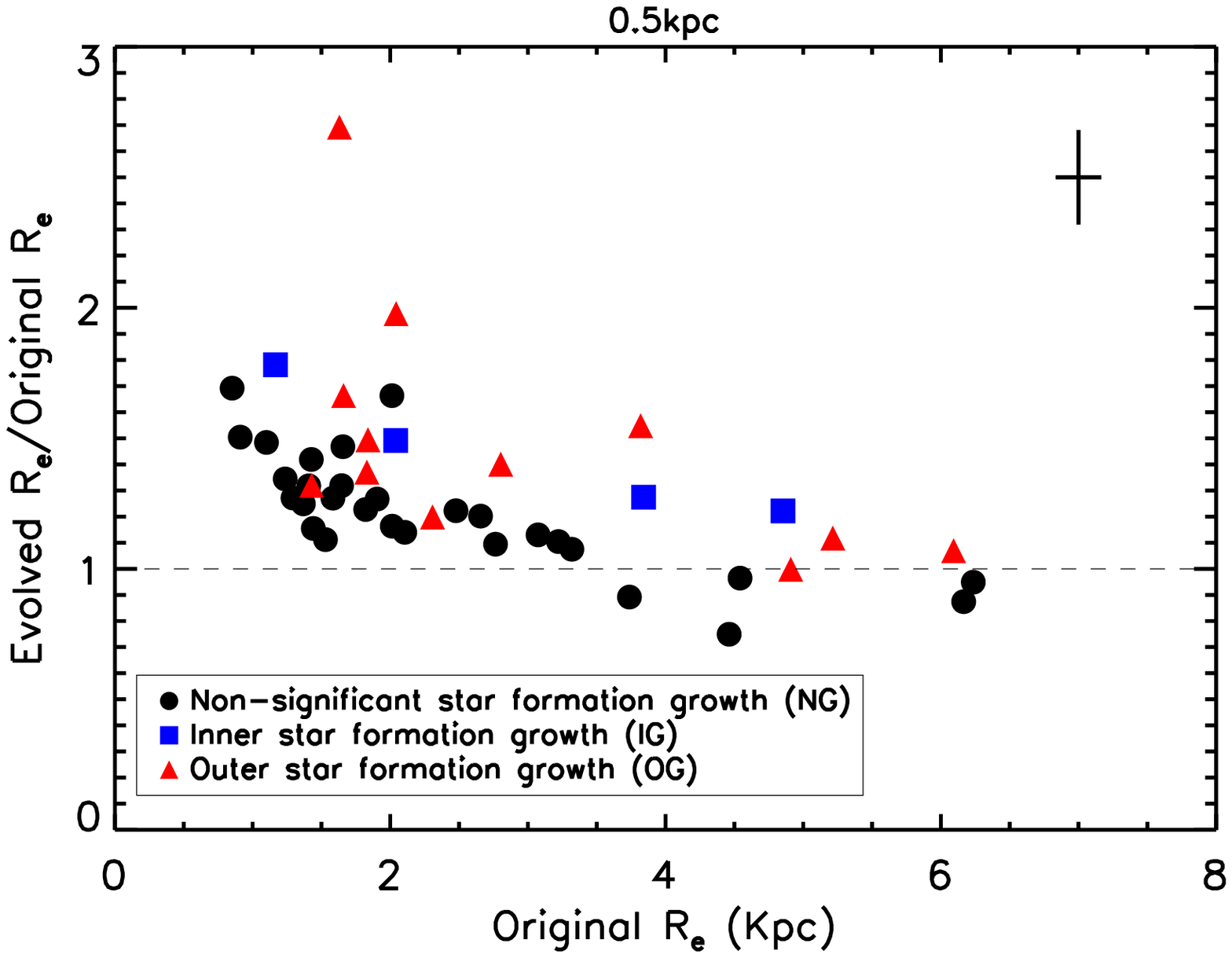}} 
\\
 \subfloat[]{\includegraphics[scale=0.5]{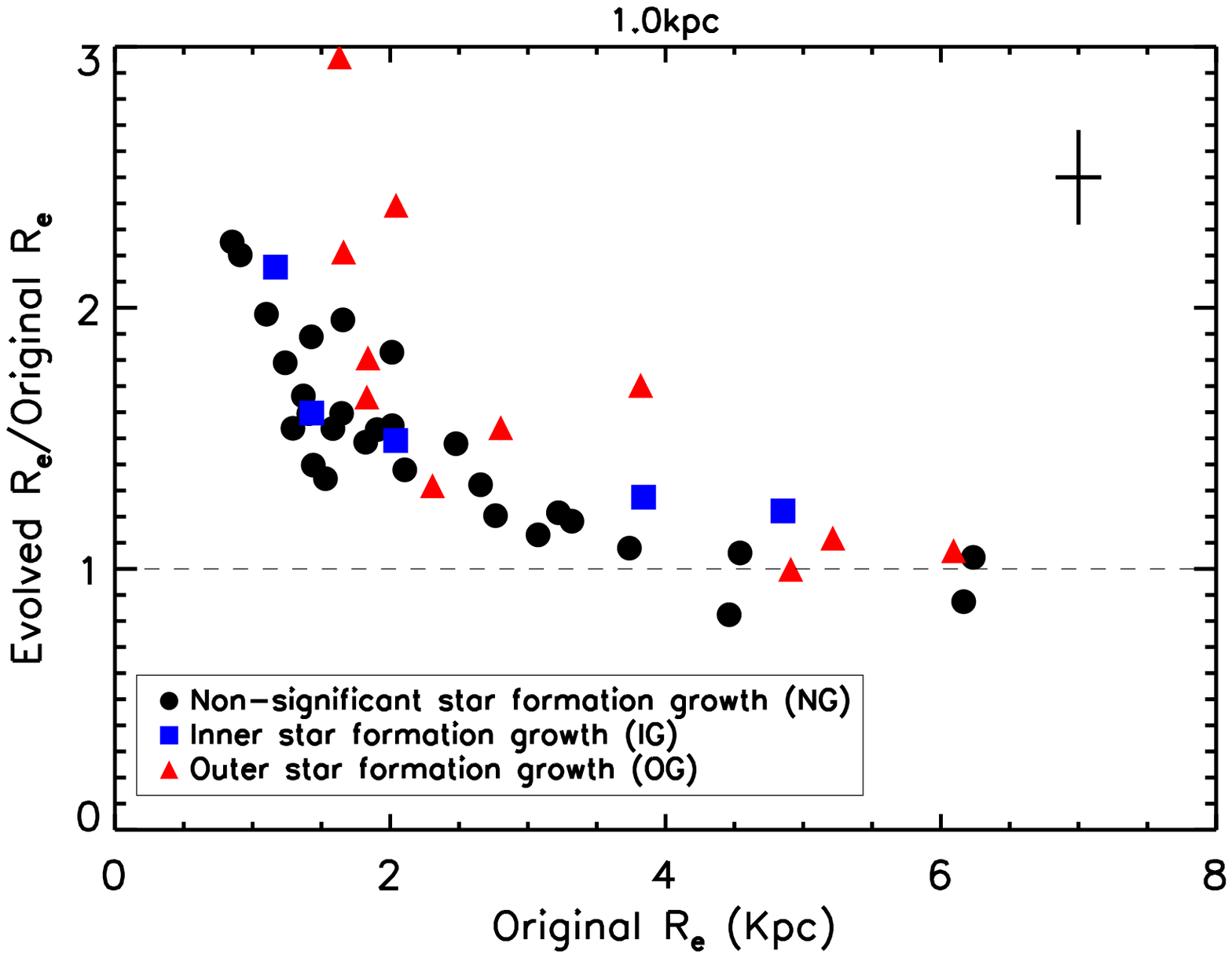}} 
\caption{The evolution of the effective radius due to various stellar migration models, the Gaussian form for stellar migration is shown in Equation 8. The values of $\sigma$ used are (a) 0.1kpc, (b) 0.5kpc (c) 1.0kpc. The black circles denote galaxies classified as having non\--significant star formation growth (NG). The red triangles denote galaxies classified as having outer star formation growth (OG) and the blue squares denote galaxies classified as having inner star formation growth (IG). The typical error bars are shown in the top right hand corner.}
\label{migre}
\end{figure*}

We show above that the star formation within the massive galaxies is not sufficient to produced a large growth in effective radius. We now investigate stellar migration as a method that may also be at work.

Recent theoretical work suggests that it may be common for stars to migrate radically across significant distances within spiral galaxies (Sellwood \& Binny 2002, Ro\u{s}kar et al. 2008).  These works showed that stellar migration happens via processes in the spiral arms of disk galaxies. However, we cannot reliably distinguish disk-like galaxies in out sample using a S\'{e}rsic index cut at $n=2.5$ because we cannot rule out that some of the galaxies with $n > 2.5$ do not have spiral like features (e.g. Buitrago et al. 2011). Therefore we add into our evolution models a simple stellar migration model to all the galaxies in the sample in order to gauge the effect this would have on the size and structural changes of the evolved galaxy profiles. 

In order to simulate this effect on the galaxies in this sample we apply a Gaussian distribution function centred on each individual radial bin across each galaxy in the sample. This distributes the total stellar mass (the stellar mass added via star formation and the in situ stellar mass) across the galaxy according to the summed Gaussian distribution:

\begin{equation}
M_{*,mig}(R,t) =\displaystyle\sum\limits_{i=R_{min}}^{R_{max}} M_{*}(i,t)\frac{1}{\sigma \sqrt{2\pi}}  e^{-\frac{(R-i)^{2}}{2\sigma^{2}}}
\end{equation}
where $M_{*}(i,t)$ is the total stellar mass at radius $i$ and at time $t$ from equation 6 and $\sigma$ is the width of the Gaussian distribution. The $R_{min}$ and $R_{max}$ are the total range of the galaxy radial distribution. Simulations from Ro\u{s}kar et al. (2008) showed that the radial migration of stars in a Milky Way type disk can change by several kpc over the lifetime of the galaxy. To simulate this with some scatter we run a series of different widths to represent a wide range of migrations with, $0.01\rm{kpc} < \sigma < 1.0\rm{kpc}$. The new stellar mass distribution is then fit with a new S\'{e}rsic profile. Figure \ref{mig} shows the effect of increasing levels of stellar migration upon one galaxy at different levels.

As expected, we find that larger levels of migration have an increasing stronger effect on the effective radius. The maximal effect on the radius of the galaxy is for the largest width Gaussian we applied, $1.0\rm{kpc}$. This level of migration is of the same order as the effective radii for the galaxies in our sample. Therefore to test maximal size growth we use a 1.0kpc migration from this point on. It is quite possible that stellar migration may happen on a larger scale and we experimented with wider Gaussian distributions but found that we could no longer accurately simulate the galaxy evolution due to losing stellar mass outside the confines of the simulation. From the $1.0\rm{kpc}$ Gaussian we find the effective radius grows by $54 \pm 19\%$ of the original effective radius. This represents an evolved effective radius $\sim 3.4$ times larger than with just the star formation evolution alone. The effect of this migration on the S\'{e}rsic indices on average are consistent with a small change, with $\Delta n = -1.1\pm 1.3$, similar to the non migration case. This is similar to the insignificant change in $n$ we found in the profiles without the stellar migration.

In Figure \ref{mign} we show how the S\'{e}rsic index changes using different stellar migration models.  When this is applied to our different galaxy classes we find that the NG galaxies have a increased effective radius of $48\pm 7\%$ over the initial effective radius. 

The IG galaxies show a increase in the effective radius of $55 \pm 15\%$. Compared to the increased radii from star formation alone this is a $\sim1.5$ times larger result. OG galaxies have an effective radius the largest increase with migration of $71\pm18\%$. This increase in effective radius is $\sim1.9$ times larger than the non\--migration case for the OG class of galaxies in this sample. This is likely due to these galaxies producing more stellar mass in the outer regions than the other classes buy definition, and therefore having a larger amount of stellar mass already at large radii to move during migration.
 
In Figure \ref{migre} we show how the effective radius changes due to the addition of the star formation and  stellar migration. The galaxy classes that have the highest star formation rates are affected the most by stellar migration due to having more new stellar mass to migrate, with the non\--changing, non\--significant star formation galaxies lying close to the non\--changing line and the outer and inner star formation growth galaxies lying above.
 
\section{Discussion}

\subsection{Size Evolution}

As stated previously, recent studies over the last few years have found evidence for a dramatic size evolution of massive galaxies over the past 10 billion years (e.g Daddi et al. 2005; Truijllo et al. 2007; van Dokkum et al. 2010; Buitrago et al. 2008, 2011). Current estimates for this size growth argue that massive galaxies may grow in size on average up to a factor of 3 for disk-like galaxies, while for spheroid-like objects this evolution reaches even a factor of 5 since $z=3$ (Buitrago et al. 2008).

In this paper we have shown that the effective radius of massive galaxies is altered by the star formation present, growing on average by $16 \pm 5\%$ from $z=3$ to $z=0$. This value is only $\sim 3-5\%$ of the total increase in the size of massive galaxies from observational studies (e.g. Buitrago et al 2008). This indicates that the star formation has a very minor contribution to the observable overall size evolution at $z<3$.

When we apply a simple model of stellar migration to the new stellar mass created via star formation to the present day we find that the size of these massive galaxies is influenced to a greater extent. The effective radius increases by $54 \pm 19\%$. This increase would represent $11-18\%$ of the total size evolution that massive galaxies undergo between $z>1$ and 0. This result shows that the effects of stellar mass added via star formation, and any subsequent stellar migration, plays a minor role in massive galaxy size evolution and only contributes roughly a tenth of the total size growth needed to explain the observed size evolution. This implies that other evolution mechanisms must also be at work to produce the remaining $\sim80\%$ of the observed size growth over cosmic time. From also examining the total size growth in the other models of evolution (see \S 5.4.1) we also find that the maximal size increase we can obtain can only produce $\sim54\%$ of the total observed size growth.

Recent studies have found that minor and major mergers have a large influence on the size evolution of massive galaxies. These mergers could explain the majority of the remaining $\sim 80\%$ of the observed size growth unaccounted for by the SF via increasing the total stellar mass of the galaxies (Bluck et al. 2011). Our results are consistent with this view that something other than SF produces the change in the sizes of massive galaxies.

\subsection{Structural Properties}

Recent studies have shown that the massive galaxy population at $z \ge 1.5$ is dominated by disk like galaxy morphologies with $n<2$ (e.g., Buitrago et al. 2011; Weinzirl et al. 2011). This is in contrast to the local universe where the massive galaxy population is almost entirely dominated by spheroids (e.g. Baldry et al. 2004; Conselice et al. 2006). This transformation is also seen through changes in the S\'{e}rsic index of these galaxies from a low value of $n$ at $z>1$ to a high value of $n$ at $z<1$.    

In this study we show that due to the star formation present within the massive galaxies at $z>1.5$ the S\'{e}rsic index has an insignificant change over cosmic time, $\Delta n = -0.9 \pm 0.9$. When we introduce the effects of stellar migration to the mass added via star formation the change in S\'{e}rsic index is again negligible over cosmic time with, $\Delta n = -1.1 \pm 1.3$. In the other methods of SF evolution we find that the change in $n$ is very similar. This implies that with both star formation and stellar migration the change to the S\'{e}rsic index is minimal. Also, this does not agree with observations of the general increase of $n$ over time. Therefore SF alone cannot account for the observed morphological change which appear to show that $n$ is increasing over time (e.g Buitrago et al. 2011).

\subsection{Spatial Location of Star Formation}

In this study we find that the structural properties of our massive galaxies remain largely unchanged after evolution via star formation. This unchanging $n$ shows that the location and magnitude of star formation within massive galaxies largely follows the observed initial stellar mass density profile. This is most pronounced in the case of the inner growth (IG) galaxies. In this class of galaxies the observed stellar mass profile is much smaller than the stellar mass profile added via star formation. Therefore, for this class of galaxy to retain its original S\'{e}rsic index the stellar mass produced via star formation over evolution to the present day would have to be produced in amounts which largely reflect the already present stellar density i.e. high density regions would have a higher star formation rates than lower density regions. This was also seen in other ways in Trujillo et al.(2007), Buitrago et al. (2008) and Cassata et al. (2010, 2011).

The measured $\sim16\%$ growth of the effective radii of our massive galaxies due to star formation alone, without any stellar migration, reveals that there is star formation located in the outer regions of our massive galaxies. This is most pronounced in the OG galaxies by definition. In these galaxies the surface stellar mass density of the inner region remains roughly constant over star formation evolution with the outer regions increasing in stellar mass density. Thus in our simulated star formation evolution the observed high redshift galaxy would become surrounded by an envelope of new stellar material over time. With the addition of stellar migration this effect becomes more pronounced with newly created stellar mass migrating outwards. Recent work examining the stellar mass density profiles of high redshift, $z>2$, and low redshift, $z=0$, massive galaxies has shown that the density in the core region of low redshift galaxies is comparable to the density of the compact high redshift galaxies (Hopkins et al. 2009; van Dokkum et al. 2010; Carrasco et al. 2010). The compact high redshift galaxies have become surrounded by an envelope of lower density material from $z>2$ to $0$. This is similar to what we find in the OG class of galaxies.

The models that we use in this study do not account for any new gas that can be accreted at later times, at $z<1.5$, and at early times at  $z > 3$ where we also do not observe our sample. This new gas and possible new star formation is likely to have a different radial distribution from the current in situ gas, with most of the new gas being at larger radii (Keres et al. 2005; Dekel et al. 2009). Therefore the distribution of star formation that we observe at high redshift is mostly likely the result of previous events of gas accretion (see Conselice et al. 2012).  However, not all the gas accreted may convert into stars immediately, and this gas may remain in the outer portions of these galaxies and may form into stars at an epoch later than our observations at $z < 1.5$, which in principle may increase the sizes of these systems at a later time, or alter their S\'{e}rsic
indices.   

\subsection{Model Limitations}

In this study we have taken a snapshot of our massive galaxy sample over 2 Gyr in time, and derived the resulting evolution based on a derived star formation model. Thus we do not take into account any post\--observation star formation events in our basic model. However this is likely a  fair assumption due to observations of the majority of massive galaxies at $z<1.4$ having old stellar populations and red colours (e.g. McCarthy et al. 2004; Daddi et al. 2005; Saracco et al. 2005; Bundy et al. 2006; Labb\'{e} et al. 2006; Conselice et al 2007; Mortlock et al. 2011; Gr\"{u}tzbauch et al. 2011). This would imply that the SF we observe at $z>1.5$ is the last major burst of SF in massive galaxies. The effect of new star formation events would increase the total amount of stellar mass added to the host galaxy. The galaxy's structural properties and size could also be affected by these events, depending on the location and magnitude of this star formation as discussed in the previous section. 

Conversely, we also do not take into account any feedback mechanisms that would negatively affect star formation rates. Examples of such processes are AGN and supernovae feedback. Massive galaxies can spend up to 1/3 of their lifetimes in an AGN phase (Hickox et al. 2009, Bluck et al. 2011). This phase introduces energy into the interstellar gas and can expel it from the host galaxy (Schawinski et al. 2006), or heat it such that it cannot cool. Also ongoing star formation results in the creation of many high mass stars which can lose mass during evolution and subsequently die in supernovae, thereby lowering the total stellar mass of the galaxy. When many supernovae are present in a short time the created shock waves introduce vast amounts of energy into interstellar gas. The gas can then can be heated or ejected from the host galaxy (e.g. Bertone et al. 2007). The result of these feedback mechanisms would be a reduction of the star formation rate, and the total stellar mass within the galaxy would be lower. This decreased amount of stellar mass added via SF would also result in the stellar mass added via star formation having a decreased effect on the total size growth and morphological change.

We also use a very simple models to describe the stellar migration that is limited to the extent of the $z_{850}$ band profiles. This means that we can not accurately measure how large values of stellar migration would affect the sizes and structural properties of our massive galaxies. However even though we cannot accurately measure the S\'{e}rsic index or the effective radius of the simulated galaxies with larger values of the stellar migration, we find that the stellar mass begins to be distributed evenly over all radii, with increasing amounts of stellar mass lost outside the confines of the simulation. The amount of stellar mass added via star formation moved by migration is constant for each galaxy but is distributed over wider areas for larger values of stellar migration. This results in the stellar mass density added via star formation to individual regions of the massive galaxies dropping to increasingly smaller values. This implies that with larger values of stellar migration, the stellar mass density added via star formation would have an increasingly smaller effect on the total stellar mass density profile. Therefore even if larger values of stellar migration could be simulated in this study the change in S\'{e}rsic index and effective radius after star formation evolution and migration would be negligible.

Stellar migration has also been found, in simulations, to be most affected by spiral arms in galaxies (Ro{\v s}kar et al. 2011). $73\%$ of the sample of massive galaxies have a low S\'{e}rsic index, $n < 2.5$, implying a disk\--like morphology. Within these galaxies we may assume therefore that stellar migration via disk features may take place, but this is far from certain. A few of the galaxies in our sample have a high S\'{e}rsic index, $n >2.5$, implying an early\--type morphology, and within these galaxies stellar migration is less understood. This does not imply that stellar migration does not take place in these galaxies but it must occur by other processes than those involving disks. Also, as stated in \S4.3 we cannot reliably distinguish disk-like galaxies in our sample using a S\'{e}rsic index cut because we cannot rule out that some of the galaxies with $n>2.5$ do not have spiral like features (e.g. Buitrago et al. 2011: Mortlock 2012, private communication).

\subsubsection{Evolutionary models}

In this paper we extrapolate the star formation evolution using a exponentially declining star formation model based on SED derived $\tau$ values. This value can be uncertain so we explore different models of evolution that the star formation could follow down to $z=0$. Firstly we do not investigate an exponentially increasing SFR evolution model because previous studies (e.g. Papovich et al. 2011) show that galaxies at $z<3$ are not well described by this SF history. Therefore we investigate the SF evolution models of: constant SFR to $z=0$, constant SFR to $z=1.5$, maximum valid tau and minimum valid tau.

\begin{itemize}

\item{Constant $\rm{SFR_0}$ to $z=0$: This model of evolution assumes that the massive galaxies we observe at $z>1.5$ have a very large reservoir of gas and can continue the observed SFR over the next ~10Gyr. This evolutionary method produces galaxies in the local universe with very high star formation rates compared to the galaxies we observe (e.g. Conselice et al 2007). This combined with the fact that over the course of their evolution these galaxies will have accumulated significant amounts of stellar mass with the average massive galaxy in this sample increasing its total stellar mass by $\sim1500\%$. This large amount of stellar mass added to the galaxies increases the value of $R_{e}$ by $80\pm20\%$. This is an increase of a factor of 5 over the derived tau model in effective radius growth, but still only $16-27\%$ of the observed size evolution. This model of evolution  is highly unlikely due to the many features of this model that we do not observe in the local universe, such as very large stellar mass growth leading to very massive galaxies with stellar masses over $10^{13}M_{\odot}$ (e.g. Brammer et al. 2011;  Conselice et al 2011; Mortlock et al. 2011, all find that the stellar mass growth at the massive end of the luminosity function is on the order of $200\%$ from $z>1.5$ to 0) and very high star formation rates of 100's of solar masses per year.}

\item{Constant $\rm{SFR_0}$ to $z=1.5$: This model of SF evolution is based on the observation that the majority of massive galaxies at $z<1.4$ have old stellar populations and red colours (e.g. Conselice et al 2007: Mortlock et al. 2011, Gr\"{u}tzbauch et al. 2011). This would imply that these galaxies have turned off their SF before $z=1.5$. To model this we employed a constant observed SFR until $z=1.5$ at which point the SFR is reduced to 0. In this evolution scenario the total stellar mass of the massive galaxies is increased by $126\pm20\%$. The effective radii in this model are increased on average by $37\pm19\%$. This is a factor of $\sim2.5$ larger then the increase from the derived tau model. This is still insignificant to the total observed size increase. This model has a very similar effect on the change in $n$, $\Delta n = -1.1\pm 1.1$, as the derived tau model.}

\item{Maximum valid tau to $z=0$: In this model of evolution we use the largest value of tau derived for our galaxy sample, $\tau = 2.71\times10^{9} \rm{yr}$. We apply this exponentially declining rate to all the galaxies in the sample. In this scenario we obtain a large average increase in total stellar mass of the sample of $377\pm172\%$. The change in the effective radii of this model is on average $R_{e} = 57\pm33\%$ a factor of $\sim3.8$ larger than the derived tau model of evolution. This increase in effective radius is still only $\sim11-19\%$ of the observed size evolution. The change in $n$ for this model, $\Delta n = -1.5\pm1.7$ is similar to change for the derived tau model.}

\item{Minimum valid tau to $z=0$: This model is similar to the previous model. Except that the minimum valid tau, $\tau = 1.2\times10^{8} \rm{yr}$, is used to extrapolate the SF. This would give the shortest time scale that the SF would occur. In this model the average galaxy in the sample increases its stellar mass by only $\sim17\%$. This very small increase in mass is accompanied by an equally small change in $R_{e}$, average $\Delta R_{e} = 3\pm1\%$, and $n$, $\Delta n = -0.4 \pm 0.6$. }
\end{itemize}

From this investigation of different models of SF evolution to $z=0$ we find that the value in the increase of the effective radii of the massive galaxies can at no point fully explain the total observed size increase. The valid models of SF evolution that we applied can only produce a factor of $\sim 3.8$ times larger than the size increased we obtained from using the derived tau model at maximum. The change in S\'{e}rsic index in all the models are within the error consistent with the answer obtained from the derived tau model used in this paper.

\subsubsection{Dust Gradients}

In this paper we assume that the dust obscuration is constant across the radius of individual galaxies. From studies of local and distant studies this may not be the case. Colour gradients in the local universe have been shown to correspond to age and dust gradients (e.g. Boquien et al. 2011; Smith et al. 2012). 

We apply a dust gradient to our sample of massive galaxies that allows the attenuation due to dust to vary within the given error across each galaxy. This is done in two ways. A positive dust gradient with higher attenuation towards the outer regions of the galaxy, and a negative dust gradient with higher dust attenuation towards the central regions of the galaxy. 

In the positive gradient case we find that the average increase in the effective radius was $68\pm36\%$ larger than the original measured effective radius. This is a factor of $\sim4.5$ larger change than the growth in $R_{e}$ we obtain from using a radially constant dust correction. From this gradient the change in $n$ is largely the same as before but with a much larger scatter, $\Delta n = -0.9\pm2.0$. The positive gradient could contribute a maximum of $\sim23\%$ to the $300-500\%$ size growth.

In the negative gradient case we find that the average increase in $R_{e}$ is minimal, $\Delta R_{e}=7\pm3\%$. This small increase in the effective radius is accompanied by a change in $n$ that is very similar to most other cases, $\Delta n = -1.0\pm1.0$. This negative gradient case would seem to produce a very small increase in the effective radii of our sample and only contribute a maximum of $\sim2\%$ to the total observed size growth.

Neither of the gradient cases that we applied to the sample are able to fully explain the observed size growth or observed change in S\'{e}rsic index.

\section{Summary}

We investigate the resolved star formation properties of a sample of 45 massive galaxies ($M_{*}>10^{11}M_{\odot}$) within a redshift range of $1.5 \le z \le 3$ detected in the GOODS NICMOS Survey, a HST $H_{160}$\--band imaging survey. We derive the star formation rate as a function of radius using rest frame UV data from deep $z_{850}$ ACS imaging. The star formation present at high redshift is then extrapolated to $z=0$, and we examine the stellar mass produced in individual regions within each galaxy. We also construct new stellar mass profiles of the in situ stellar mass at high redshift from S\'{e}rsic fits to rest-frame optical, $H_{160}$\--band, data. We combine the two stellar mass profiles to produce an evolved stellar mass profile. 

We then fit a new S\'{e}rsic profile to the evolved profile, from which we examine what effect the resulting stellar mass distribution added via star formation has on the structure and size of each individual galaxy. In summary:

\begin{itemize}

\item{We find three different profiles of star formation within the massive galaxies in this sample, Non\--significant Star Formation Growth (NG), Outer Star Formation Growth (OG) and Inner Star Formation Growth (IG) (see \S3.3). With most of this sample of massive galaxies falling in to NG class using the derived tau model of evolution.}

\item{We find that the star formation we observe at high redshift, and its effects on galaxy sizes, is not large enough to fully explain the observed size evolution of effective radius of $\sim300\--500\%$. Star formation alone alone can only produce an increase in effective radius on the order of $\sim16\%$ over the whole sample. This value can vary as much a a factor of 4.5 by using different evolution mechanisms but is always insufficient to fully explain the observations.} 

\item{We find that over the whole sample of massive galaxies the stellar mass added via star formation has a slight effect on the S\'{e}rsic index of the evolved galaxy profile such that they decrease. This indicates that the star formation within these galaxies follows the same radial distribution as the original stellar mass profile. This also implies that star formation evolution has a minimal effect on structural evolution between $z\sim3$ and the present day.}

\item{The increase in effective radius can be enhanced by adding in the effects of stellar migration to the stellar mass created via star formation. This increases the total effective radius growth to $\sim55\%$, which is still however much smaller than the total observed size increase.}

\end{itemize}

We conclude that due to the lack of sufficient size growth and S\'{e}rsic evolution by star formation and stellar migration other mechanisms must contribute a large proportion to account for the observed structural evolution from $z>1$ to the present day. Recent studies by Bluck et al. (2011) have found that minor and major mergers have a large influence on the size of massive galaxies possibly contributing the remaining $80\%$ of size growth needed to explain the observed trends. Large surveys such as CANDELS and future telescopes such as JWST and E-ELT will provide the quality of data that is required to explore the star formation locations of lower mass galaxies and probe resolved star formation at higher redshifts for similarly massive galaxies.

\section*{Acknowledgements}
We thank the GNS team for their support and work on this survey and assistance on this paper, especially Amanda Bauer, Asa Bluck, and Ruth Gr\"{u}tzbauch. We acknowledge funding from the STFC and the Leverhulme trust for supporting this work. We would also like to thank the referee for their constructive report and useful comments.

\label{lastpage}
\end{document}